\documentclass[twocolumn,twocolappendix]{aastex631}

\usepackage{amsmath}
\usepackage{empheq}
\usepackage{mathrsfs}
\usepackage{textcomp}
\usepackage{enumitem}   
\usepackage{gensymb}

\usepackage{graphicx}
\usepackage[caption=false]{subfig}
\usepackage{multirow}
\usepackage{longtable}
\usepackage[flushleft]{threeparttable}
\usepackage{booktabs} % To thicken table lines
\usepackage{CJK}
\usepackage{placeins} 
\usepackage{float} 

\hypersetup{colorlinks, linkcolor={blue}, citecolor={blue}, urlcolor={blue}}

\definecolor{DarkOrange}{RGB}{204, 85, 0}
\definecolor{LincolnGreen}{RGB}{17, 102, 0}
\definecolor{srcgold}{RGB}{204, 153, 0}
\definecolor{srcred}{RGB}{214, 39, 40}
\definecolor{srcblue}{RGB}{31, 119, 180}
\definecolor{srcgreen}{RGB}{44, 160, 44}
\definecolor{srcpurple}{RGB}{148, 103, 189}

\usepackage{xspace}

\newcommand\nicer{\textit{NICER}\xspace}
\newcommand\chandra{\textit{Chandra}\xspace}
\newcommand\swift{\textit{Swift}\xspace}
\newcommand\xmm{\textit{XMM-Newton}\xspace}

\newcommand\srge{\textit{SRG}/eROSITA\xspace}

\newcommand\Rin{$R_{\rm in}$\xspace}

\newcommand\Tp{$T_{p}$\xspace}

\newcommand\Rout{$R_{\rm out}$\xspace}

\def\msun{M_\odot}

\begin{document}
\setlength{\abovedisplayskip}{6pt plus 2pt minus 2pt}\setlength{\belowdisplayskip}{6pt plus 2pt minus 2pt}\setlength{\abovedisplayshortskip}{2pt}\setlength{\belowdisplayshortskip}{2pt}

\pagenumbering{arabic}

\title{Compact Accretion Disks in the Aftermath of Tidal Disruption Events:\\
Parameter Inference from Joint X-ray Spectra and UV/Optical Photometry Fitting}

\author[0000-0002-5063-0751]{M. Guolo}
\affiliation{Bloomberg Center for Physics and Astronomy, Johns Hopkins University, 3400 N. Charles St., Baltimore, MD 21218, USA}

\author{A. Mummery}
\affiliation{School of Natural Sciences, Institute for Advanced Study, 1 Einstein Drive, Princeton, NJ 08540, USA }

\author{S. van Velzen}
\affiliation{Leiden Observatory, Leiden University, Postbus 9513, 2300 RA Leiden, NL }

\author[0000-0003-3703-5154]{S. Gezari}
\affiliation{Department of Astronomy, University of Maryland, College Park, MD, 20742-2421, USA}

\author[0000-0002-2555-3192]{M. Nicholl}
\affiliation{Astrophysics Research Centre, School of Mathematics and Physics, Queens University Belfast, Belfast BT7 1NN, UK}

\author[0000-0001-6747-8509]{Y. Yao}
\affiliation{Miller Institute for Basic Research in Science, 206B Stanley Hall, Berkeley, CA 94720, USA}
\affiliation{Department of Astronomy, University of California, Berkeley, CA 94720-3411, USA}

\author[0000-0003-2495-8670]{M. Karmen}
\affiliation{Bloomberg Center for Physics and Astronomy, Johns Hopkins University, 3400 N. Charles St., Baltimore, MD 21218, USA}

\author[0009-0007-8764-9062]{Y. Ajay}
\affiliation{Bloomberg Center for Physics and Astronomy, Johns Hopkins University, 3400 N. Charles St., Baltimore, MD 21218, USA}

\author[0000-0002-4043-9400]{T. Wevers}
\affiliation{Astrophysics \& Space Institute, Schmidt Sciences, New York, NY 10011, USA}

\author[0000-0002-2249-0595]{N. LeBaron}
\affiliation{Department of Astronomy, University of California, Berkeley, CA 94720-3411, USA}
\affiliation{Berkeley Center for Multi-messenger Research on Astrophysical Transients and Outreach (Multi-RAPTOR), University of California, Berkeley, CA 94720-3411, USA}

\author[0000-0002-2249-0595]{R. Chornock}
\affiliation{Department of Astronomy, University of California, Berkeley, CA 94720-3411, USA}
\affiliation{Berkeley Center for Multi-messenger Research on Astrophysical Transients and Outreach (Multi-RAPTOR), University of California, Berkeley, CA 94720-3411, USA}
%%%%% ABSTRACT is essential a summary of the paper %%%%% 
\begin{abstract}
We present a multi-wavelength analysis of 16 tidal disruption events (TDEs)—including an off-nuclear event in an ultra-compact dwarf galaxy—selected for having available thermal X-ray spectra during their late-time UV/optical plateau phase. We show that at these stages, the full spectral energy distribution—X-ray spectra and UV/optical photometry—is well described by a compact, yet standard accretion disk, the same disk which powers the X-rays at all times. By fitting up to three epochs per source with a fully relativistic disk model, we show that many system properties can be reliably recovered, including importantly the black hole mass ($M_{\bullet}$). These accretion-based $M_{\bullet}$ values, spanning three orders of magnitude in the sample, are consistent with galactic scaling relations, but are significantly more precise (typical uncertainty of $ \sim\pm  0.2$ dex) and physically motivated. Expected accretion scaling relations ($L_{\rm Bol}^{\rm disk}/ L_{\rm Edd} \propto T_p^4 \propto M_{\bullet}^{-1}$), TDE-specific physics correlations ($L_{\rm{plat}} \propto M_{\bullet}^{2/3}$ and $R_{\rm out}/r_g \propto M_{\bullet}^{-2/3}$) and black hole–host galaxy correlations ($M_{\bullet}$-$M_{\rm gal}$ and $M_{\bullet}$-$\sigma_{\rm *}$) naturally emerge from the data and, for the first time, are self-consistently extended into the intermediate-mass (IMBH, $M_{\bullet} < 10^{5} M_{\odot}$) regime. We discuss the implications of these results for TDE physics and modeling. We also review and discuss different methods for $M_{\bullet}$ inference in TDEs, and show that approaches based on physical models of the early-time UV/optical emission are not able to recover black hole–host galaxy correlations, implying $M_{\bullet}$ inferred by these methods may not be reliable.
\end{abstract}

\vspace{1em}

\section{Introduction}
Tidal disruption events (TDEs) occur when a star is scattered onto a near-radial orbit around a massive black hole and approaches within the radius at which the tidal forces exerted by the black hole exceed the star's own self-gravity. The star is then disrupted, forming a stream of debris whose subsequent evolution ultimately powers a bright flare observed across the electromagnetic spectrum \citep{Rees1988,Gezari_2021}. These rare transients provide clean laboratories for fundamental studies of black hole accretion and turbulence \citep[e.g.,][]{Cannizzo1990,Balbus2018,Mummery2025_var}, as well as unique probes of the demographics of otherwise quiescent black holes in the local Universe \citep{Frank1976,Stone2016,Yao2023,Mummery_vanVelzen2024,Yao2025}.  

Importantly, TDEs are particularly sensitive to black holes at the low-mass end of the black hole mass function, a regime in which TDE observations will prove crucial for understanding black hole growth and seed formation \citep{Kormendy2013,Shankar2016,Greene2020}. If they can be well understood, TDEs will also inform broader questions in galaxy evolution, such as the black hole occupation fraction in dwarf galaxies \citep{Silk2017,Bradford2018,Zou2025} and the dynamical evolution of dense stellar systems \citep{MillerHamilton2002,Stone2016}. This is because TDEs offer a (perhaps the most) promising avenue through which to study accreting intermediate-mass black holes (IMBHs; $10^3$--$10^5\,M_\odot$), a population that remains poorly constrained observationally \citep{Greene2020}, with only a handful of strong candidates reported \citep[e.g.,][]{Farrell2009,Soria2017,Lin2018,Jin2025}. Confirming the masses of TDEs associated with IMBH candidates can shed light on their demographics, formation pathways, and the mass-scale (in)dependence of accretion physics, bridging the gap between stellar-mass and supermassive black holes. This is particularly important for off-nuclear IMBHs, which may otherwise remain undetectable, as gas accretion onto such systems is expected to be negligible in the local Universe \citep{ricarte2021}.  

Realizing the potential of TDEs as probes of black hole demographics depends critically on two requirements:  
(i) a robust understanding of the physical processes powering their multi-wavelength emission; and  
(ii) methods capable of \textit{reliably and consistently} recovering key system parameters, especially black hole mass ($M_{\bullet}$).  

While TDEs were first identified as high-energy (X-ray) transients powered by the accretion of disrupted stellar debris \citep[e.g.,][]{Bade1996,Komossa_99}, wide-field optical time-domain surveys have since become the dominant discovery channel. A ``typical''\footnote{This classification reflects observational capabilities rather than intrinsic properties: optical discoveries dominate numerically because of survey efficiency, not because of an intrinsically higher rate. In fact, X-ray emission is physically better understood than the early-time optical flare, and the intrinsic rates in the two channels are consistent \citep[e.g.,][]{Yao2023,Sazonov2021}.} TDE is now usually discovered as a luminous optical flare \citep[e.g.,][]{van_Velzen_21,Yao2023}, which may or may not coincide with rapid X-ray brightening \citep[e.g.,][]{Guolo2024}. Optical luminosities peak at $L \sim 10^{42}$--$10^{44}\,\mathrm{erg\,s^{-1}}$ and decline on month-long timescales \citep[e.g.,][]{Hammerstein2023a,Yao2023}. On longer timescales (years), a pronounced plateau dominates their UV/optical light curves \citep{vanVelzen2019,Mummery2020}; it is most easily detected in the UV but is also observed in the optical \citep{Mummery2024}. This plateau marks the transition from the poorly understood early-time flare to a long-lived, disk-dominated phase \citep{Cannizzo1990,Mummery2020,Mummery2024}. %For instance, ASASSN-14li remains significantly detected more than a decade post-disruption.  

The physical origin of the early-time UV/optical flare observed from these optically selected TDEs remains uncertain---although it is known that it cannot arise directly from a compact disk---and there is currently no consensus in the community. The two common interpretations are emission from stream--stream shocks \citep[e.g.,][]{Ryu2023} or the reprocessing of high-energy radiation by a wind/envelope \citep[e.g.,][]{Dai2018}; see \citet{Roth2016} for a review. Parameter inference using analytical prescriptions based on these models \citep[e.g.,][]{Ryu2020,Mockler2019}, particularly inference of the black hole mass $M_{\bullet}$ (which can be independently probed using galaxy properties), was initially considered successful but relied on individual sources or very small samples. Later population-level studies showed that these models have not yet succeeded in reproducing either (i) established correlations between black hole mass and host-galaxy properties \citep[e.g.,][]{Ramsden2022,Hammerstein2023a} or (ii) empirical relations between observables (e.g., peak UV/optical luminosity) and black hole mass \citep{Mummery2024,Mummery_vanVelzen2024}, casting doubt on their accuracy in inferring physical parameters.

\begin{figure*}
    \centering
    \includegraphics[width=0.48\linewidth]{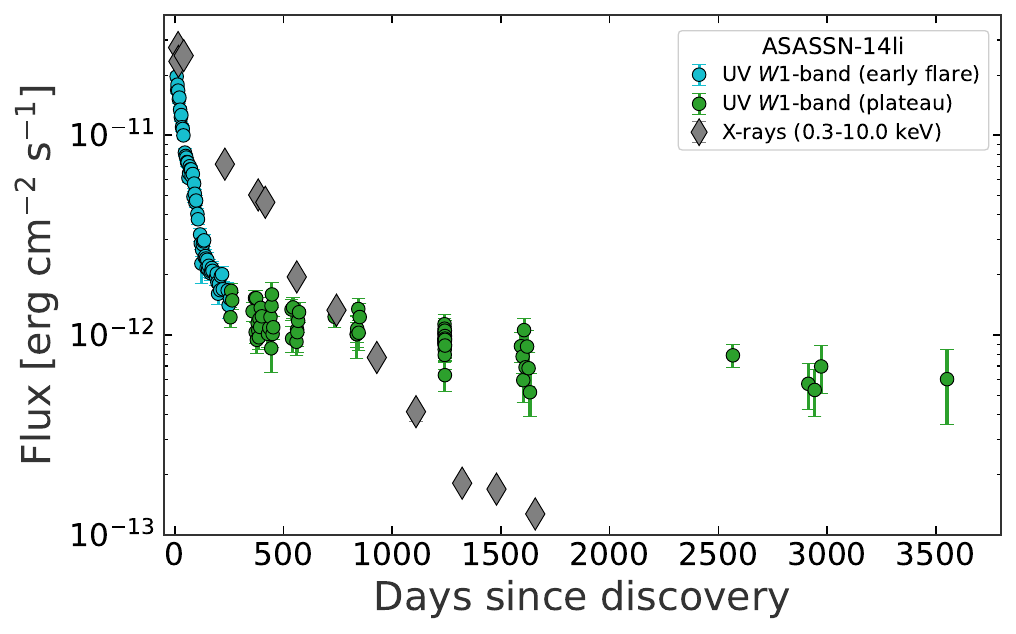}
     \includegraphics[width=0.49\linewidth]{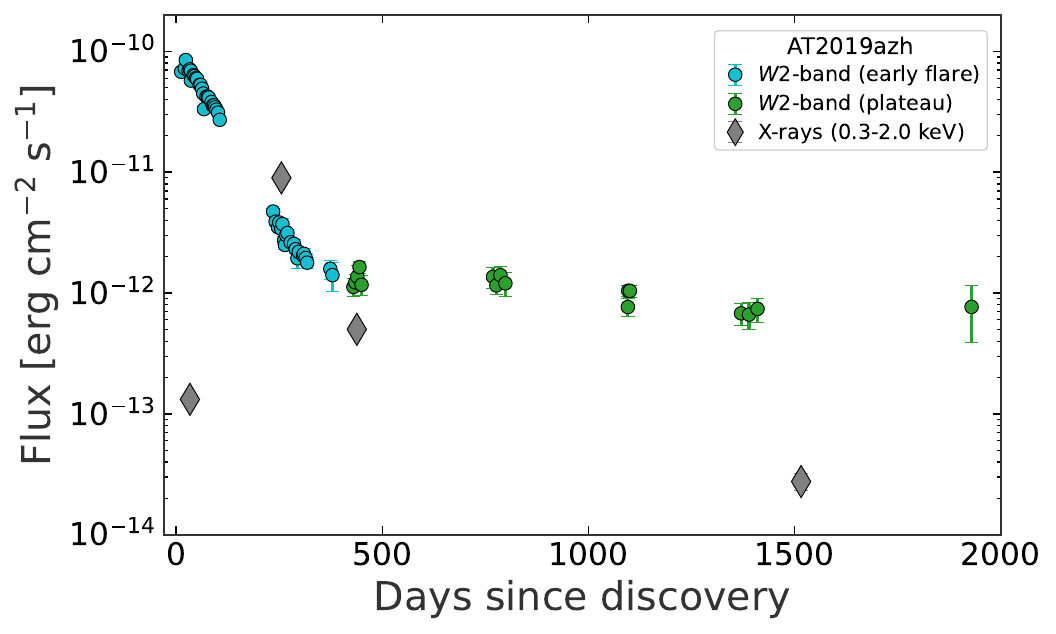}
    \caption{Examples of host-subtracted light curves of well observed TDEs, ASASSN-14li (left) and AT2019azh (right). Each panel shows a single UV band, $W1$ for ASASSN-14li and $W2$ for AT2019azh, together with the X-ray light curve. Cyan points mark the early-time flare and green points the plateau phase. The transition is taken from the exponential-plus-constant fit, at the epoch where the exponential component falls below the typical flux uncertainty. This two-phase UV evolution is ubiquitous to TDEs, independently of whether the X-rays are prompt or delayed. Fluxes shown are not corrected for any absorption or extinction.}
    \label{fig:lcs}
\end{figure*}

By contrast, \citet{Mummery2024} recently showed that the observed luminosity reached during the late-time ``plateau'' phase --  originating physically from the simultaneous cooling and expansion of the disk -- correlates strongly with host galaxy mass (used as a proxy for black hole mass), with the  observed correlation emerging naturally from time-dependent accretion disk theory \citep{Mummery2020, Mummery2024}. Meanwhile, X-ray emission has been firmly established as originating from the innermost regions of a newly formed accretion disk \citep{Mummery2020,wen22_disk_spectrum}. Observed as very soft thermal radiation \citep{Auchettl2017,Saxton_20,Guolo2024}, this X-ray emission should, in principle, encode information about the central black hole. Attempts to use this thermal X-ray emission as a mass diagnostic have had some success---both spectrally \citep[e.g.,][]{Wen2020,Wen2021,Mummery2023,Jin2025} and in the time-domain regime \citep[e.g.,][]{Mummery2020,Mummery2021b}---but have remained limited by parameter degeneracies, sparse temporal coverage, and the small number of TDEs with high-quality X-ray data.

These findings naturally suggest that the X-ray and UV/optical “plateau’’ emission arise from the same accretion disk, in agreement with the luminosity function results of \citet{Mummery_vanVelzen2024}. However, no detailed spectral analysis of the TDE population has been performed with the aim of testing this key prediction of TDE accretion theory. While some individual sources have been analyzed in this spirit, this has typically been limited to integrated luminosities in the time-domain regime  \citep[i.e., optical/X-ray light curve fitting, e.g.,][]{Mummery2020,Goodwin2022,Goodwin2024,Nicholl2024, Chakraborty25,Mummery2024fitted}, which does not utilize the full spectral information available in multi-wavelength multi-epoch data. If the optical--X-ray TDE accretion disk paradigm is correct, it implies that X-ray \textit{spectra} and UV/optical photometry during the ``plateau'' phase should be simultaneously described by a single disk model. 

Until recently, testing this paradigm directly on the population level was not possible: models with the expected characteristics of TDE disks—compact (i.e., small radial extent), thermal, and typically lacking persistent coronal emission—were not available in forms suitable for fitting simultaneous X-ray spectra  (e.g. via \texttt{XSPEC}; \citealt{Arnaud1996}) and UV/optical photometry. This limitation was overcome by \citet{Guolo_Mummery2025}, who implemented such models. Early applications to individual sources \citep{Guolo_Mummery2025,Wevers2025,Guolo2025} demonstrated their feasibility, paving the way for broader ensemble studies.

The aim of this paper is to show that joint fitting of X-ray spectra and plateau-phase UV/optical photometry with a single disk model is feasible across a diverse sample of TDEs, and that such fits provide compelling evidence for a common disk origin of both (plateau UV/optical––X-ray) emission components. We demonstrate that this approach yields robust inferences of key system parameters—including (for the first time on the population level) black hole masses, spins, and disk sizes. With these spectral fits we demonstrate that both general accretion disk scaling relations and also those which are specific to TDE disks naturally emerge. Moreover, we show that host–black hole correlations are independently recovered, consistent with other accretion-based relations \citep[e.g., $L_{\rm plat}$--$M_{\bullet}$;][]{Mummery2024}, but at substantially higher precision. This gain in precision arises from the data --  multi-wavelength fits simultaneously probe the inner and outer disk, thereby breaking degeneracies and reducing intrinsic scatter.

The structure of this paper is as follows: in \S\ref{sec:data}, we describe our sample and data. Our adopted model and fitting procedures are described \S\ref{sec:model}. In \S\ref{sec:results} we present our results, which are discussed in broader context in \S\ref{sec:discussion}. Our conclusions are presented in \S\ref{sec:conclusions}. We adopt a standard $\Lambda$CDM cosmology with a Hubble constant $H_0=73\,{\rm km\,s^{-1}\,Mpc^{-1}}$ \citep{Riess2022}. A Bayesian statistics framework is considered throughout the paper, converging posterior for inferred parameters are reported as medians and the uncertainties correspond to the  68\% credible intervals, while upper (lower) limits on one side converged posteriors are 90\% (10\%) credible intervals.

\begin{deluxetable*}{lccccCCc}\label{tab:sample}
\tablecaption{Sample Information \label{tab:props}}
\tablehead{
\colhead{Source} & \colhead{$t_0$} & \colhead{$z$} &
\colhead{$E(B-V)_G$} & \colhead{$N_\mathrm{H,G}$} & \colhead{$\log_{10}(M_{\rm gal})$} & \colhead{$\sigma_{\star}$ } & \colhead{Reference} \\
\colhead{} & \colhead{(MJD)} & \colhead{} & \colhead{} & \colhead{(cm$^{-2}$)} & \colhead{($M_{\odot}$)} & \colhead{(km s$^{-1}$)} & \colhead{}
}
\startdata
AT2019qiz & 58536 & 0.015 & 0.09 & $6.3 \times 10^{20}$ & 10.0 \pm 0.2 & 72 \pm 2 & (1, 2, 3) \\
GSN\,069 & 55391 & 0.018 & 0.02 & $2.3 \times 10^{20}$ & 9.8 \pm 0.1 & 63 \pm 4  & (4, 5, 6) \\
AT2021ehb & 59276 & 0.018 & 0.12 & $9.9 \times 10^{20}$ & 10.2 \pm 0.1 & 93 \pm 5 & (7, 8) \\
ASASSN-14li & 56983 & 0.021 & 0.02 & $1.9 \times 10^{20}$ & 9.7 \pm 0.2 & 81 \pm 2 & (9, 10, 11) \\
AT2019azh & 58533 & 0.022 & 0.04 & $4.1 \times 10^{20}$ & 9.9 \pm 0.1 & 68 \pm 2 & (12, 13) \\
AT2022dsb & 59627 & 0.023 & 0.19 & $1.1 \times 10^{21}$ & 10.6 \pm 0.3 & 84 \pm 4 & (14, 15, this work) \\
AT2022lri & 59665 & 0.032 & 0.015 & $1.6 \times 10^{20}$ & 9.6 \pm 0.1 & 33 \pm 2 & (16, 17) \\
ASASSN-15oi & 57248 & 0.048 & 0.06 & $4.8 \times 10^{20}$ & 10.1 \pm 0.1 & 61 \pm 7 & (18, 19, 20) \\
AT2018cqh & 58283 & 0.048 & 0.02 & $2.4 \times 10^{20}$ & 9.5 \pm 0.1 & 53 \pm 10 & (21, 22) \\
AT2019dsg & 58582 & 0.051 & 0.08 & $6.6 \times 10^{20}$ & 10.5 \pm 0.1 & 87 \pm 4 & (23, 24) \\
3XMM J2150-05 & 53681 & 0.055 & 0.03 & $2.8 \times 10^{20}$ & 7.3 \pm 0.4 & ... & (25, 26) \\
AT2023cvb & 60016 & 0.071 & 0.19 & $7.5 \times 10^{20}$ & 10.5 \pm 0.2 &  79 \pm 5 & (27, this work) \\
AT2024aepd & 60646 & 0.084 & 0.04 & $3.4 \times 10^{20}$ & 10.1 \pm 0.1 & 79 \pm 7 & (28, 29) \\
AT2019vcb & 58803 & 0.089 & 0.015 & $1.5 \times 10^{20}$ & 9.7 \pm 0.1 & 42 \pm 8 & (30, 31, 32, this work) \\
AT2020ksf & 58951 & 0.092 & 0.04 & $3.6 \times 10^{20}$ & 9.9 \pm 0.1 & 56 \pm 2 & (33, 34) \\
AT2020wjw (eRA J2344‒35) & 59090 & 0.10 &  0.013 &  $1.1 \times 10^{20}$ & 10.5 \pm 0.1 &  146 \pm 25 & (35, 36, this work) \\
\enddata
\tablecomments{$t_0$ corresponds to the first detection of the TDE, in any wavelength.
Galaxy extinction color-excess $E(B-V)_G$ values are from \citet{Schlafly_2011}, and Galactic hydrogen-equivalent column density $N_\mathrm{H,G}$ from the HI4PI survey \citep{HI4PI2016}.
(1, 2, 3) \citet{Siebert2019,Nicholl2020,Nicholl2024};
(4, 5, 6) \citet{Saxton2011,Guolo2025,Guolo2025b};
(7, 8) \citet{Gezari2021,Yao2022};
(9, 10, 11) \citet{Jose2014,Miller2015,Holoien2016_14li};
(12, 13) \citet{Hinkle2021,van_Velzen_21};
(14, 15) \citet{Fulton2022,Malyali2024};
(16, 17) \citet{Yao2022,Yao2024_22lri};
(18, 19, 20) \citet{Brimacombe2015,Holoien2016_15oi,Gezari2017};
(21, 22) \citet{Saxton2011,Guolo2025};
(23, 24) \citet{Short2019,Cannizzaro2021};
(25, 26) \citet{Lin2018,Lin2020};
(27) \citet{Yao2022};
(28, 29) \citet{Charalampopoulos2025,WU2026};
(30, 31, 32) \citet{Dahiwale2020,Guolo2024,Bykov2024};
(33, 34) \citet{Gilfanov2020,Wevers2024};
(35, 36) \citet{Homan2023,Baldini2026}.
}
\end{deluxetable*}

\begin{deluxetable*}{lcCcccc}

\tablecaption{SED Fitting Data Summary \label{tab:SED}}
\tabletypesize{\scriptsize}
\tablehead{
\colhead{Source} & \colhead{Epoch} & \colhead{$\Delta t$} & 
\multicolumn{2}{c}{X-ray Data} & \multicolumn{2}{c}{UV/optical Data} \\
\colhead{} & \colhead{} & \colhead{(days)} & \colhead{Mission/Instrument} & \colhead{Obs-ID} & \colhead{Telescope/Observatory} & \colhead{Bands}
}
\startdata
AT2019qiz & E1 & 815\pm115 & \swift/XRT & 00012012043 & ZTF, \swift/UVOT & $r, g, W2$  \\
  & E2 & 1970\pm 180 & \xmm/PN& 0942560101 & ZTF, \swift/UVOT & $r, g,  W2$  \\
\hline
 & E1 & 141\pm 1 & \xmm/PN& 0740960101 & \nodata  & \nodata  \\
GSN\,069 & E2 & 1160\pm 10 & \xmm/PN & 0657820101 &{\it HST}/STIS  & G140L + G230L \\ 
& E3 & 3080\pm 10 & \xmm/PN& 0823680101 & {\it HST}/STIS   & G140L + G230L  \\
\hline
AT2021ehb & E1 & 205\pm 25 & \swift/XRT & 014217011-8 & ZTF, \swift/UVOT & $r, g, W1, M2, W2$  \\
& E2 & 550\pm 30 &\xmm/PN& 0902760101  & ZTF, \swift/UVOT  & $r, g, W1, M2, W2$  \\
\hline
 & E1 & 15\pm 1 & \xmm/PN& 0694651201 & \nodata  & \nodata  \\
ASASSN-14li & E2 & 376\pm 8 & \xmm/PN& 0770980501 &\swift/UVOT & $ U, W1, M2, W2$  \\
& E3 & 1281\pm 43 & \xmm/PN& 0770981001 & \swift/UVOT & $ W1, M2, W2$  \\
\hline
 & E1 & 255\pm 1 & \xmm/PN& 0823810401 & \nodata  & \nodata  \\
AT2019azh & E2 & 410\pm 50 & \xmm/PN& 0842592601 &ZTF, \swift/UVOT & $r, W1, M2, W2$ \\
& E3 & 1435\pm 120 & \xmm/PN& 0902761101 & ZTF,\swift/UVOT & $r, W1, M2, W2$ \\
\hline
 AT2022dsb & E1 & 1245\pm 85 & \swift/XRT& 00015054021-7 & ZTF, \swift/UVOT & $r, g, W1, M2, W2$  \\
\hline
 & E1 & 230\pm 1 & \swift/XRT& 00015378004 & \nodata  & \nodata  \\
AT2022lri & E2 & 670\pm 50 & \xmm/PN& 0932390701 & \swift/UVOT & $ U, W1, M2, W2$  \\
& E3 & 965 \pm 70 & \xmm/PN& 0940800101 & \swift/UVOT & $ r, W1, M2, W2$  \\
\hline
 ASASSN-15oi & E1 & 292\pm 62 & \xmm/PN& 0722160701 & \swift/UVOT & $  W1, M2, W2$  \\
& E2 & 1425\pm 50 & \swift/XRT& 00095141001-11 &\swift/UVOT & $  W1, M2, W2$  \\
\hline
 AT2018cqh & E1 & 2442\pm 142 &\xmm/PN & 0954191001 &\swift/UVOT & $ W1, M2, W2$  \\
 \hline
  & E1 & 42\pm 1 & \nicer/XTI& 2200680101 & \nodata  & \nodata  \\
AT2019dsg & E2 & 57\pm 1 & \nicer/XTI&  2200680108 &\nodata  & \nodata  \\ 
& E3 & 540\pm 60 & \xmm/PN& 0842591901 & ZTF, \swift/UVOT   & $r, g, W1, M2, W2$   \\
\hline
3XMM J2150-05 & E1 & 179\pm 1 &\xmm/PN& 0404190101&  \nodata  & \nodata \\
& E2 & 4000 \pm 650 &\chandra/ACIS & 17862  &  CFHT, {\it HST}/WFC3 & $F775W, r, g,$  \\
\hline
AT2023cvb & E1 & 810\pm 40 &\xmm/PN& 0942540801  & ZTF, \swift/UVOT  & $r,  W1, M2, W2$  \\
\hline
AT2024aepd & E1 & 60 \pm 20 &\swift/XRT &  00019019001-15 &  \nodata  & \nodata \\
& E2 & 415 \pm 15  & \xmm/PN & 0942541901  & \swift/UVOT  & $g, M2, W2$  \\
\hline
AT2019vcb & E1 & 188\pm 1 &\xmm/PN & 0871190301 &  \nodata  & \nodata \\
& E2 & 962\pm 2 &\xmm/PN& 0882591401  &\swift/UVOT  & $W2$  \\
\hline
AT2020ksf & E2 & 242 \pm 2 &\nicer/XTI & 3639010101-501 &  \nodata  & \nodata \\
& E2 & 780\pm 10 &\xmm/PN& 0882591401  &\swift/UVOT  & $M2, W2$  \\
\hline
AT2020wjw & E1 & 110 \pm 1 & \xmm/PN & 0862770101   &  \nodata  & \nodata \\
& E2 & 1100\pm 30 &\swift/XRT &  00013946021-4 & \swift/UVOT  & $W1, M2, W2$  \\
\enddata
\tablecomments{$\Delta t$ represents the time since the first detection of the source at any wavelength. The quoted ranges define the intervals over which data are combined to form a single SED epoch, and the central value is taken as the midpoint between the earliest and latest observations included in that epoch.}
\end{deluxetable*}

\section{Sample and Data}\label{sec:data}

\begin{figure*}
\centering
\def\gridline#1{\vskip1pt\hbox to\hsize{#1}\vskip1pt}
% ============================
% Flux panels (top block)
% ============================
\gridline{\fig{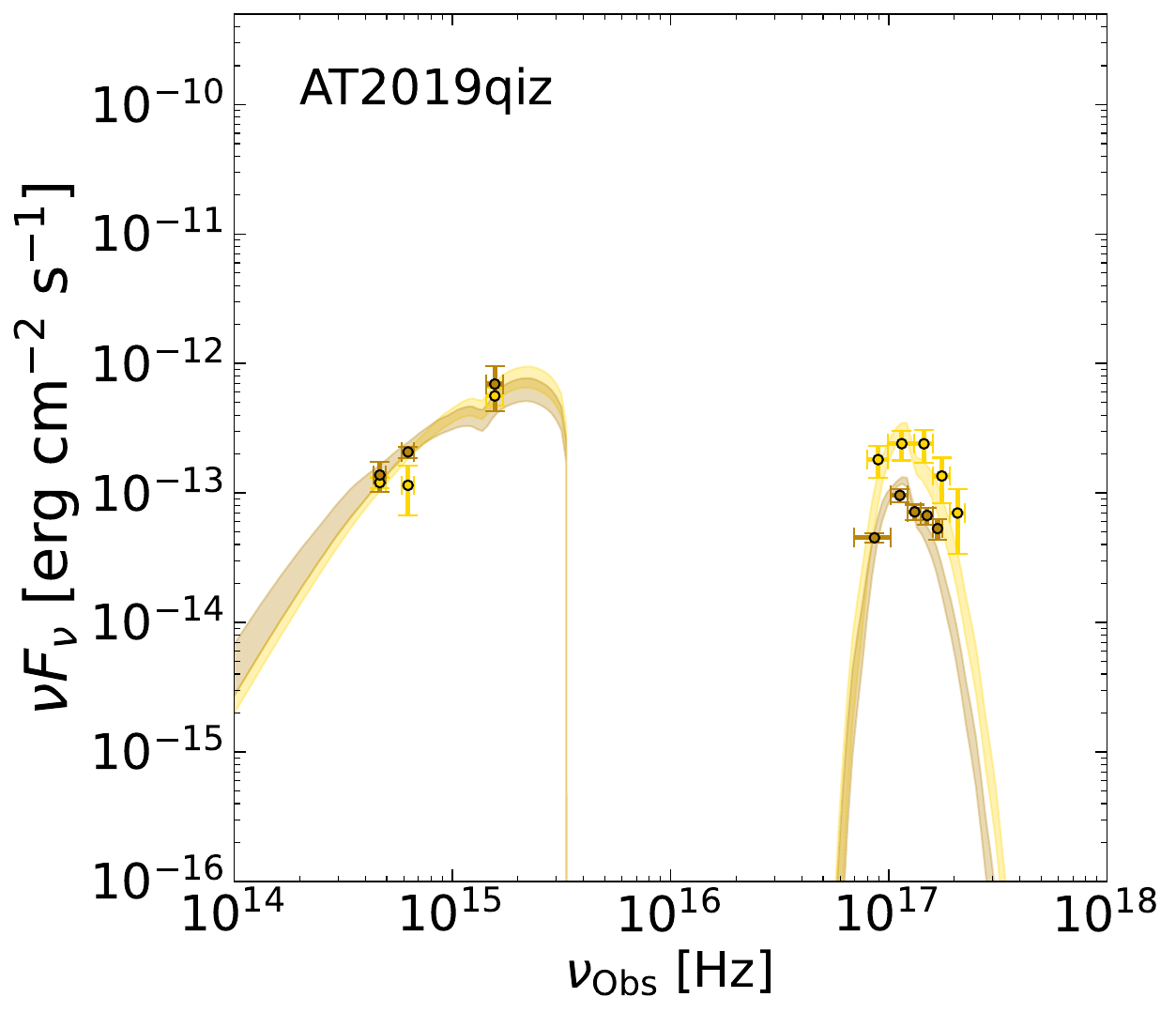}{0.245\textwidth}{}
          \fig{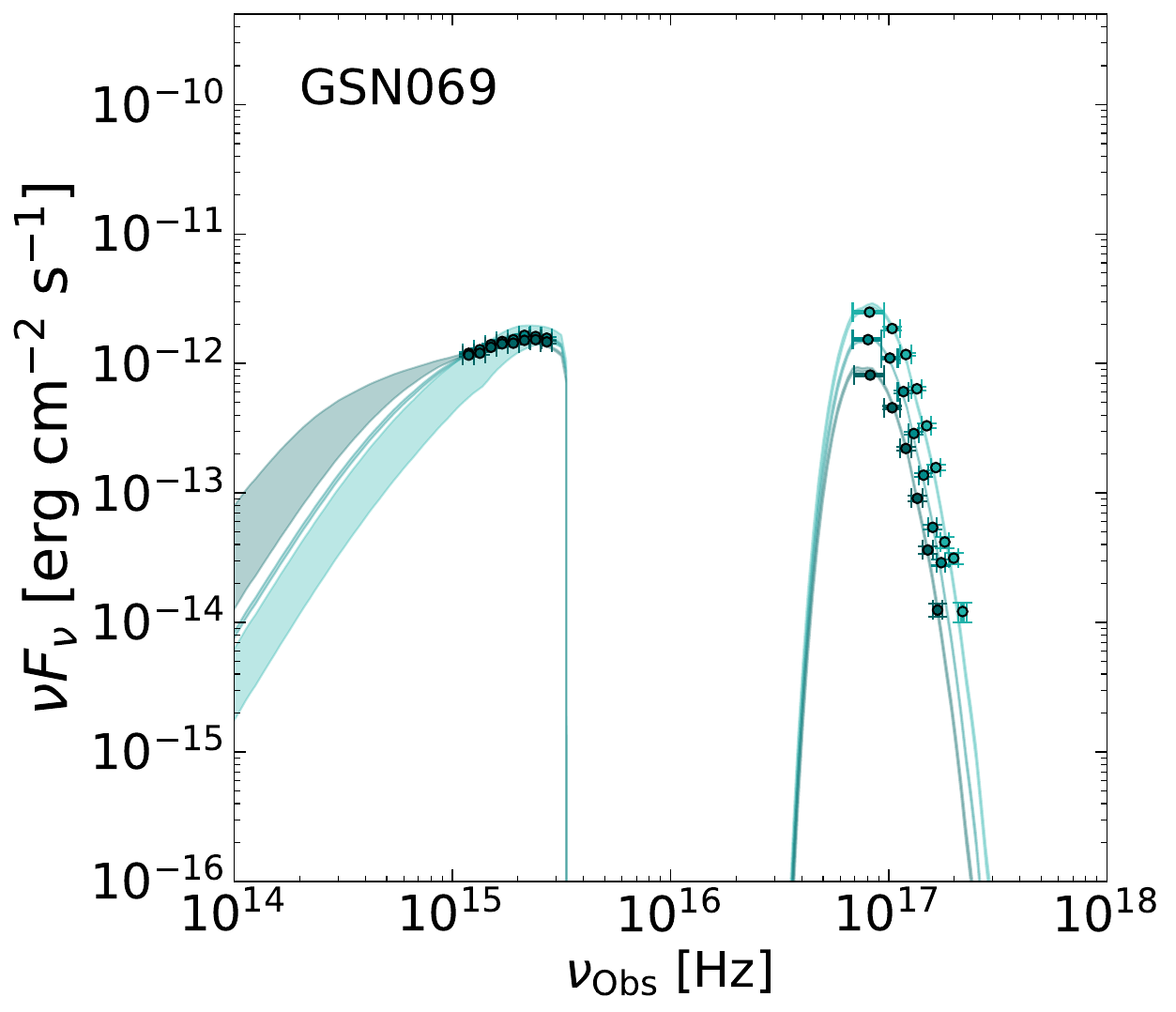}{0.245\textwidth}{}
          \fig{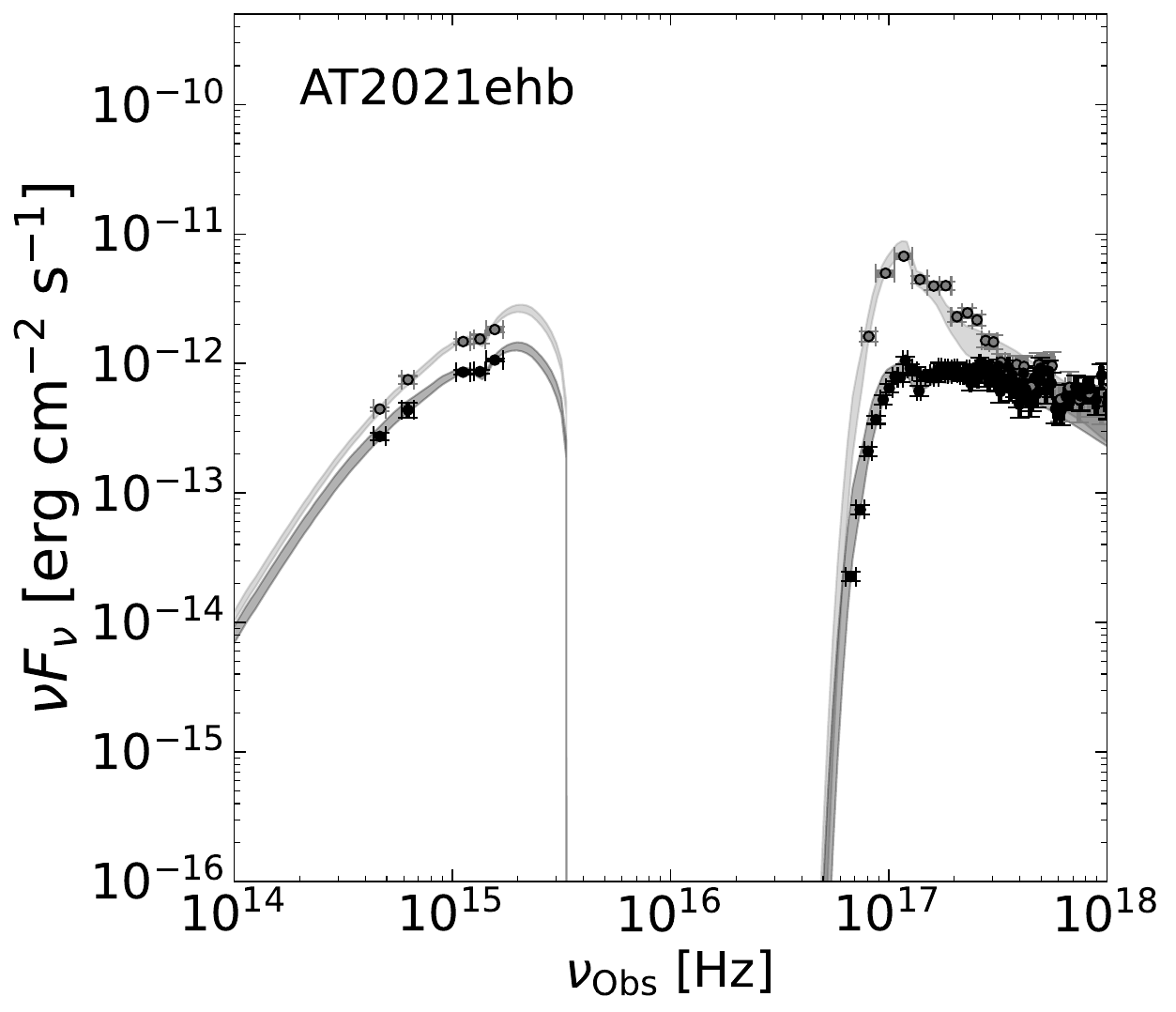}{0.245\textwidth}{}
          \fig{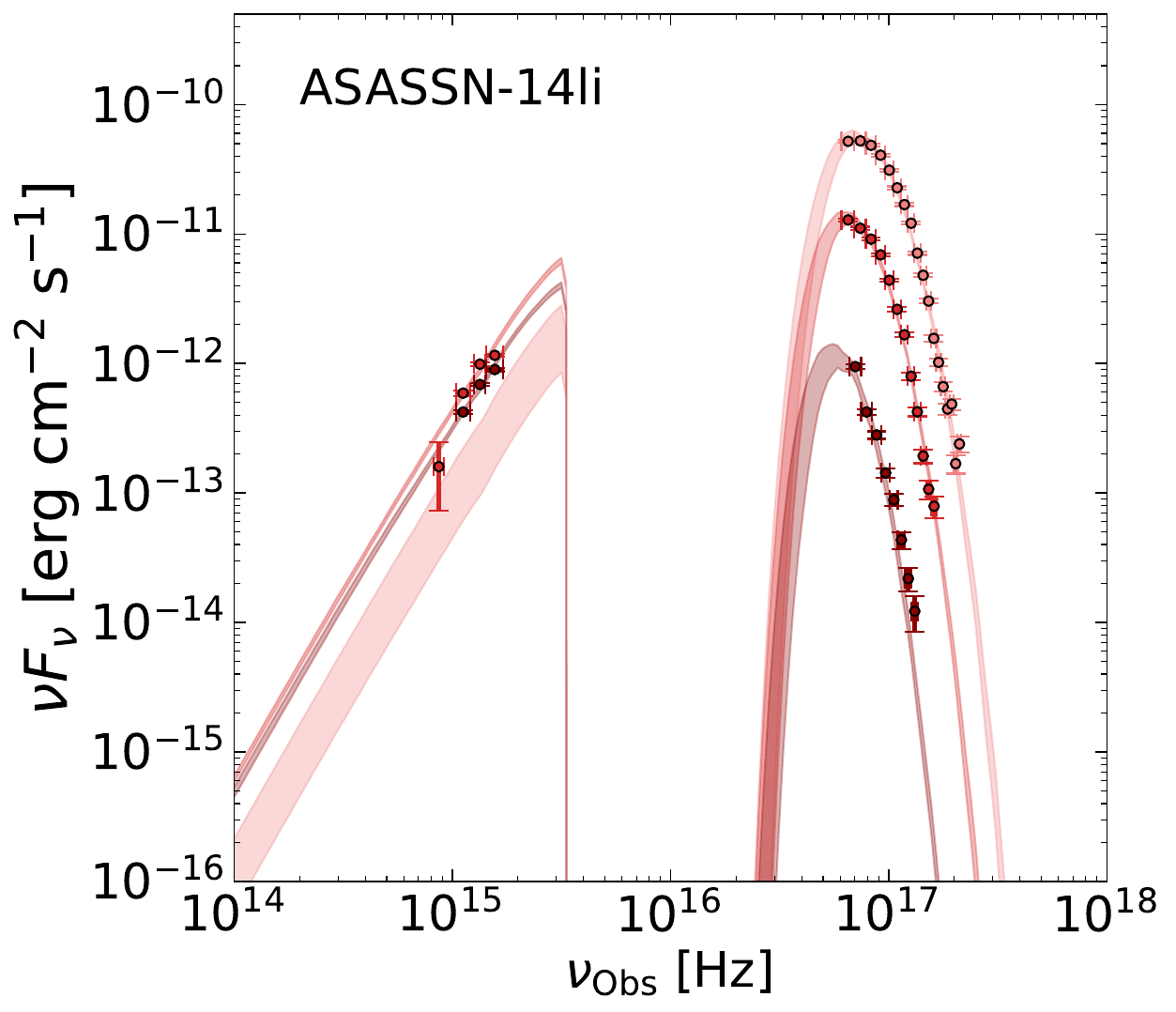}{0.245\textwidth}{}}\vspace{-0.65cm}

\gridline{\fig{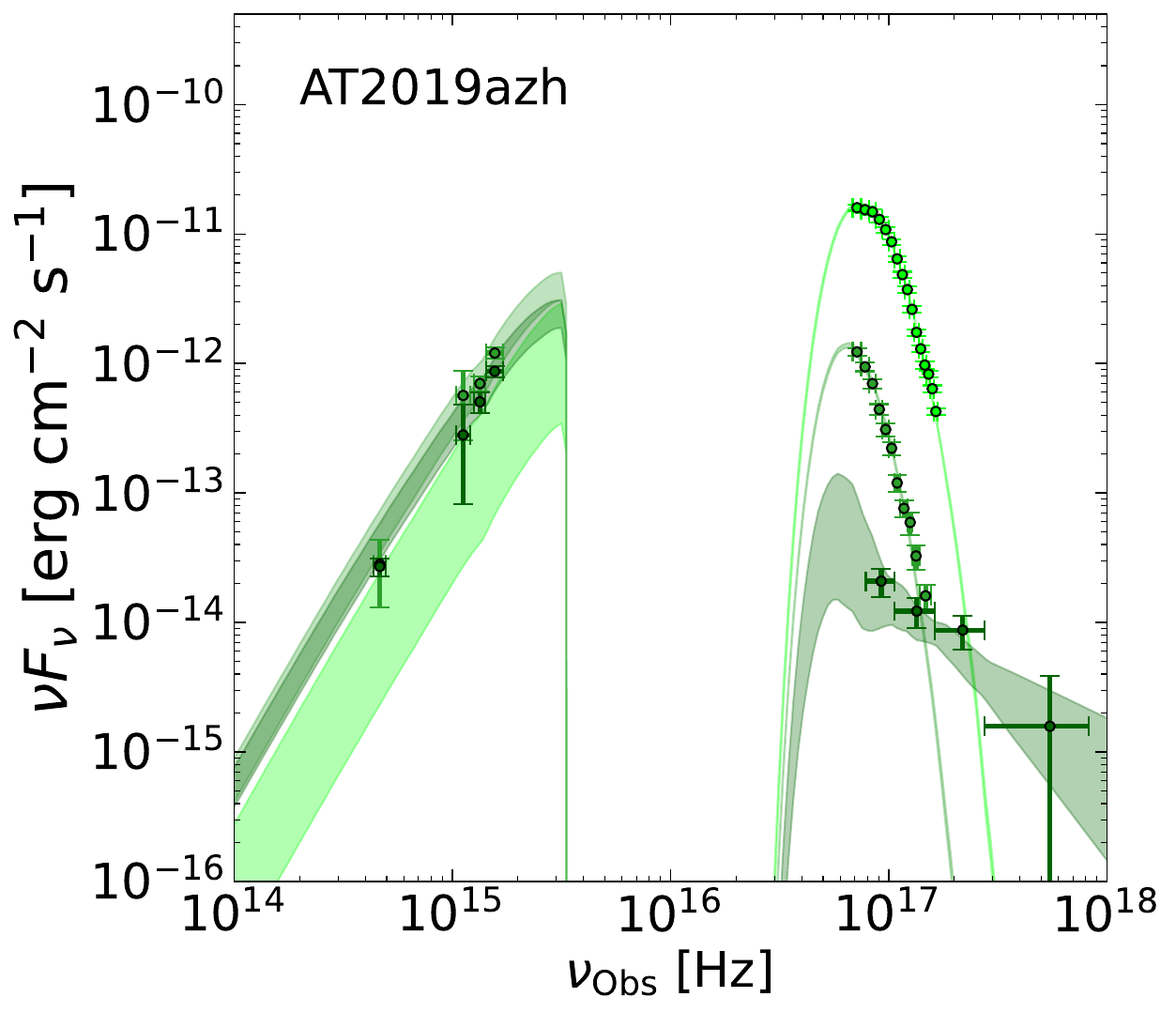}{0.245\textwidth}{}
          \fig{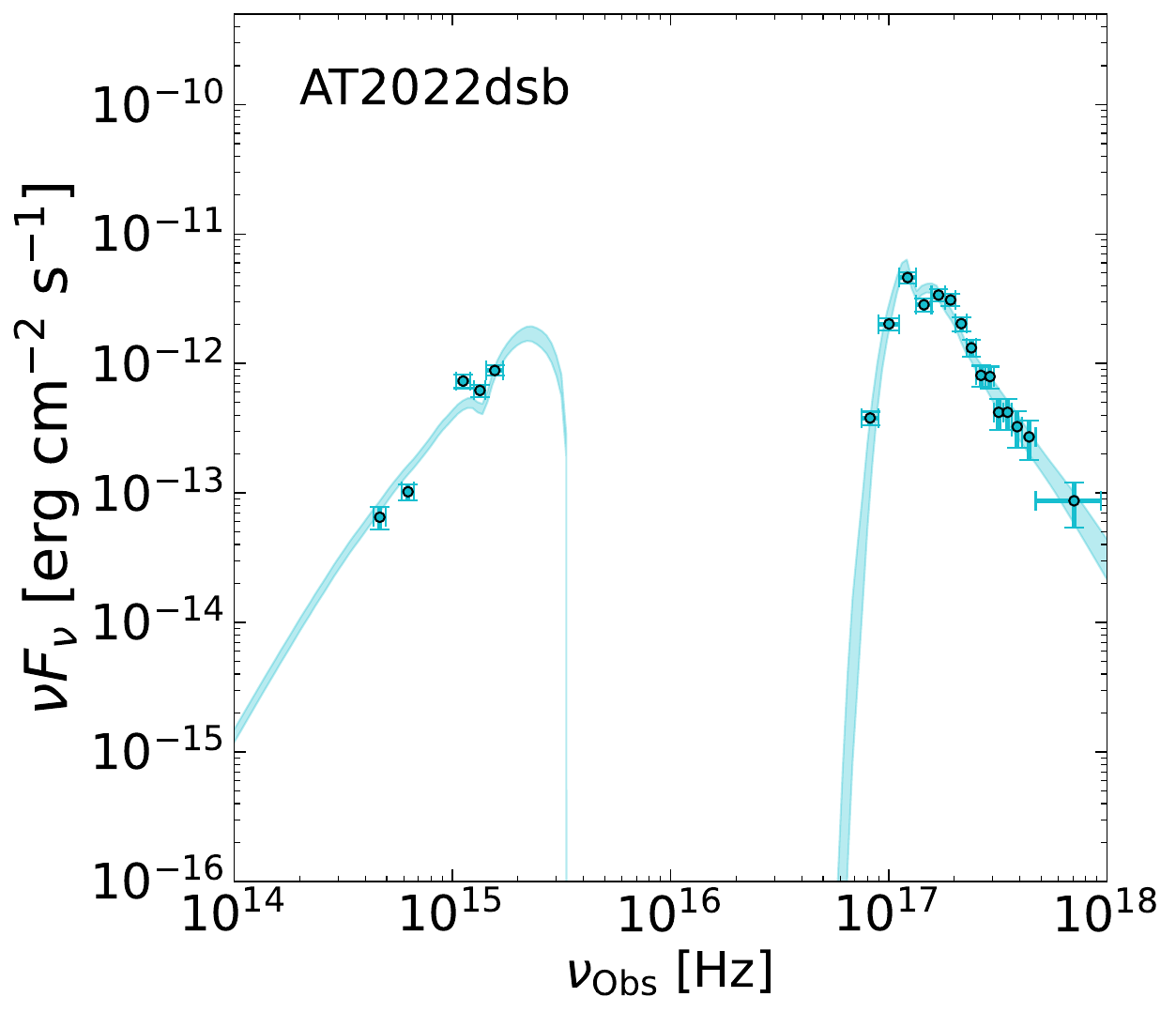}{0.245\textwidth}{}
          \fig{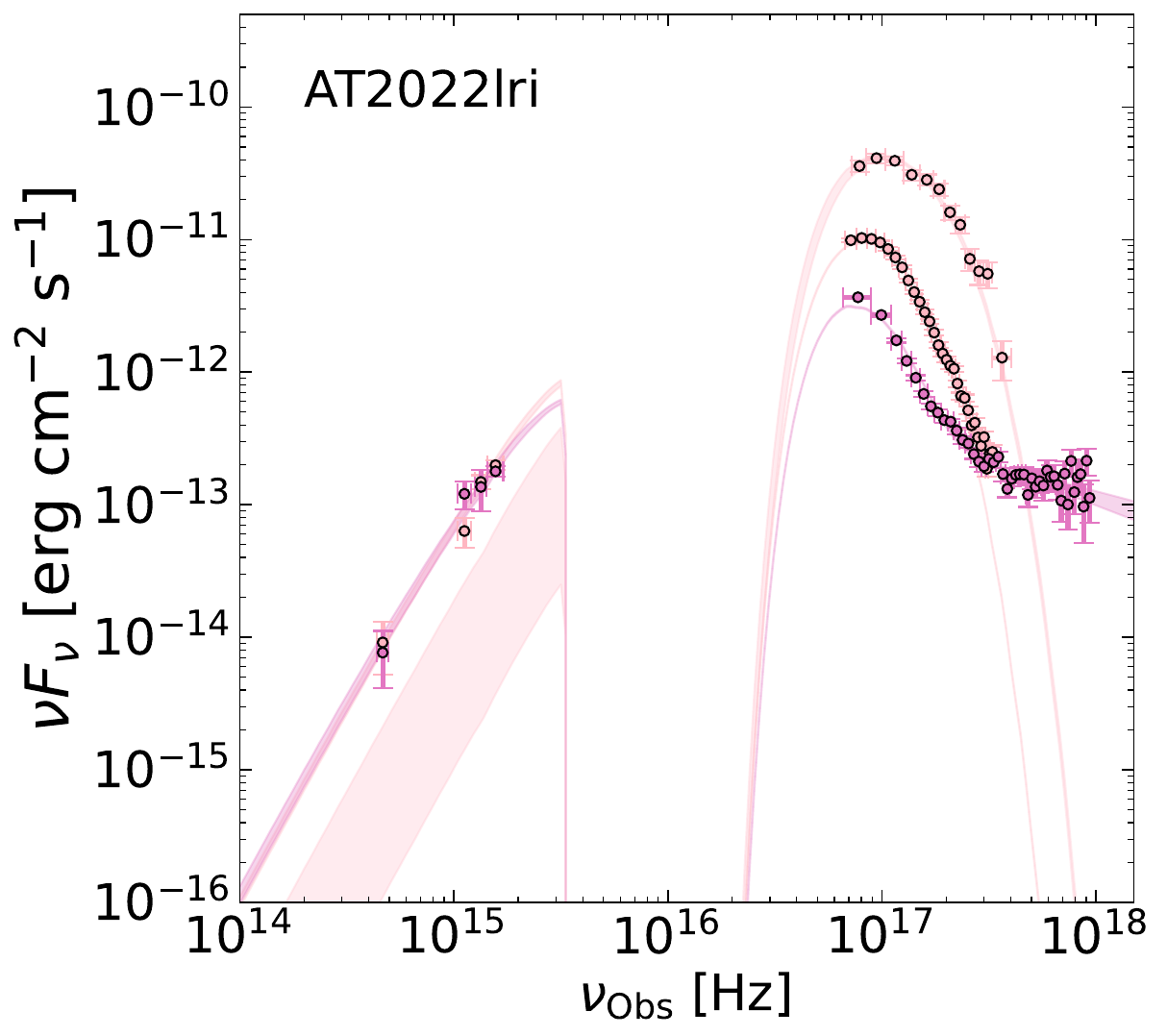}{0.245\textwidth}{}
          \fig{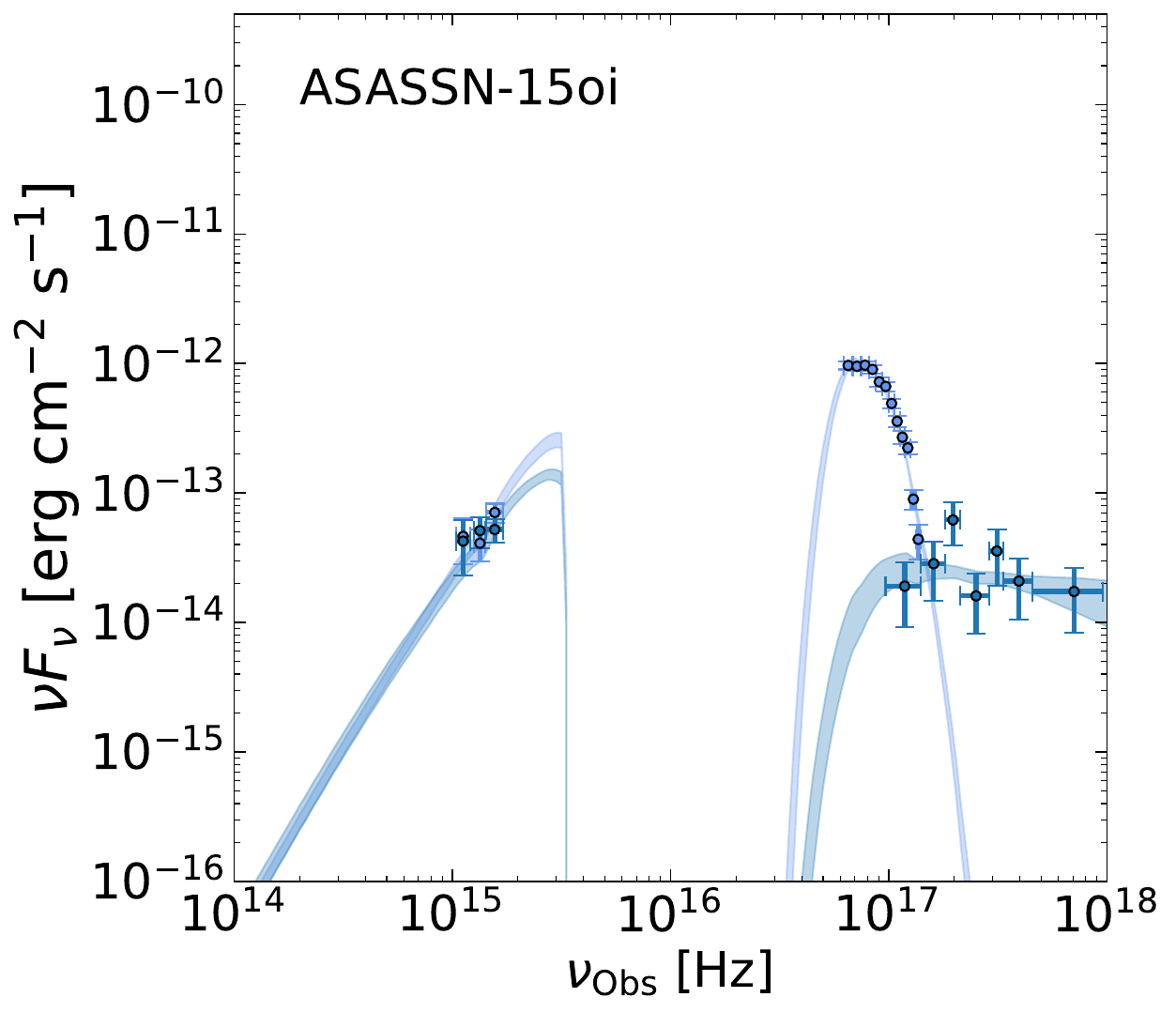}{0.245\textwidth}{}}\vspace{-0.65cm}

\gridline{\fig{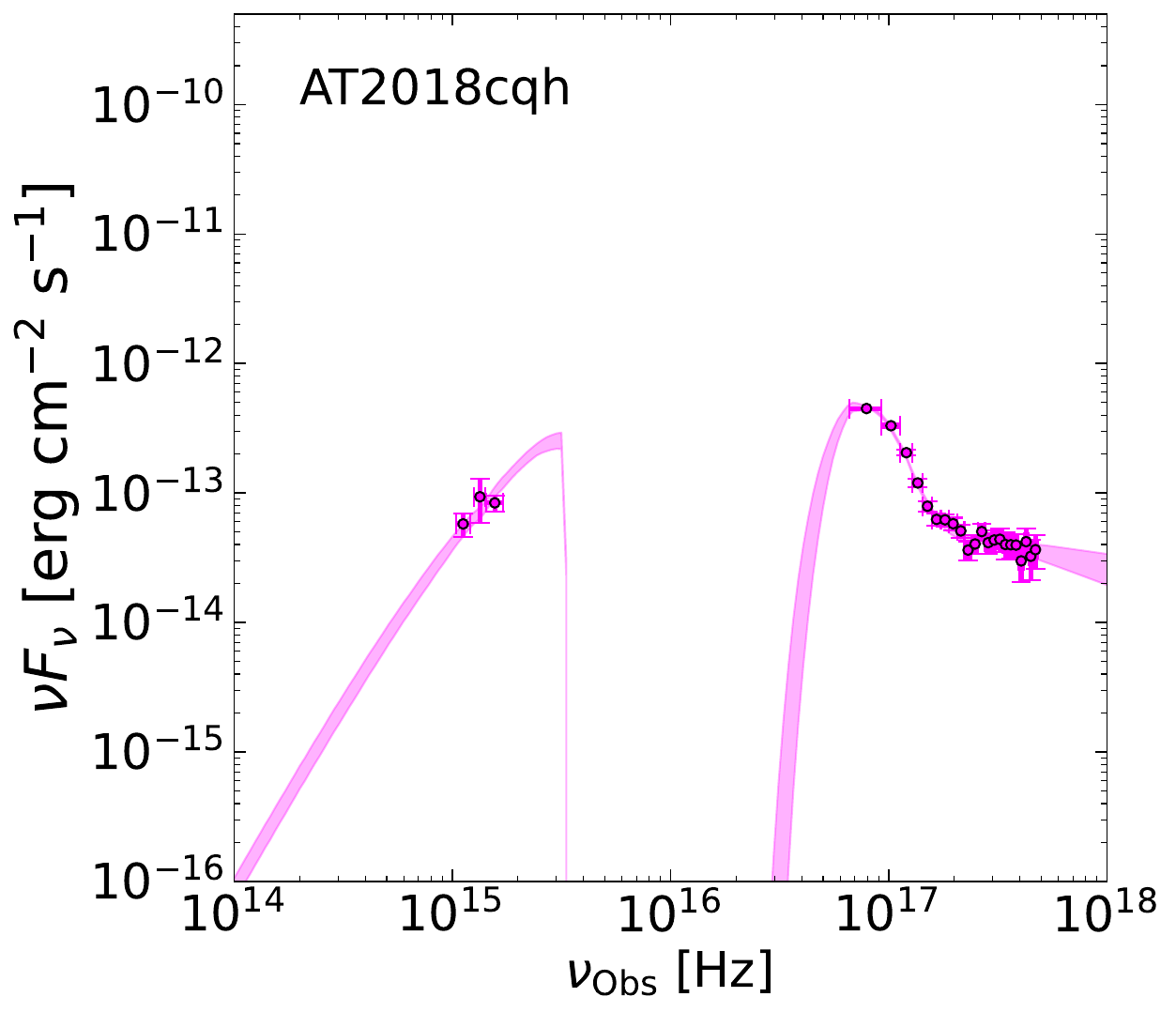}{0.245\textwidth}{}
          \fig{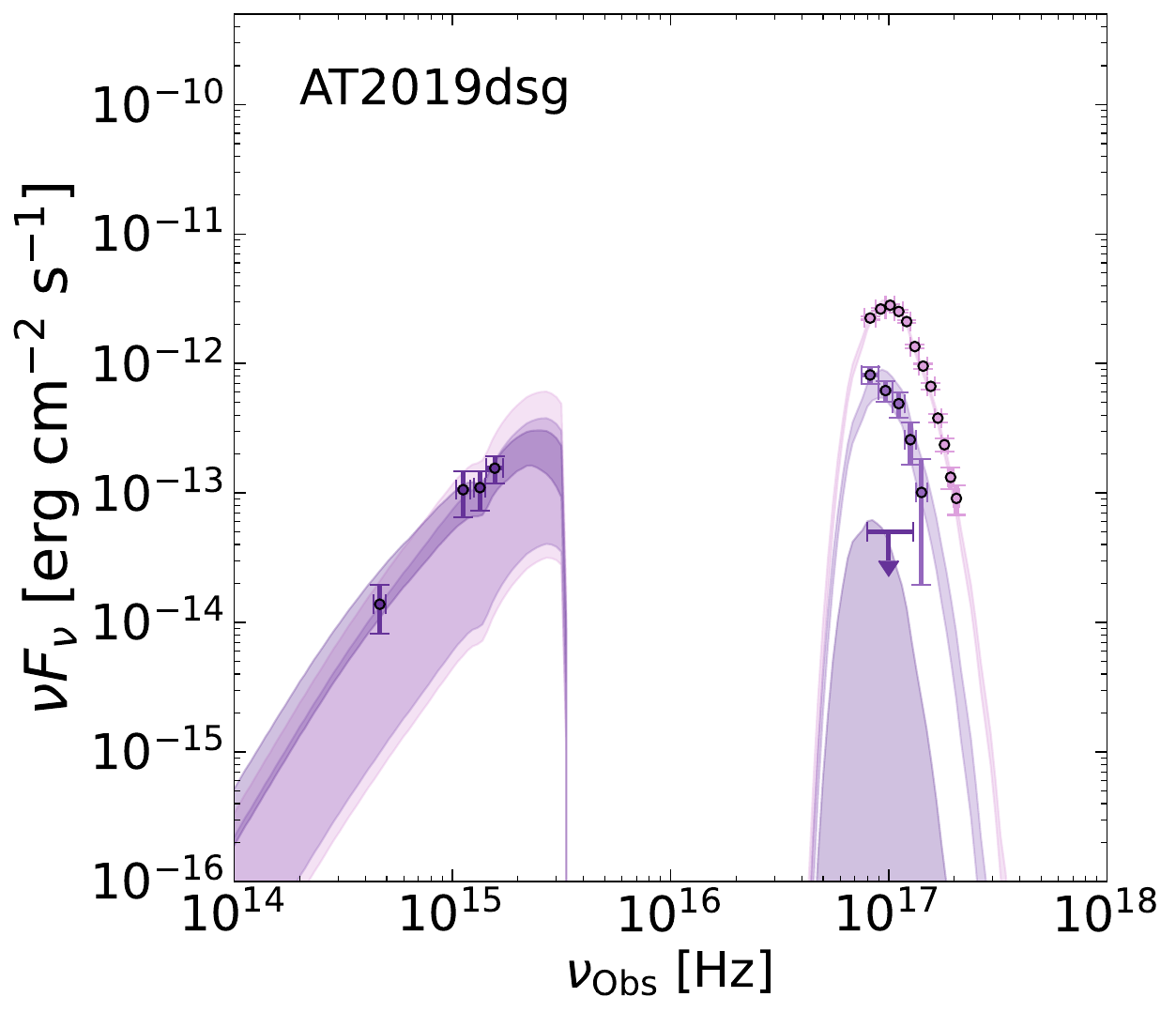}{0.245\textwidth}{}
          \fig{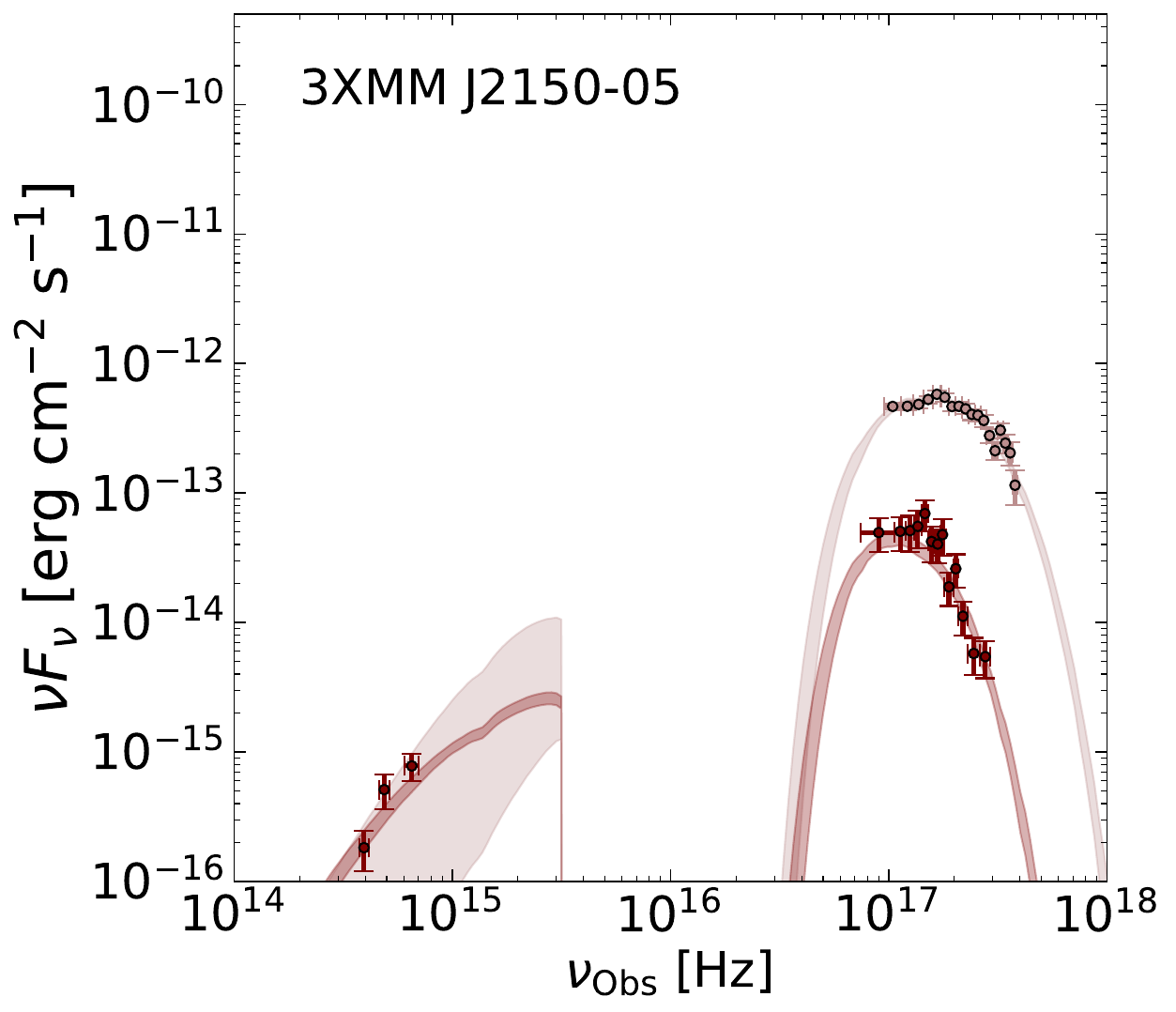}{0.245\textwidth}{}
          \fig{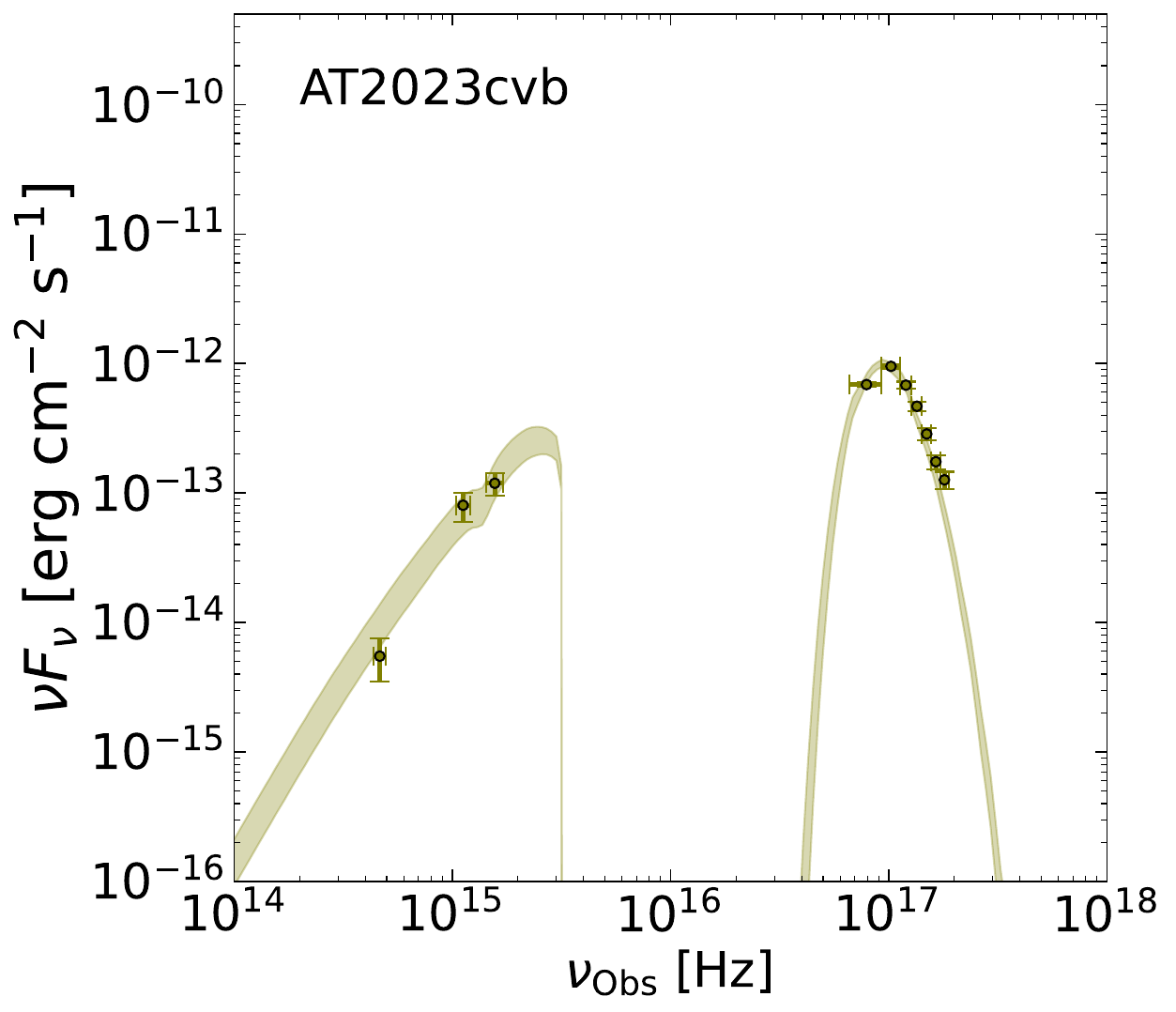}{0.245\textwidth}{}}\vspace{-0.65cm}

\gridline{\fig{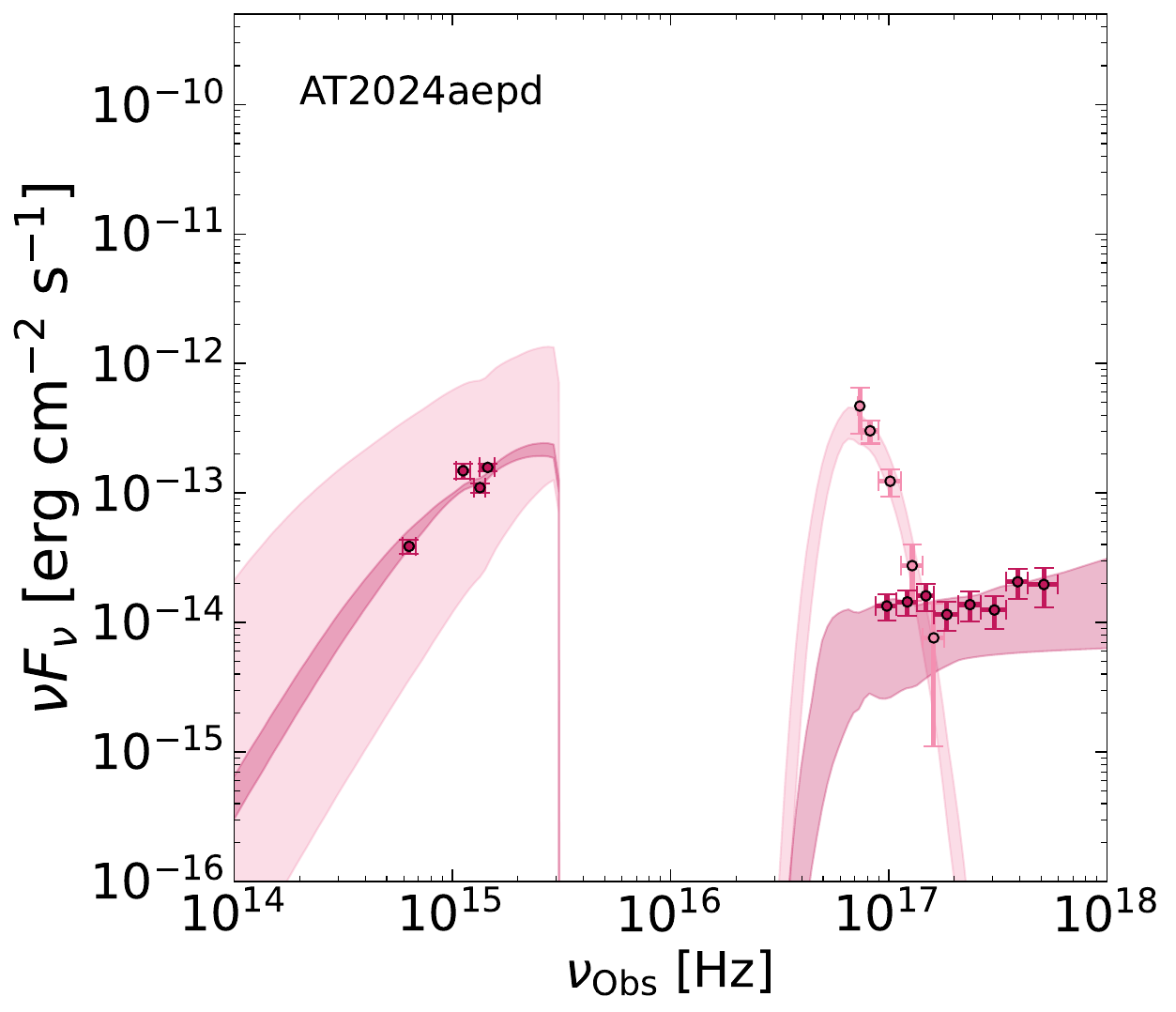}{0.245\textwidth}{}
        \fig{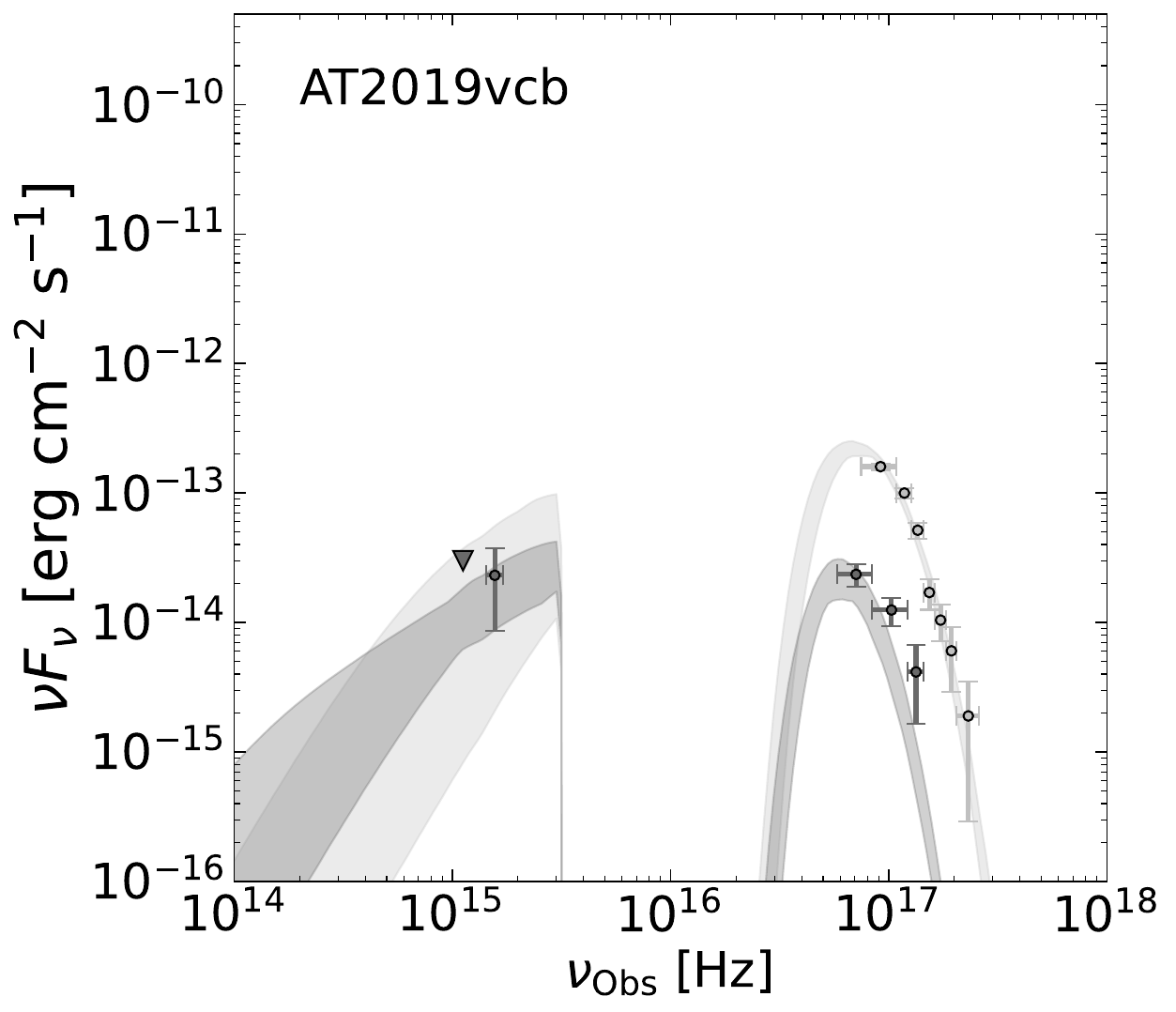}{0.245\textwidth}{}
          \fig{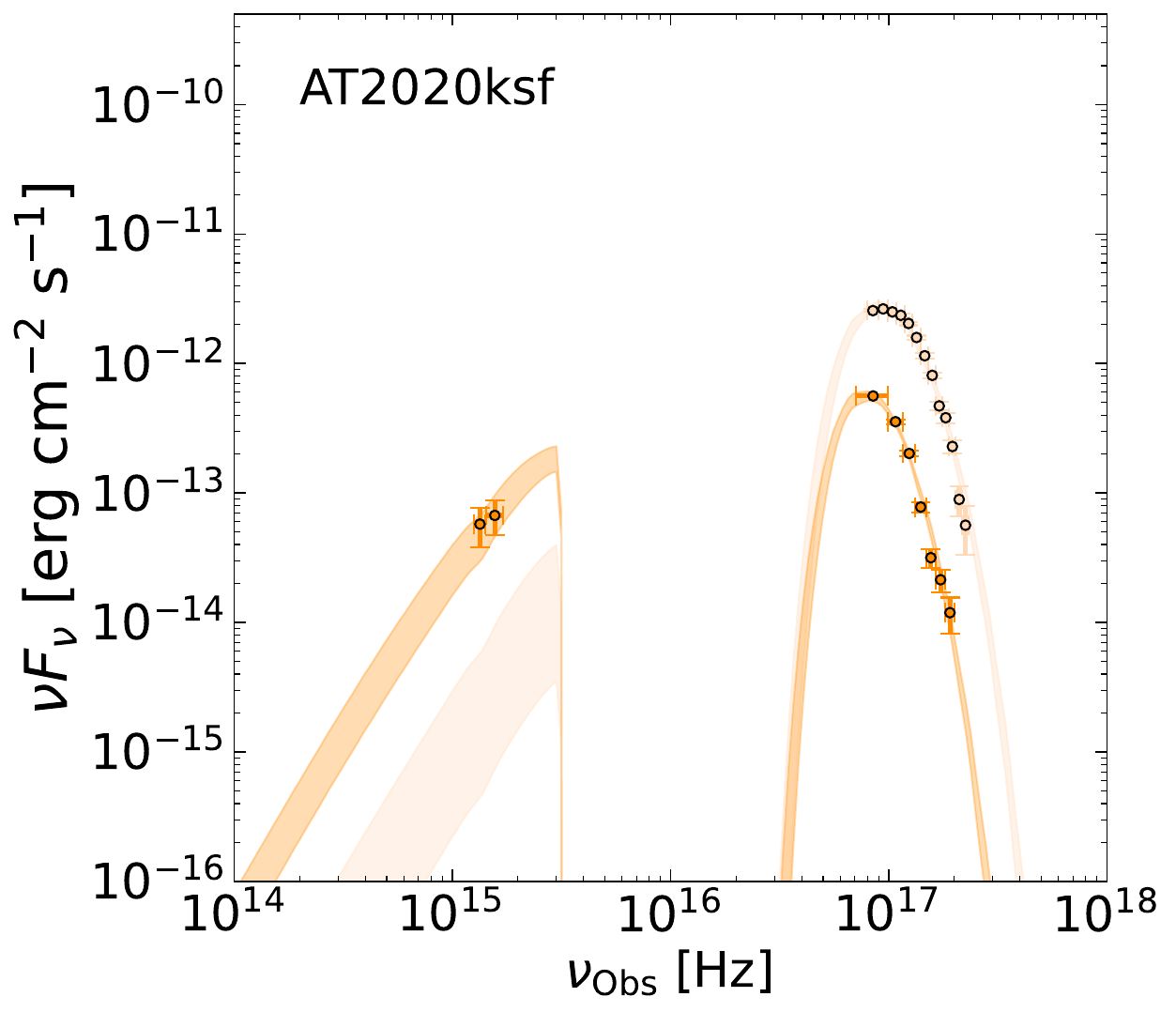}{0.245\textwidth}{}
           \fig{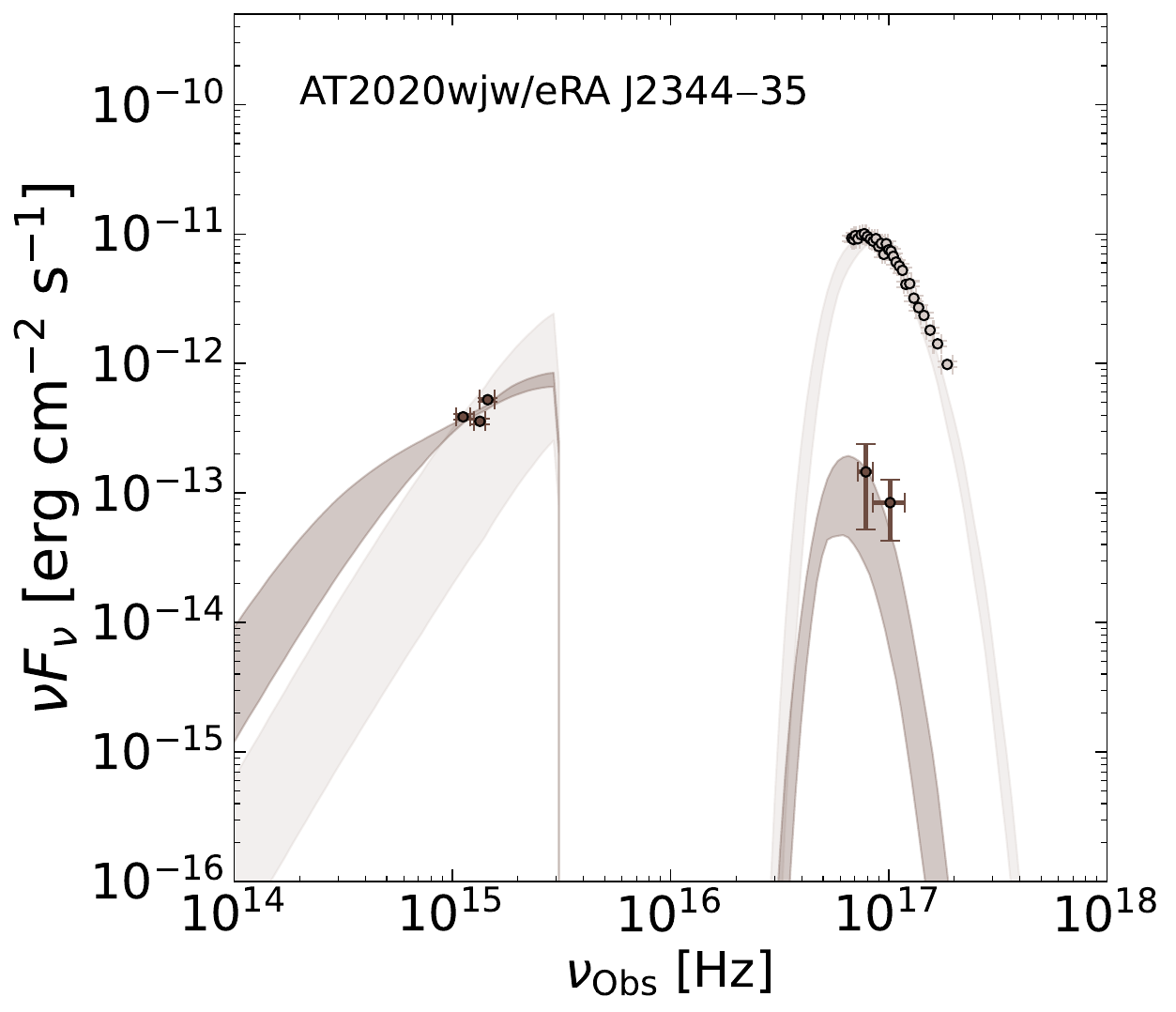}{0.245\textwidth}{}} % two empty slots for alignment

\caption{Results of the spectral modeling for our sample. Each panel shows the observed (i.e., uncorrected for absorption or extinction) UV/optical photometry and the unfolded X-ray spectrum. For sources with X-ray–only epochs (see Table~\ref{tab:SED}), the UV/optical flux of the disk component is displayed for illustrative purposes only, as the outer disk parameters are unconstrained. Contours show the 68\% credible interval of the model posterior. }
\label{fig:SED_fit1}
\end{figure*}

\begin{figure*}
\centering
\def\gridline#1{\vskip1pt\hbox to\hsize{#1}\vskip1pt}
% ============================
% Luminosity panels (4-column layout)
% ============================
\gridline{\fig{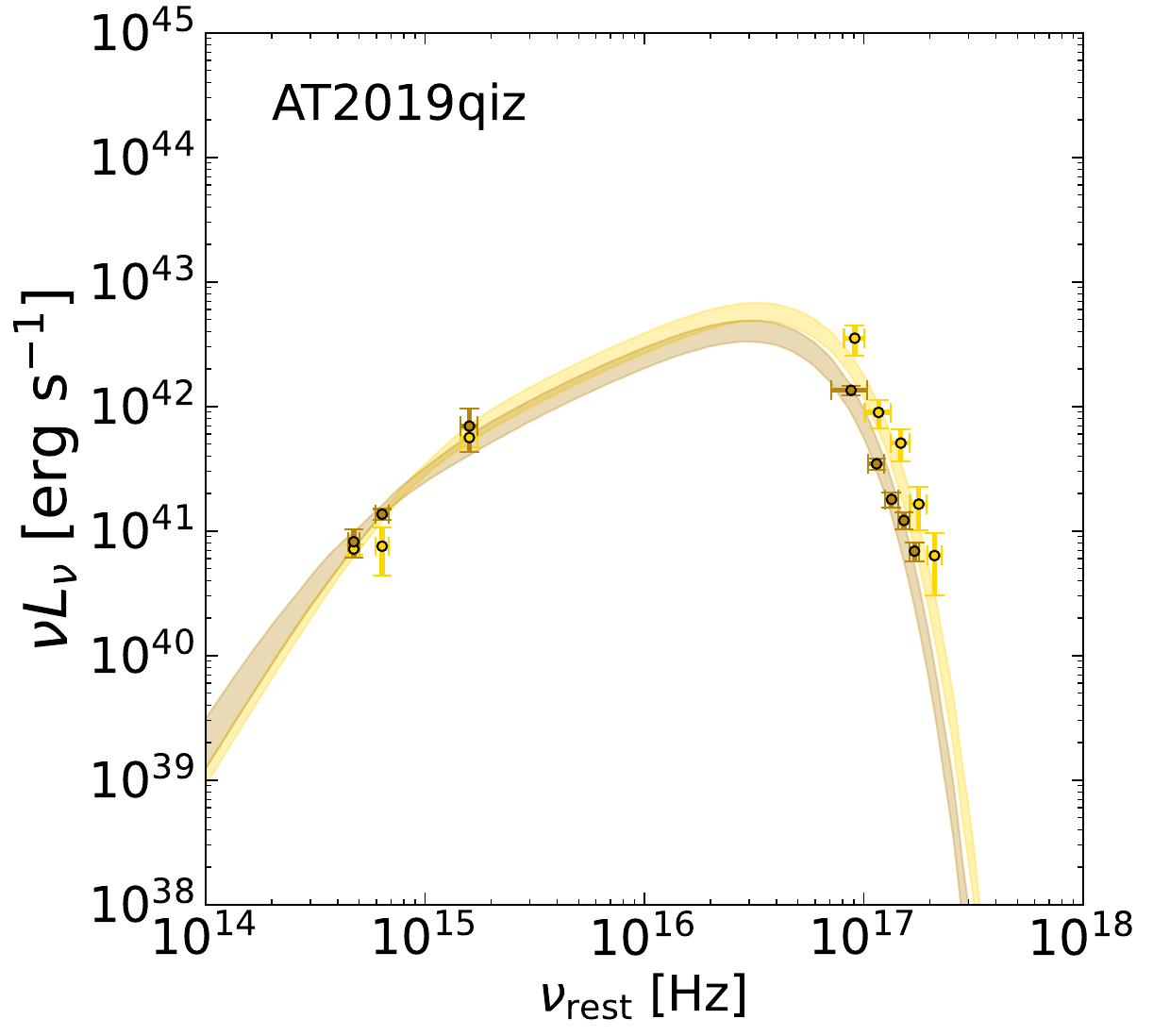}{0.245\textwidth}{}
          \fig{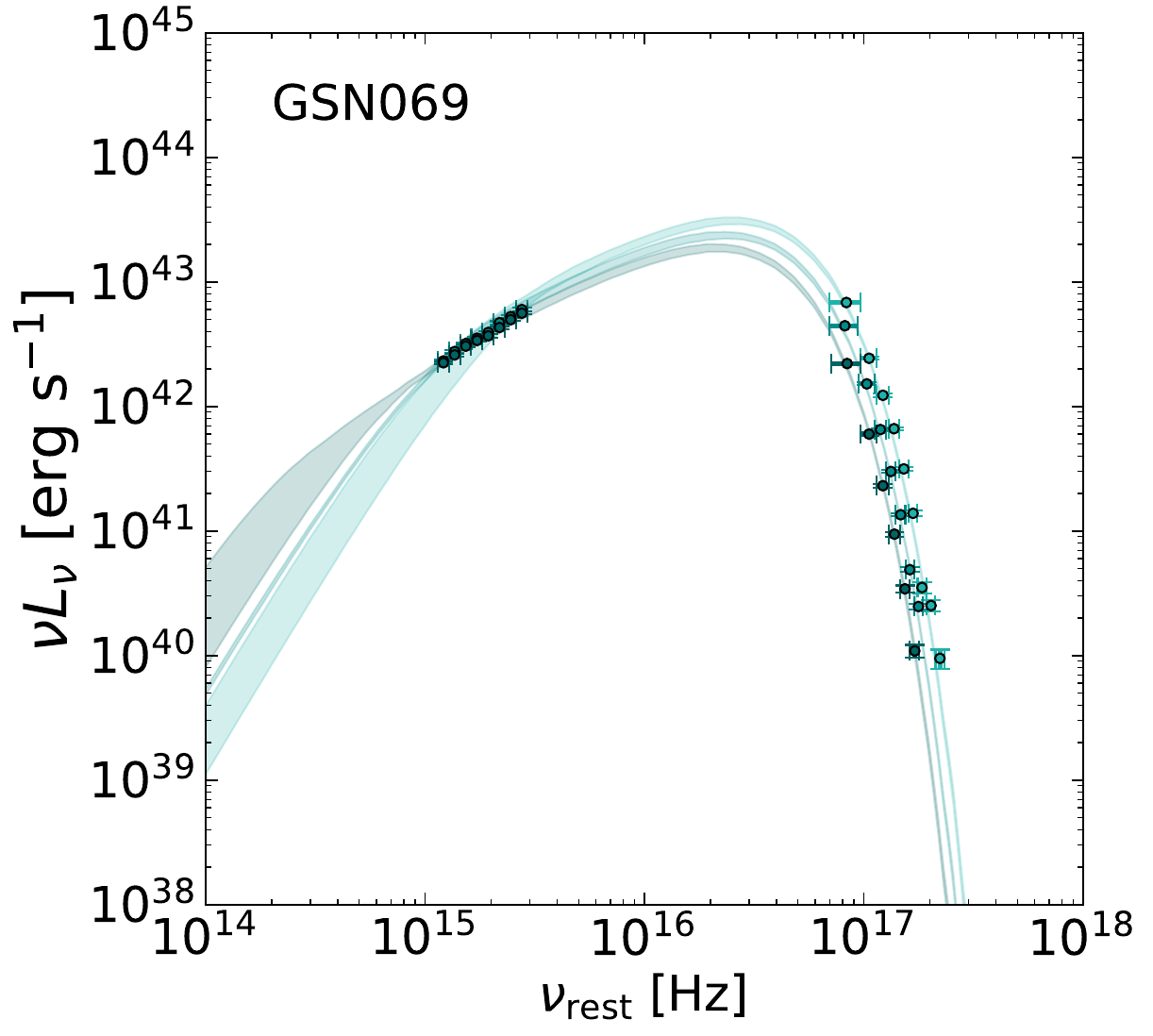}{0.245\textwidth}{}
          \fig{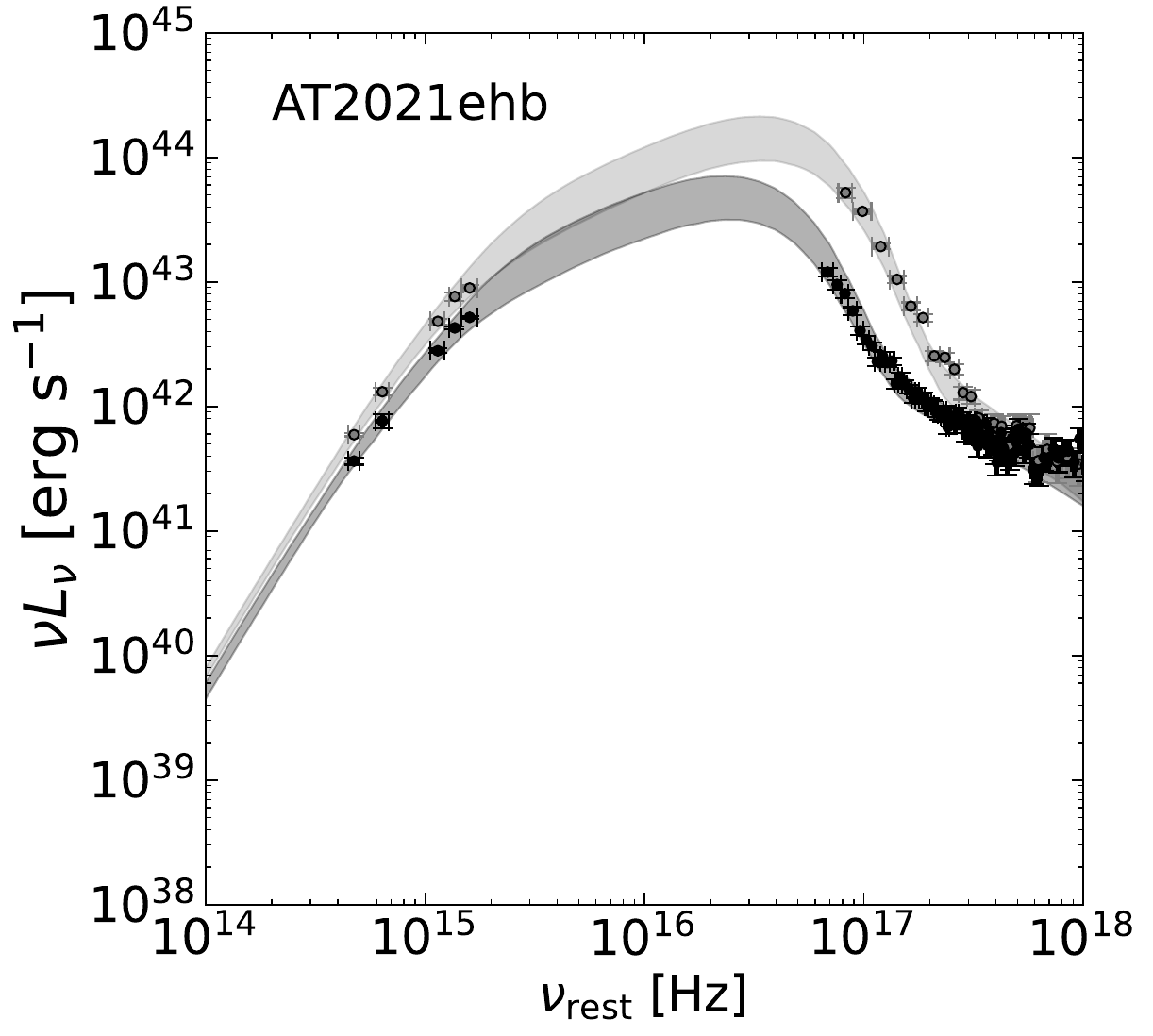}{0.245\textwidth}{}
          \fig{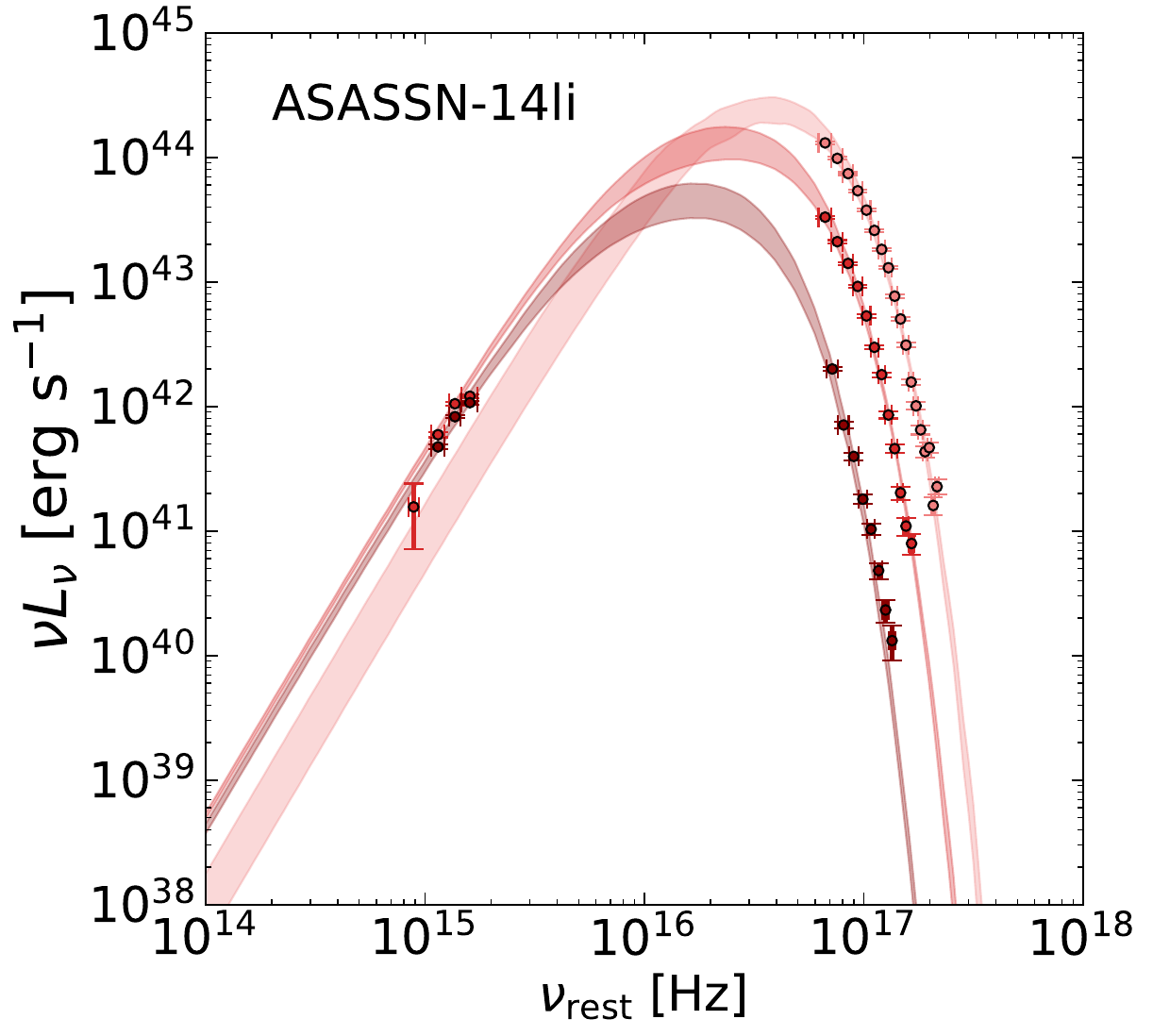}{0.245\textwidth}{}}\vspace{-0.65cm}

\gridline{\fig{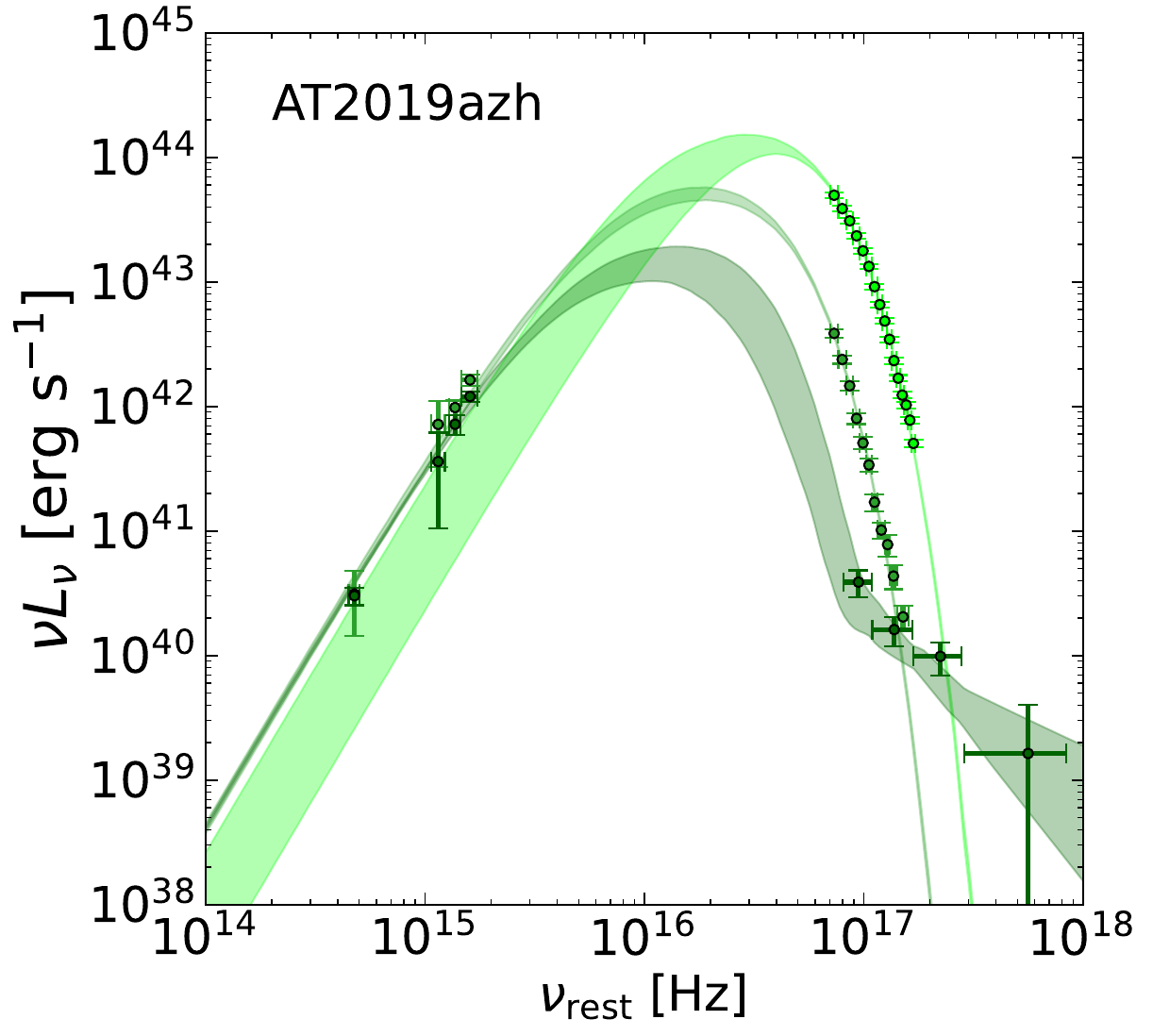}{0.245\textwidth}{}
          \fig{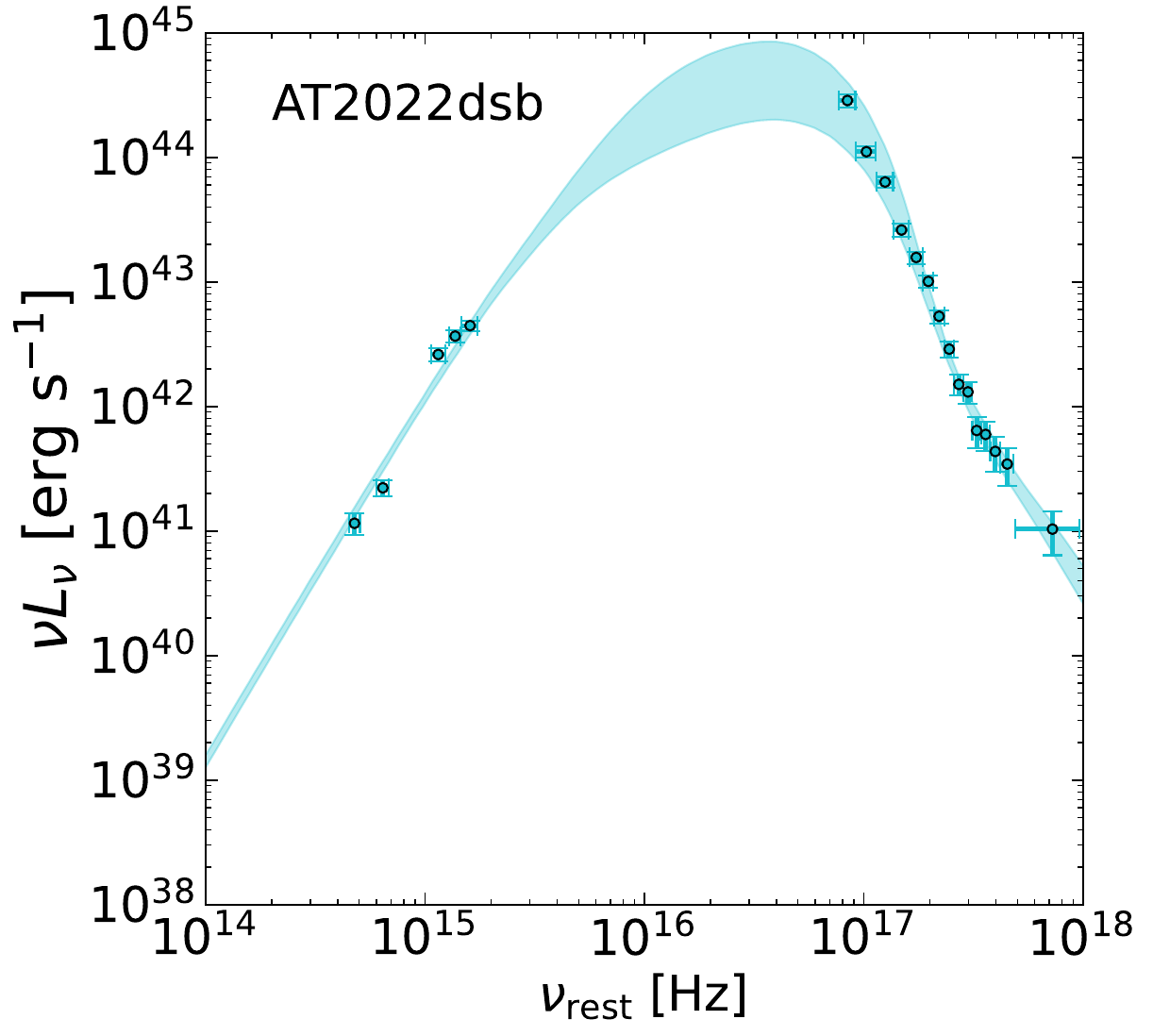}{0.245\textwidth}{}
          \fig{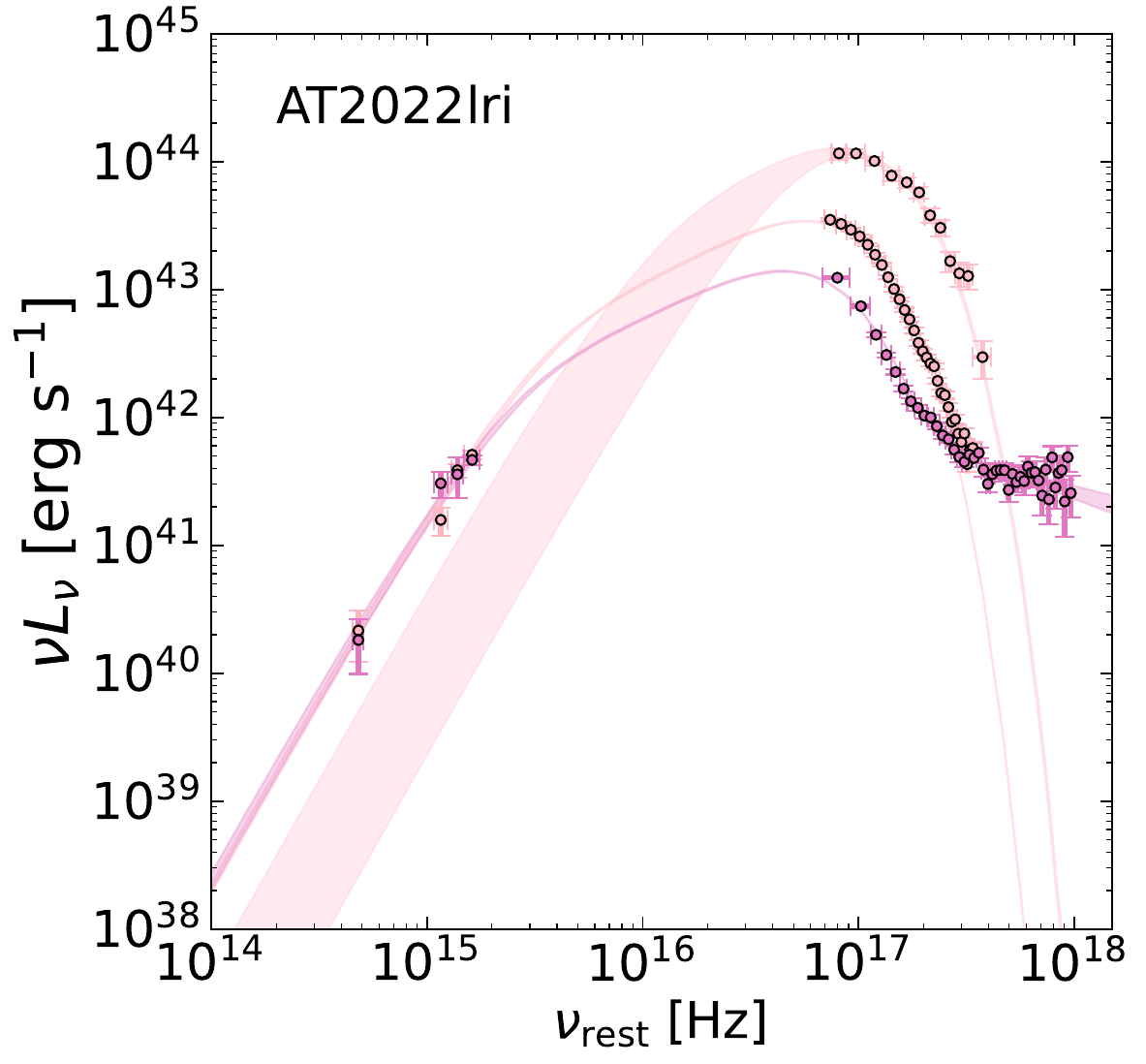}{0.245\textwidth}{}
          \fig{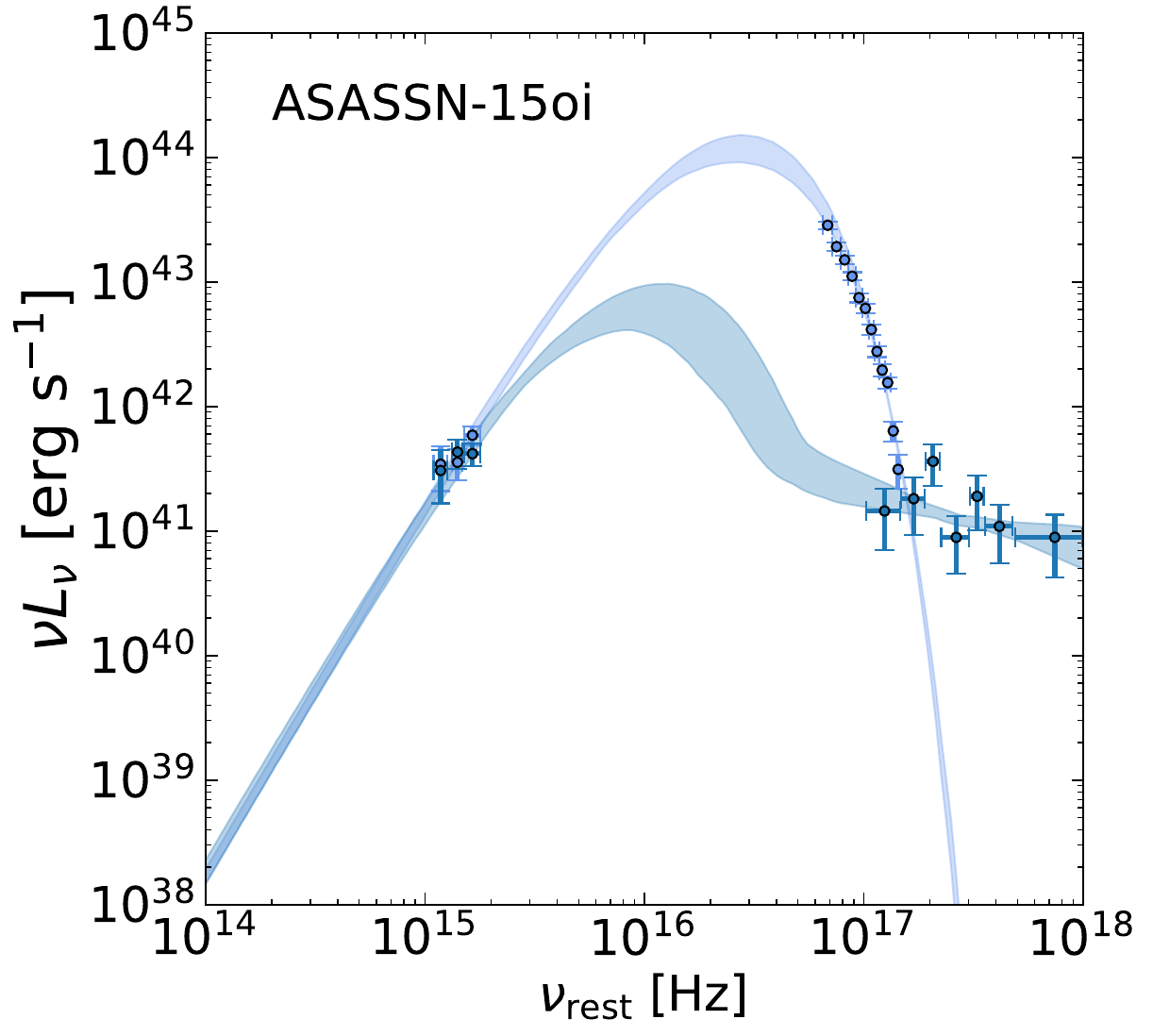}{0.245\textwidth}{}}\vspace{-0.65cm}

\gridline{\fig{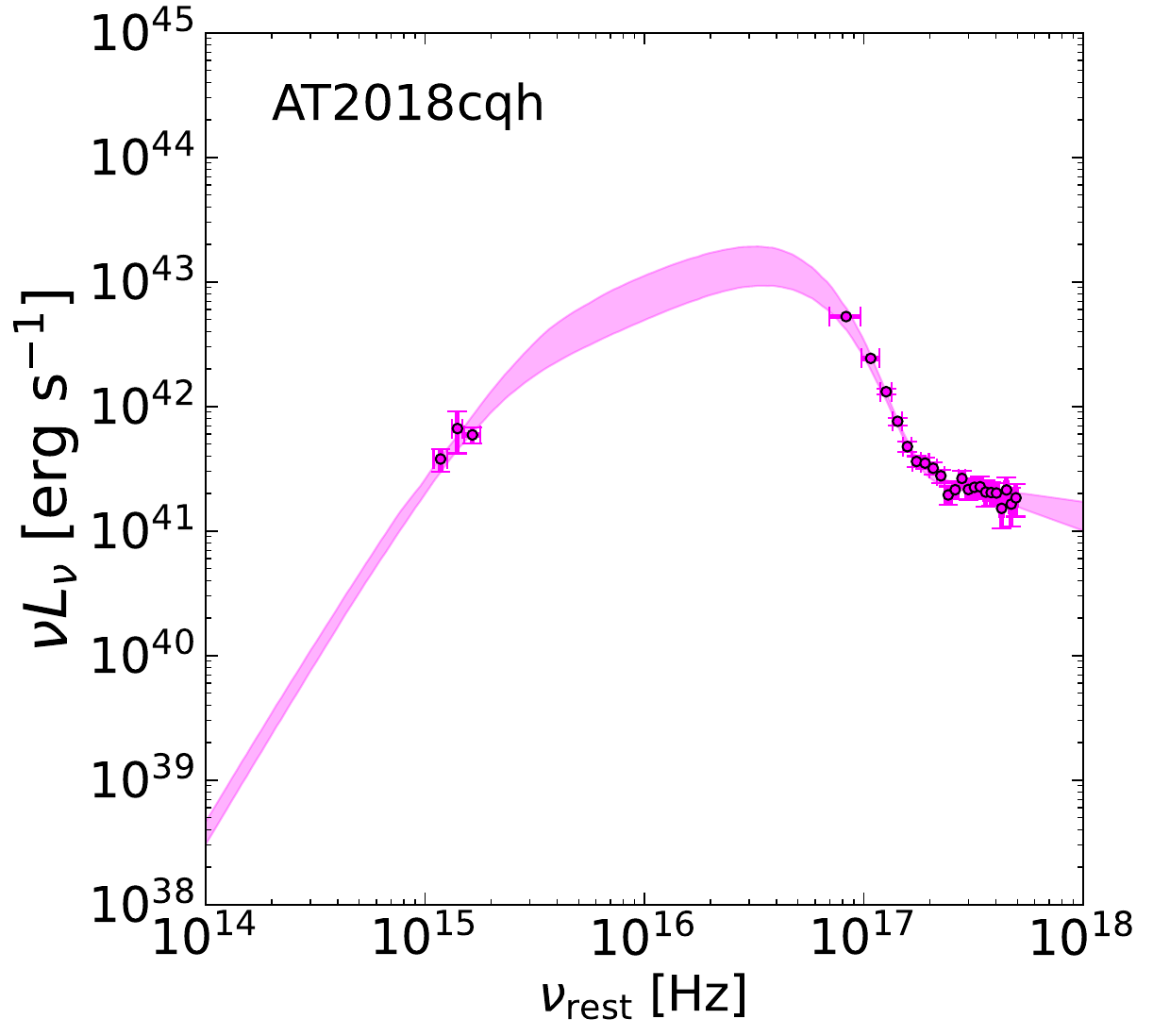}{0.245\textwidth}{}
          \fig{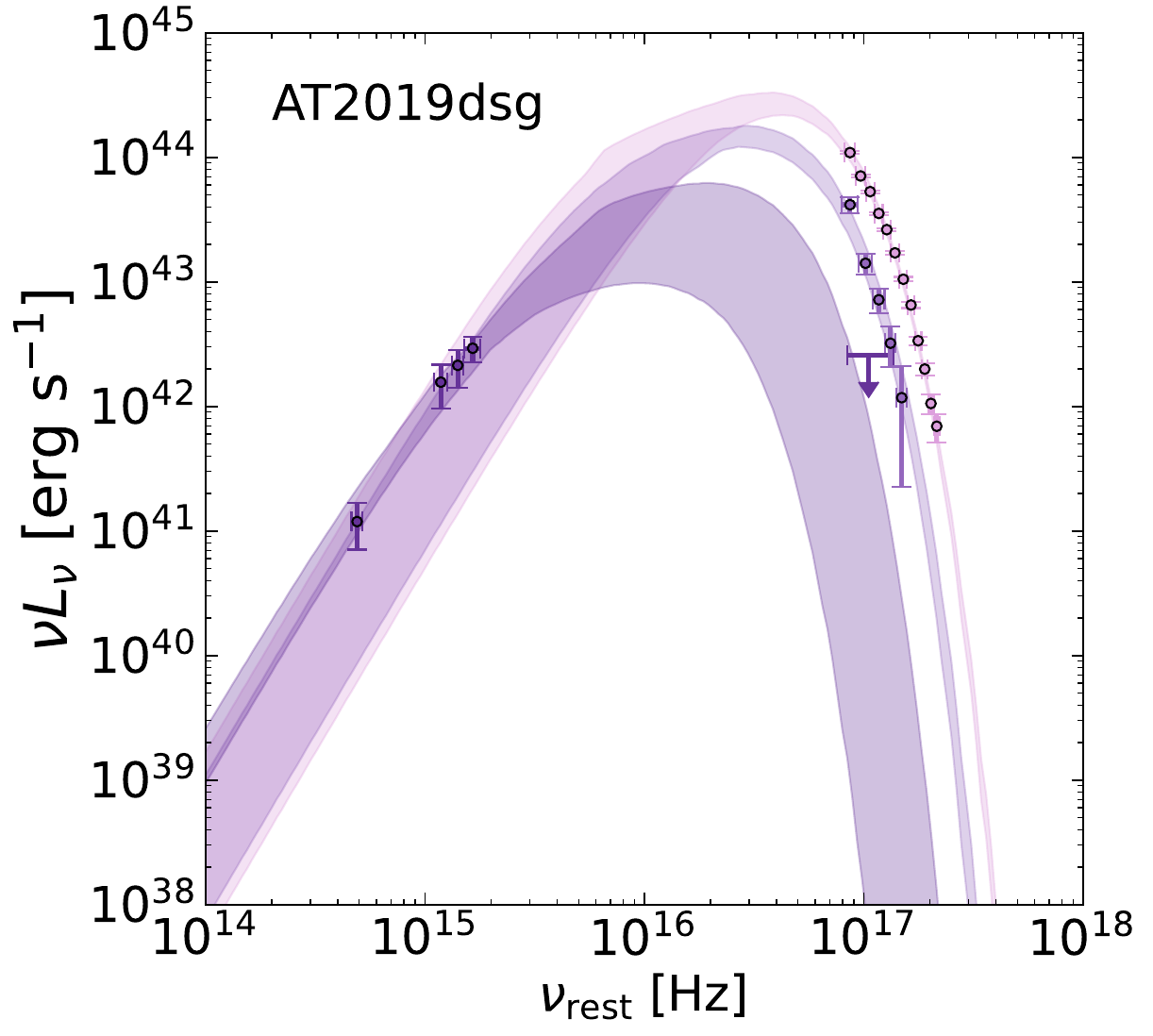}{0.245\textwidth}{}
          \fig{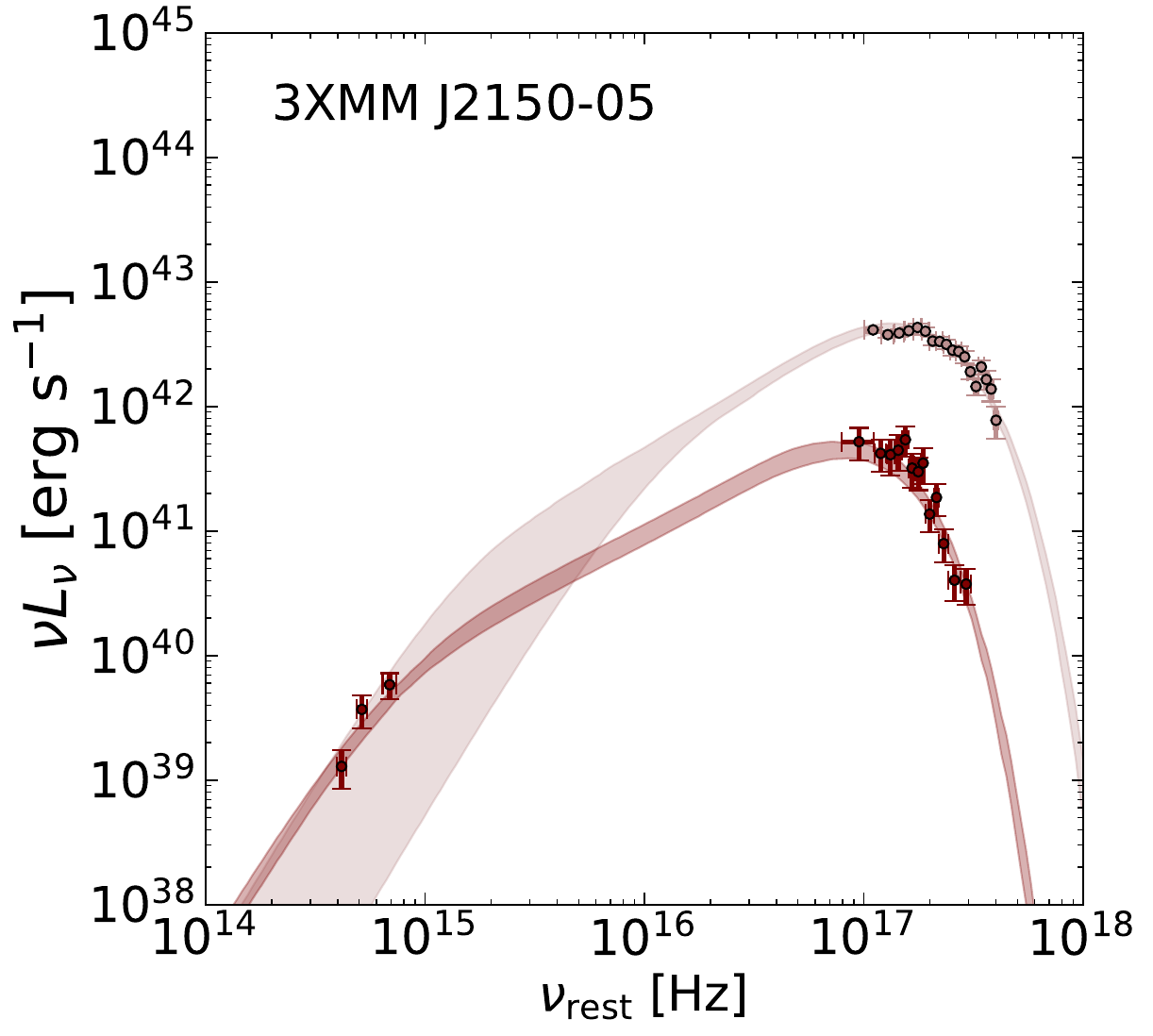}{0.245\textwidth}{}
          \fig{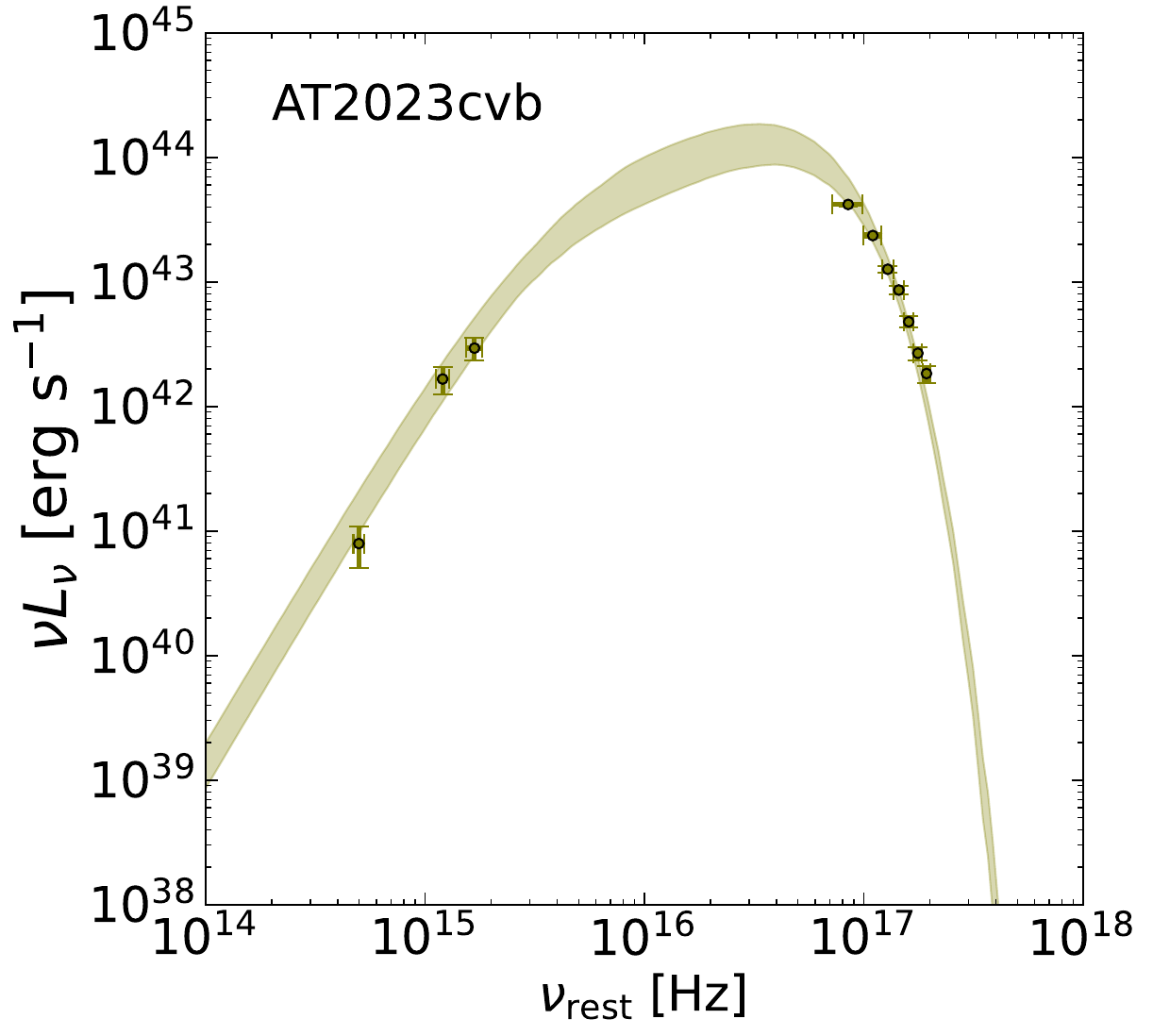}{0.245\textwidth}{}}\vspace{-0.65cm}

\gridline{\fig{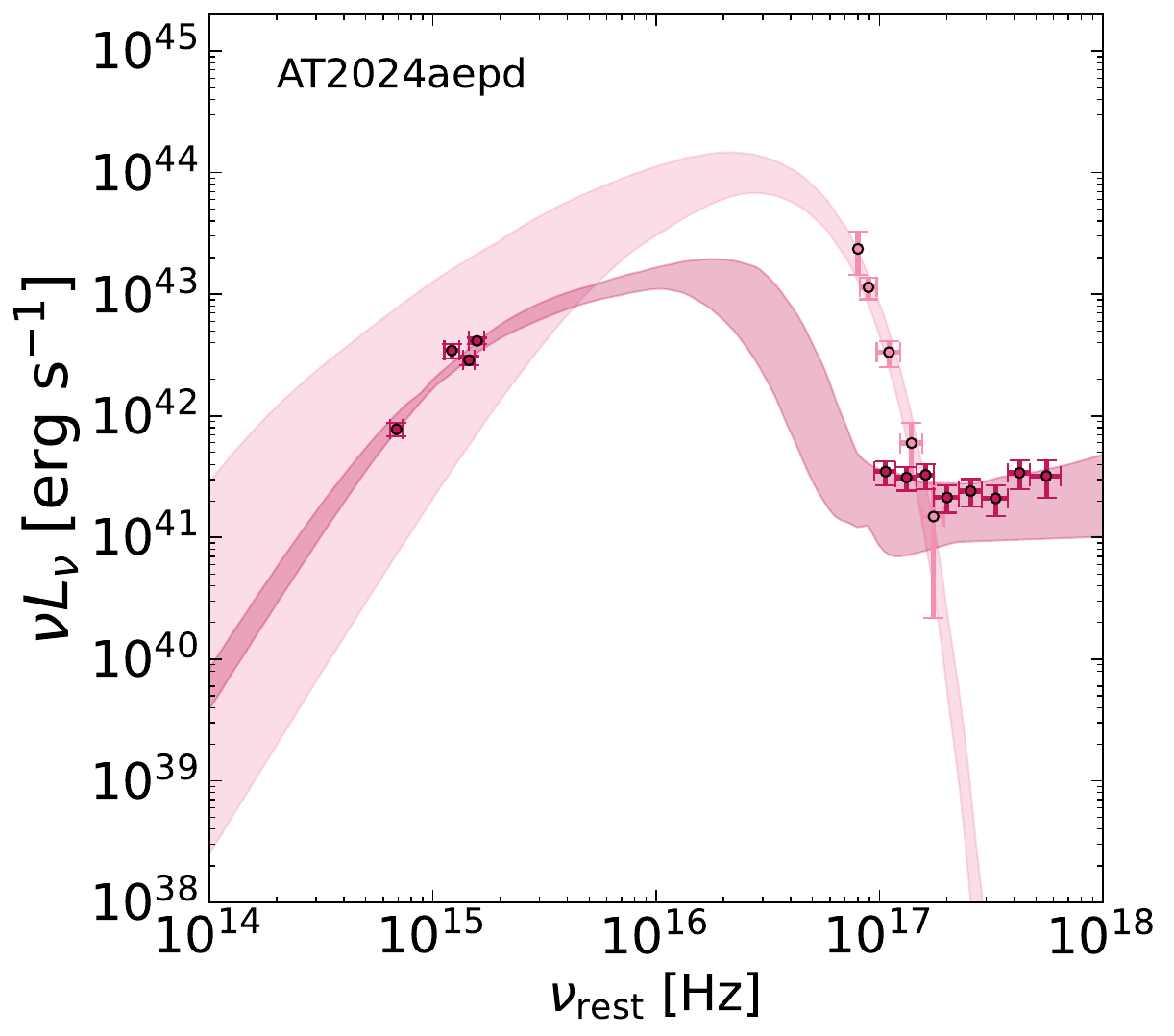}{0.245\textwidth}{}
        \fig{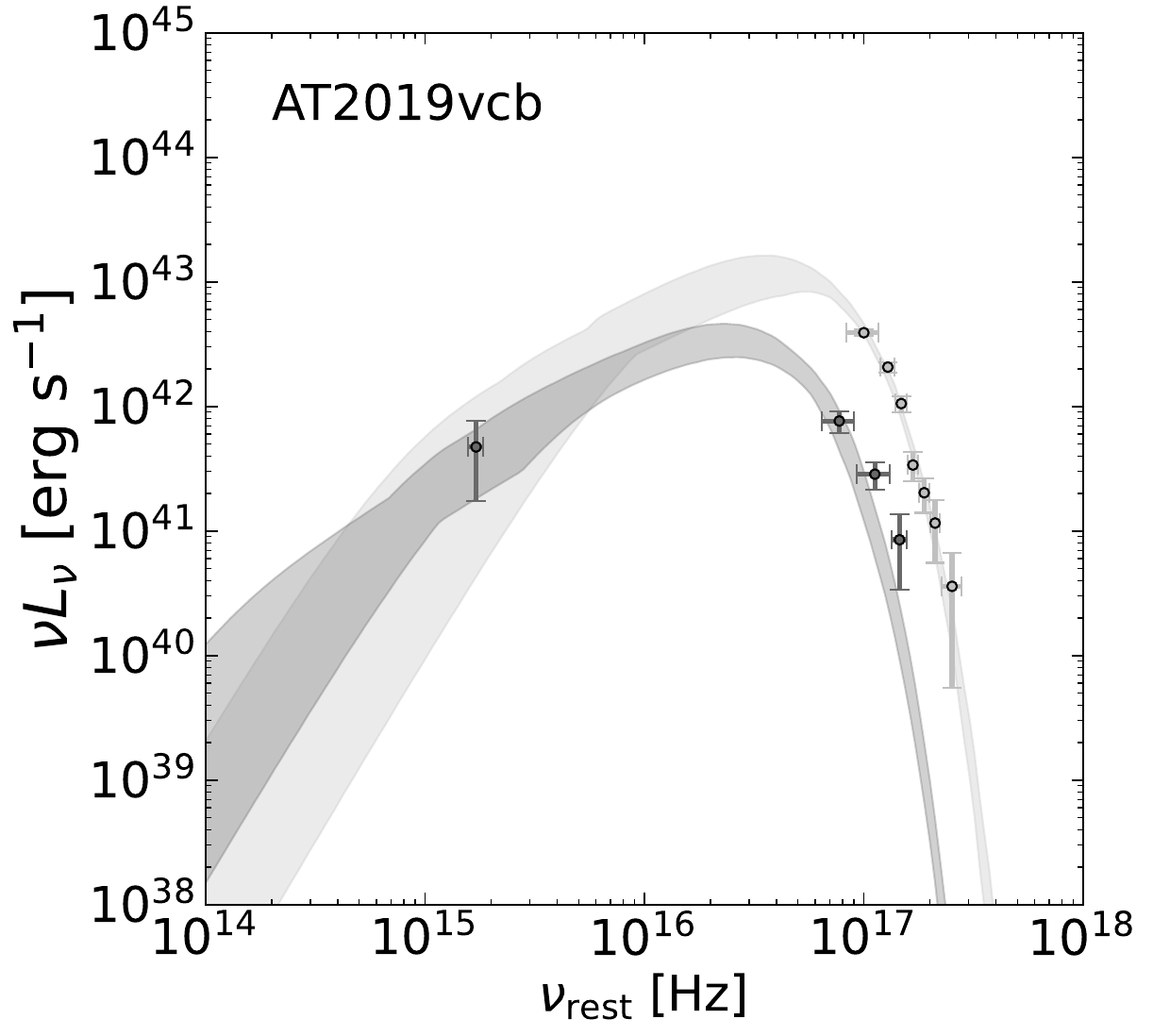}{0.245\textwidth}{}
          \fig{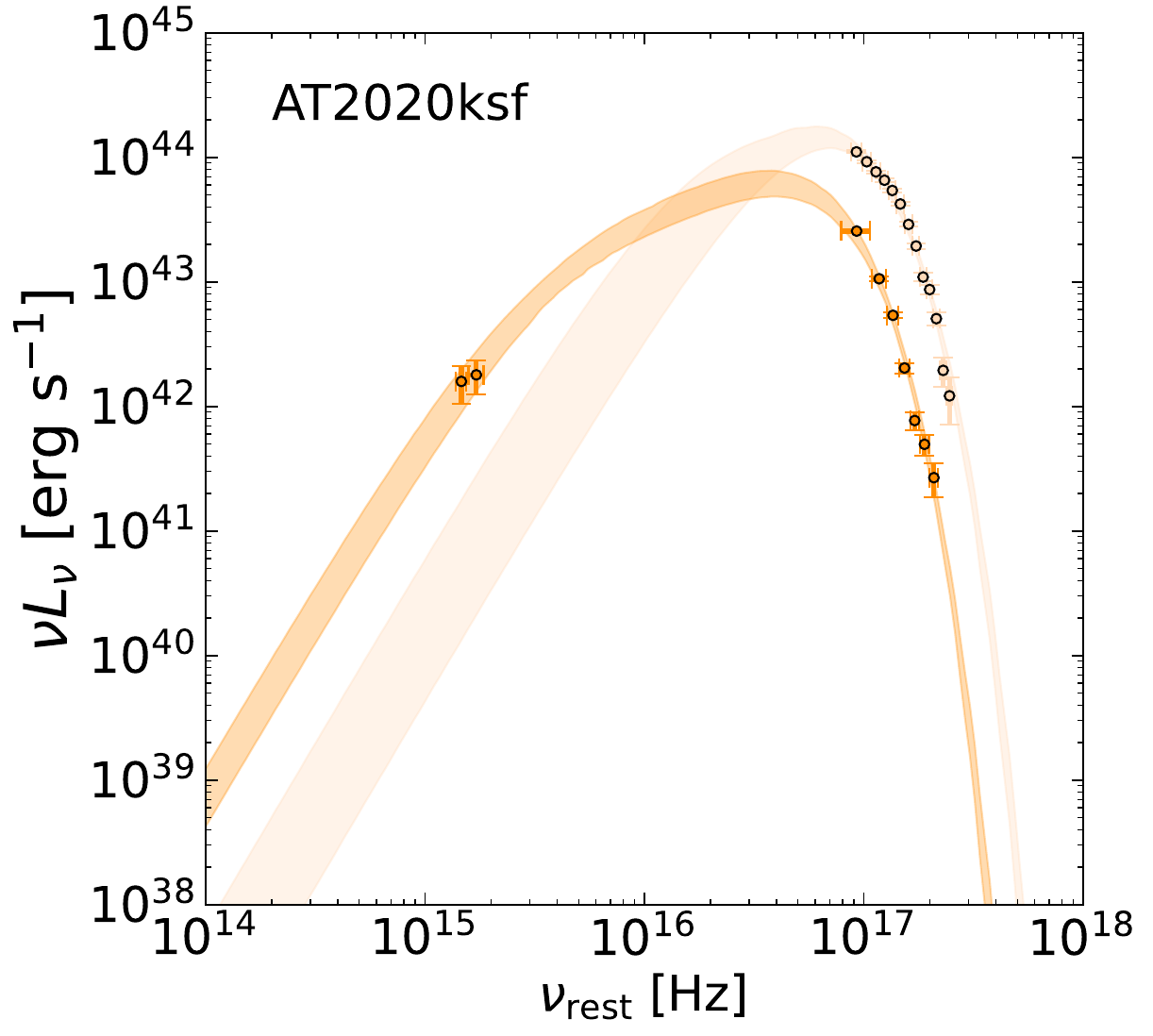}{0.245\textwidth}{}
          \fig{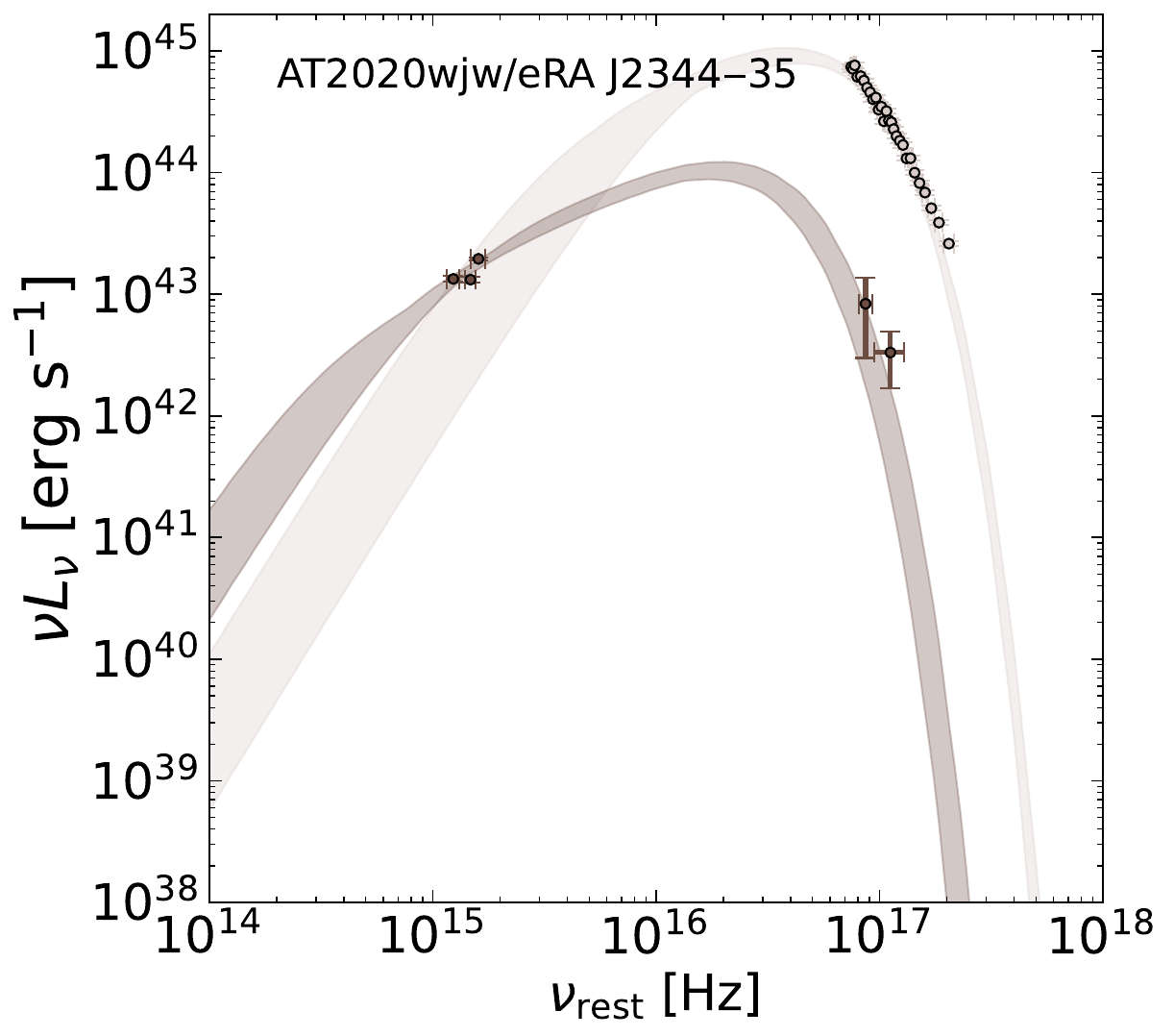}{0.245\textwidth}{}} % two empty slots for alignment

\caption{Results of the spectral modeling for our sample. Each panel shows the intrinsic rest-frame (i.e., corrected for absorption and extinction) UV/optical photometry and the unfolded X-ray spectra. For sources with X-ray–only epochs (see Table~\ref{tab:SED}), the UV/optical flux of the disk component is displayed for illustrative purposes only, as the outer disk parameters are unconstrained. Contours show the 68\% credible interval of the model posterior. }
\label{fig:SED_fit2}
\end{figure*}

Our goal is to analyze the full spectral energy distribution (SED) of TDEs in the late-time ``plateau'' phase using disk models. It is therefore important to define our labeling of the different emission phases carefully. Following, e.g., \citet{van_velzen19_late_time_HST,Mummery2024,Mummery_vanVelzen2024}, we divide UV/optical TDE light curves into two phases, which we refer to as the ``early-time phase'' and the ``late-time plateau phase.'' These phases are illustrated in Fig.~\ref{fig:lcs} and are distinguished primarily by the rate of flux decline: rapid in the early phase and slow or nearly constant in the plateau phase. To determine the transition between these phases, we follow \citet{Mummery2024} and model the temporal evolution of each UV/optical light curve with a Gaussian rise, when observations of the rise are available, an exponential decay, and a constant plateau component. We define the beginning of the plateau phase as the epoch at which the fitted exponential component becomes fainter than the typical flux uncertainty at that epoch (see Appendix~\ref{app:data}). This criterion ensures that the contribution of the decaying early-time flare to the plateau flux is negligible. For sources whose UV/optical light curves are too sparsely sampled for this procedure, we classify observations obtained more than two years after the X-ray peak as belonging to the plateau phase.

To be amenable to the type of spectral modeling we wish to perform, any source in our sample must satisfy three observational requirements: (i) have at least one reasonably good (at an absolute minimum, 50 background-subtracted counts), thermally dominated X-ray spectrum; (ii) have a detected UV/optical plateau; and (iii) have some X-ray data during the plateau phase (even if only an upper limit). These requirements naturally restrict us to relatively nearby TDEs that have been detected in X-rays and to sources that were not discovered very recently (as the plateau phase typically becomes observable $\gtrsim1$--$2$ yr post-discovery). Using all publicly available data, combined with ongoing late-time follow-up programs with XMM-Newton (0942540 and 094256, PI Guolo; 094080, PI Yao), we identified 16 sources that meet these criteria, which are listed in Table~\ref{tab:sample}.

Our sample is not selected according to a uniform discovery survey or wavelength, but rather according to the availability of long-term, high-quality data. Most sources were discovered by wide-field optical surveys, including the Zwicky Transient Facility (ZTF; \citealt{Bellm2019}), the All-Sky Automated Survey for Supernovae (ASAS-SN; \citealt{Shappee2014}), and the Asteroid Terrestrial-impact Last Alert System (ATLAS; \citealt{Tonry2018}). Five events---GSN 069, 3XMM J2150$-$05, AT2018cqh, AT2020ksf, and AT2020wjw/eRA J2344-35---were instead discovered in X-rays, either by the XMM-Newton Slew Survey \citep{Saxton2008} or by eROSITA \citep{Predehl2021}, the soft X-ray telescope aboard the Spectrum-Roentgen-Gamma (SRG) mission. Nevertheless, all five have late-time UV/optical detections during their plateau phase, and four of them (all except GSN 069, for which no meaningful constraints are available) were also detected in the optical during their early-time flare. It is worth noting that 3XMM J2150$-$05 \citep{Lin2018} originates from an ultra-compact dwarf galaxy (UCD; $M_{\rm gal}\sim10^7\msun$) located on the outskirts of a massive ($M_{\rm gal}\sim10^{11}\msun$) lenticular galaxy, is one of the best-known off-nuclear IMBH candidates \citep{Lin2020,Greene2020,Wen2021}, and was the first off-nuclear TDE ever discovered.

%Available data for all sources were collected from the literature and public archives.
On the X-ray side, the majority of the spectra used here were obtained with the \textit{European Photon Imaging Camera} (EPIC-pn; \citealt{Struder2001}) onboard \textit{XMM-Newton} \citep{Jansen2001}, with complementary observations from the \textit{Neil Gehrels Swift Observatory} X-Ray Telescope (\swift/XRT; \citealt{Burrows2005}), the \textit{Neutron Star Interior Composition Explorer} X-ray Timing Instrument (\nicer/XTI; \citealt{Gendreau2016}), and the \textit{Chandra X-ray Observatory} Advanced CCD Imaging Spectrometer (\chandra/ACIS-S; \citealt{Garmire2003}).

Host-subtracted UV/optical photometry, from which we constructed the spectral energy distributions (SEDs), was obtained in most cases from the \texttt{manyTDE}\footnote{https://github.com/sjoertvv/manyTDE} database \citep{Mummery2024}. The data-reduction procedures adopted in this database follow standard practices in the literature and include, for example, subtraction of the host-galaxy contribution using stellar-population models constrained by observations in other filters when pre-TDE imaging is unavailable for direct image subtraction. For a subset of sources, we relied on individual reductions and analyses presented in previous dedicated studies. Details of the data-reduction procedures, along with references for each dataset, are provided in Appendix~\ref{app:data}, but follow standard procedures in the TDE literature. Host-galaxy masses ($M_{\rm gal}$) and nuclear stellar velocity dispersions ($\sigma_{\star}$) were also collected from the literature (again, mostly from the \texttt{manyTDE} compilation), as well as from individual studies.

We also present new measurements of $\sigma_{\star}$ for AT2019vcb, AT2022dsb, AT2023vcb, and AT2020wjw. The corresponding data and measurement procedures are described in Appendix~\ref{app:data}. With these additions, $\sigma_{\star}$ is now available for all sources in our sample except 3XMM~J2150$-$05. All host-galaxy properties are summarized in Table~\ref{tab:sample}.

The light curves serve as our starting dataset, which we use to determine the phase of the UV/optical evolution (early-time flare versus plateau); however, our final analysis relies on combined spectral energy distributions (SEDs). For each source, we selected between one and three SED epochs, spaced as widely as possible to capture the time evolution of the disk properties. Up to two of these epochs were chosen during the plateau phase of the UV/optical light curve, for which contemporaneous X-ray spectral data were available. For each such epoch, we constructed a median UV/optical SED from the light curves using the procedure described in Appendix~\ref{app:data}. These median SEDs (without extinction correction) were loaded into the X-ray spectral-fitting framework using the \texttt{ftflx2xsp} tool in HEASoft v6.33.2 \citep{Heasarc2014}, which generates the appropriate response files.

The corresponding X-ray spectra were, in most cases, obtained either from deep single exposures or by stacking multiple shorter observations to achieve a high signal-to-noise ratio. For sources for which high-quality early-time X-ray spectra were available---taken after or near the X-ray peak, but not during the rise (the reasons for excluding rising data are clarified in Section~\ref{sec:model})---we included them as an additional ``X-ray-only'' epoch.\footnote{For readers concerned about this point, the peak X-ray luminosity function and the UV/optical plateau luminosity function can be reproduced from one another without invoking any additional absorption or reprocessing, as recently shown by \citet{Mummery_vanVelzen2024}. This demonstrates, at least at the population level, that the two components arise from the same disk---an idea that goes back to \citet{Mummery2020}---and here we show that this also holds for individual sources.} We did not include contemporaneous UV/optical data in these cases, since the early-time optical flare is known not to originate from direct disk emission. Nevertheless, as initially argued at the individual-source level by \citet{Mummery2020} and at the population level by \citet{Mummery_vanVelzen2024}, the TDE-disk paradigm assumes that the X-ray emission at peak light is produced by the same accretion disk that powers the late-time UV/optical emission. By including early-epoch X-ray data in our sample, we can directly test this key prediction of the framework. Additionally, adding a brighter X-ray spectrum, which usually has a high signal-to-noise ratio, helps constrain parameters intrinsic to the black hole (e.g., spin). All X-ray spectra were binned using the optimal scheme of \citet{Kaastra2016}, and a 1\% systematic uncertainty was added to all data to account for any differences in the intrinsic flux calibration of different X-ray missions.

Even for sources with sufficient data to construct more SED epochs, we restricted our analysis to a maximum of three per source. This choice reflects two considerations: (i) only a small number of epochs are needed to robustly constrain intrinsic black hole and disk properties, with little additional gain in precision (e.g., for $M_{\bullet}$, because at this point systematic uncertainties and model degeneracies dominate over statistical uncertainty) from including more; and (ii) practical constraints. Because our fitting is performed simultaneously across all epochs and wavelengths, within a Bayesian framework (i.e., a full parameter-space search), and using fully relativistic, computationally expensive numerical photon ray tracing, the computational cost---which is already of the order of many CPU hours per source with $\leq3$ epochs---scales steeply with the number of epochs. The SED epochs, time ranges, and data used in both wavelength bands are presented in Table~\ref{tab:SED}.

\vspace{0.5cm}
\section{Model and Fitting}\label{sec:model}

As we shall model emission from accretion disks around black holes with a wide range of masses, accretion rates, and temperatures, all of which evolve between the distinct epochs considered here, our approach should include sufficient free parameters and physical detail to explain the data and capture the specifics of TDE disks while also allowing parameter degeneracies to be broken. At the same time, we should remain cautious about overfitting, ensuring that no parameters beyond those constrained by the data are introduced.  

In the context of TDEs---but also applicable to other thermal X-ray accretion sources---the \texttt{kerrSED} model \citep{Guolo_Mummery2025} is designed to achieve this balance. It is a relativistic, color-corrected, quasi-steady-state disk model with a vanishing-stress inner boundary condition and five free parameters: inner disk radius ($R_{\rm in}$), peak disk temperature ($T_{\rm p}$), outer disk radius ($R_{\rm out}$), black hole spin ($a_\bullet$), and inclination ($i$). Implemented in the \texttt{python} version of \texttt{XSPEC}, it allows multi-wavelength fitting and performs numerical ray tracing on the fly to capture relativistic effects. Relative to standard soft/thermal spectral models, \texttt{kerrSED} is a significant improvement: it includes relativistic photon-trajectory corrections, treats disk size as a free parameter (essential for TDEs), and applies a temperature-dependent correction for radiative-transfer effects.

The reader is referred to \citet{Guolo_Mummery2025} for full details and for a visual demonstration of the effects of each parameter on the resulting emission; here, we briefly summarize the main properties of \texttt{kerrSED}. The model essentially fits the following expression to the observed data---denoted by the subscript $o$---after convolving it with the instrumental response in the case of the X-ray spectra:  

\begin{equation}\label{eq:flux_kerr}
    F_\nu(\nu_{\rm o}) = \iint_{\cal S} g^3 f_{c}^{-4} \,
    B_\nu \!\left(\nu_{\rm o}/g , f_{c} T_{r}\right)\,
    \mathrm{d}\Theta_{o},
\end{equation}

\noindent where $\mathcal{S}(R_{\rm in},R_{\rm out})$ denotes the disk surface, defined by inner and outer radii, with the inner boundary taken to be the innermost stable circular orbit ($r_{\rm isco}$) of the black hole. The function $g(r,\phi|a_\bullet,i)\equiv\nu_o/\nu_e$ is the photon energy-shift factor, which captures the combined effects of Doppler and gravitational photon-energy shifts, and is defined as the ratio of the observed to emitted photon energy. For a given black hole spin and observer inclination angle, it depends on both the radius and azimuthal angle at which the photon was emitted within the disk. We calculate $g(r,\phi|a_\bullet,i)$ numerically using the ray-tracing algorithm described in \citet{Mummery2024fitted}, based on the code \texttt{YNOGK}, itself derived from \texttt{GEOKERR} \citep{Yang2013,Dexter2009}. The factor $f_c(T)$ is the temperature-dependent color correction \citep{Shimura1995,Hubeny2001,Davis2006,Done2012,Davis19}, which aims to model non-local thermodynamic equilibrium (non-LTE) effects such as metal opacity and electron scattering in the disk atmosphere. The function $B_\nu(\nu,T)$ is simply the Planck function, as the disk emission is quasi-thermal in its rest frame. 

The radial temperature profile $T_r(r)$ is assumed to follow the standard steady-state-like, null-stress boundary-condition profile:  

\begin{equation}\label{eq:T}
    T_r(r) = \left( \frac{r_{\rm max}^3}{1 - r_{\rm max}^{-1/2}} \right)^{1/4}
    \, T_p \, r^{-3/4} \, \left(1 - r^{-1/2}\right)^{1/4},
\end{equation}

\noindent with $r_{\rm max}=49/36$ \citep{Novikov1973,Shakura1973}.\footnote{Formally, this temperature profile holds exactly only in the asymptotic $t\gtrsim t_{\rm visc}$ limit and is unlikely to hold during the rise of a TDE's X-ray light curve, which necessarily probes sub-viscous timescales. Beyond $t_{\rm visc}$, however, this is a good approximation; see, for example, Figure 4 of \citet{Mummery2020}. This is why, in \S~\ref{sec:data}, we did not use X-ray spectra obtained during the rise of the X-ray light curve.} Finally, $\mathrm{d}\Theta_o$ is the differential solid-angle element as seen by the observer, which may be expressed as $\mathrm{d}\Theta_o=\mathrm{d}b_x\mathrm{d}b_y/D^2$, where $b_x$ and $b_y$ are the photon impact parameters at infinity \citep{Li2005}, and $D$ is the distance to the observer. For this implementation of \texttt{kerrSED}, we adopted the $f_c$ prescription from \citet{Hubeny2001}.\footnote{We tested refitting the same data for some sources using alternative color-correction prescriptions \citep[e.g.,][]{Done2012} and found that the effect on the inferred $M_{\bullet}$ was $<0.1\ \rm{dex}$. For the black hole mass range studied here, the two prescriptions appear very similar, although they are expected to diverge for disks around stellar-mass black holes with higher $T_p$, as in X-ray binaries.}

It is also useful to define quantities that are not free parameters of the model but are derived from them. For example, $M_{\bullet}$, $a_{\bullet}$, and $R_{\rm in}$ can be written interchangeably as functions of one another:  

\begin{equation}\label{eq:mbh}
     R_{\rm in} = \gamma(a_{\bullet})\frac{GM_{\bullet}}{c^2} \quad {\rm or} \quad  
     M_{\bullet} = \frac{R_{\rm in}c^2}{\gamma(a_{\bullet})G},
\end{equation}

\noindent where $\gamma(a_{\bullet})$ is the standard spin-dependent factor $r_{\rm isco}/r_g$ \citep[e.g.,][]{MTW}, with $\gamma(0)=6$, $\gamma(-1)=9$, and $\gamma(1)=1$. Throughout the paper, quoted $M_{\bullet}$ values refer to values derived from this relation and obtained by fitting SED data, unless otherwise stated.

Similarly, the bolometric disk luminosity is given by  

\begin{equation}
  L_{\rm Bol}^{\rm disk} 
     = 4\pi \sigma r_g^2 \int_{R_{\rm in}}^{R_{\rm out}} r \, T_r^4(r) \, {\rm d}r.
\end{equation}

\noindent Importantly, $L_{\rm Bol}^{\rm disk}\neq4\pi D^2\int_{0}^{\infty}F_{\nu}(\nu)\,{\rm d}\nu$, where $F_{\nu}$ is the observed flux (Eq.~\ref{eq:flux_kerr}) at a given inclination, because a relativistic disk does not emit isotropically (specifically, Doppler shifting breaks the isotropy). Lastly, $L_{\rm Edd}=1.26\times10^{38}(M_{\bullet}/M_{\odot})$ is the Eddington luminosity.

\begin{figure*}
    \centering
    \includegraphics[width=0.8\linewidth]{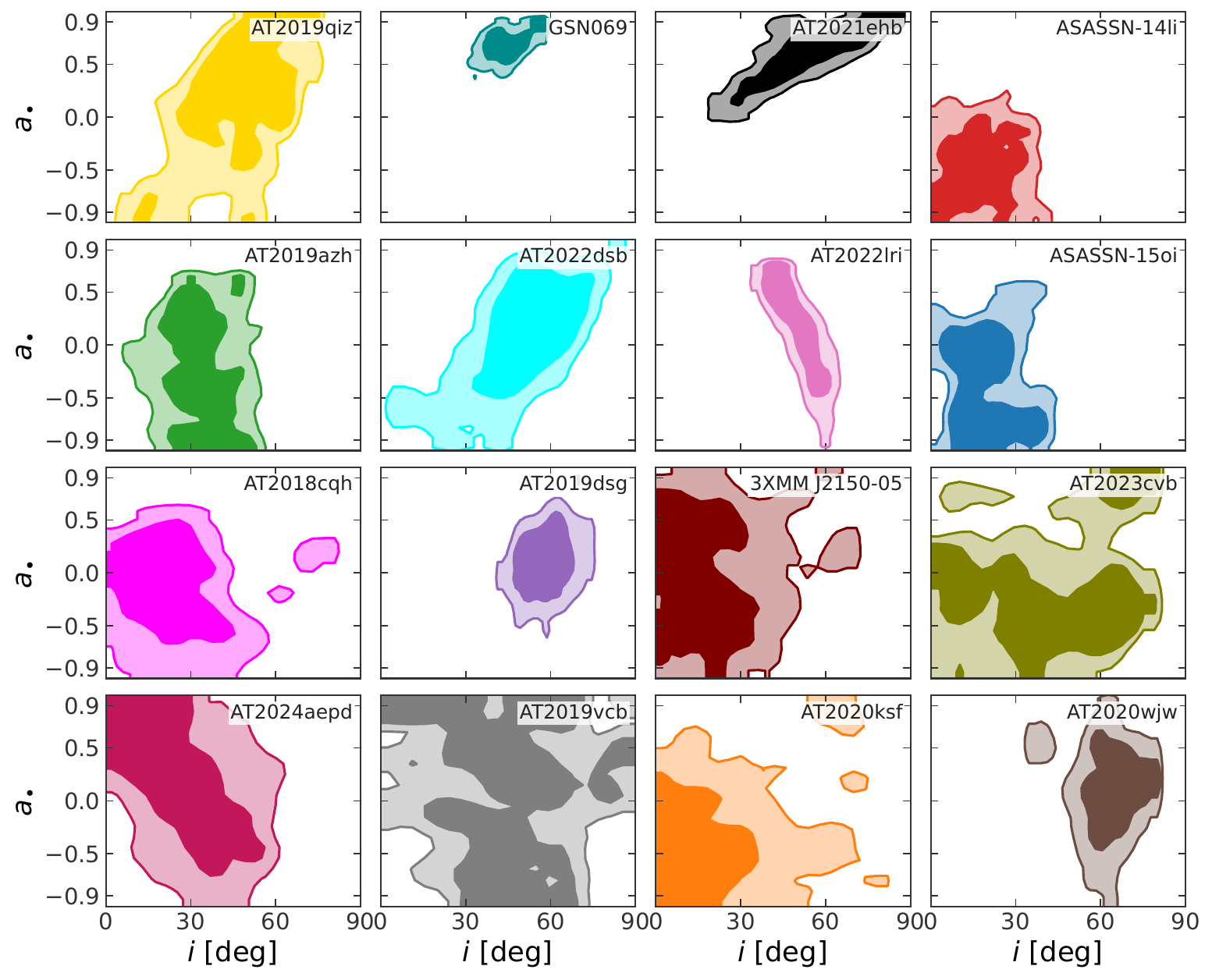}
    \caption{Probability density of the posterior in the spin ($a_{\bullet}$)–inclination ($i$) parameter plane. Contours indicate the 68\% and 99\% credible intervals. The figure illustrates the varying ability to constrain these parameters from source to source, as well as the advantage of a relativistic model over a classical one: for nearly all sources, a fraction of the prior space can be excluded.}
    \label{fig:a_i}
\end{figure*}

Before discussing the fitting procedures, it is important to clarify which model parameters are intrinsic to the system and which are dynamical, i.e., expected to evolve between epochs. By construction, $R_{\rm in}$ and $a_{\bullet}$ (and therefore $M_{\bullet}$ as well) are intrinsic properties of the black hole. They cannot vary between epochs: although allowing them to vary may produce formally acceptable statistical fits, such solutions are unphysical and inconsistent with the assumptions of the model.\footnote{At extremely low accretion rates ($\ll10^{-2}$ of the Eddington rate), this assumption may break down if TDE disks transition through different accretion states in a manner analogous to X-ray binary (XRB) systems \citep{Maccarone2003,Fender2004}. In such regimes, the inner radius of the thin disk may recede (although this is not a settled question in the XRB literature, see e.g., \citet{Parker2015,Basak2017,Connors2022}). However, for the accretion rates probed in this work, this transition is not expected to occur, and a fixed $R_{\rm in}$ provides an adequate description of the data without the need to introduce additional free parameters.} The inclination $i$ is set by the geometry of the disk--observer system. In principle, it could vary over time (e.g., due to disk precession at early times). However, in practice, degeneracies between variable-$i$ solutions and changes in $T_{\rm p}$ or $a_{\bullet}$ cannot be resolved with current data, so allowing $i$ to vary would result in more free parameters than can be constrained by the data. We therefore assume a constant $i$ across epochs. By contrast, the peak disk temperature $T_{\rm p}$ and the outer radius $R_{\rm out}$ are dynamical properties of the disk. In a system with a finite mass supply, such as a TDE, both must evolve as the disk accretes material \citep{Cannizzo1990,Mummery2020}.

\begin{deluxetable}{lcccccl}\label{tab:kerrsed_params}
\tablecaption{Model Parameter Priors.}
\tabletypesize{\scriptsize}
\tablehead{ 
  \colhead{Model} & \colhead{Parameter} & \colhead{Type$^{(a)}$} & \colhead{Range} &  \colhead{Units} &  \colhead{Prior} 
}
\startdata
\texttt{phabs} & $N_{\rm H}$         & I & $10^{19}$ - $10^{22}$      & $\rm{cm^{-2}}$     & Log-uniform \\
\hline
\texttt{reddenSF} & E(B-V)            & I & $10^{-3}$ - $1$            & \nodata            & Log-uniform$^{(b)}$  \\
\hline
\texttt{kerrSED} & $R_{\rm in}$       & I & $10^4$ - $10^9$            & $\rm{km}$          & Log-uniform  \\
\texttt{kerrSED} & $a_\bullet$        & I & $-0.998$ - $0.998$         & \nodata            & Uniform  \\
\texttt{kerrSED} & $i$                & I & $0$ - $90$                 & deg                & Uniform  \\
\texttt{kerrSED} & $T_{\rm p}$        & D & $10^5$ - $5\times10^6$     & Kelvin             & Log-uniform  \\
\texttt{kerrSED} & $R_{\rm out}$      & D & $30$ - $10^5$       & $r_g$              & Log-uniform  \\
\hline
\texttt{simPL} & $f_{\rm sc}$        & D & $10^{-3}$ - $0.5$          & \nodata            & Log-uniform \\
\texttt{simPL} & $\Gamma$            & D & $1.5$ - $4.0$              & \nodata            & Uniform \\
\enddata
\tablecomments{(a) Intrinsic (I) parameters are kept constant between epochs, while dynamical (D) parameters are allowed to vary (see \S\ref{sec:model} for details). (b) The color excess, $E(B-V)$, is either tied to $N_{\rm H}$ assuming a Galactic gas-to-dust ratio, or allowed to vary independently depending on the source (see Table~\ref{tab:best_fit}).}
\end{deluxetable}

The spectral energy distribution fitting (X-ray spectra and UV/optical photometry) is performed with version 4.0.7 of the Bayesian X-ray Analysis software (BXA; \citealt{Buchner2014}), which connects the nested-sampling algorithm \texttt{UltraNest} \citep{Buchner2019} to the fitting environment \texttt{PyXspec} \citep{Arnaud_96}. The $E_n$ epochs are loaded and fitted simultaneously, assuming Gaussian statistics. In addition to the disk model, we account for absorption and reddening by gas and dust along the line of sight, both Galactic and intrinsic to the host, while the source emission is redshifted according to the corresponding $z$. Our fiducial total model (\texttt{Model 1}) in \texttt{XSPEC} notation is \texttt{phabs$\times$redden$\times$zashift(phabs$\times$reddenSF$\times$kerrSED)}, where \texttt{redden} is the native \texttt{XSPEC} Galactic dust-extinction model \citep{Cardelli1989}, while \texttt{reddenSF} implements the law of \citet{Calzetti2000} \citep{Guolo_Mummery2025}. The Galactic extinction color excess, $E(B-V){\rm G}$, the hydrogen-equivalent column density, $N{\rm H,G}$, and the source redshift ($z$ from \texttt{zashift}) are fixed at their known values, as listed in Table~\ref{tab:sample}.
For some sources, a hard X-ray excess---thought to be produced by Comptonization in a ``corona'' of hot electrons above the disk \citep[e.g.,][]{Haardt1991}---is observed in addition to the direct disk emission and may even dominate the X-ray spectrum during some epochs. To model this component, we use the convolution model \texttt{SimPL} \citep{Steiner_09}. The corresponding model (\texttt{Model 2}), applied to such epochs, is \texttt{phabs$\times$redden$\times$zashift(phabs$\times$reddenSF$\times$(SimPL$\otimes$ kerrSED))}.

\begin{deluxetable*}{lcccccccccc}
\tablecaption{SED Fitting Parameters \label{tab:best_fit}}
\tablehead{
\colhead{Source} & \colhead{Epoch} & \colhead{Model} &  
\colhead{$\log_{10}(N_{\rm H})$} & \colhead{$E(B-V)^{(a)}$} & \colhead{$\log_{10}(R_{\rm in})$} &
\colhead{$\log_{10}(T_{\rm p})$} & \colhead{$\log_{10}(R_{\rm out})$} & \colhead{$\log_{10}(M_{\rm \bullet})$} &
\colhead{$\log_{10}(L_{\rm Bol}^{\rm disk})$} & \colhead{$\log_{10}(L_{\rm Bol}^{\rm disk}/L_{\rm Edd})$} \\
\colhead{} & \colhead{} & \colhead{} & \colhead{(cm$^{-2}$)} & \colhead{} & \colhead{(km)} &
\colhead{(K)} & \colhead{($r_g$)} & \colhead{($M_{\odot}$)} & \colhead{(erg s$^{-1}$)} & \colhead{}
}
\startdata
\multirow{2}{*}{AT2019qiz}  & 1 & 1 & \multirow{2}{*}{$21.36^{+0.05}_{-0.04}$} & \multirow{2}{*}{$<0.02$}$\dagger$ & \multirow{2}{*}{$6.66^{+0.10}_{-0.10}$} & $5.54^{+0.03}_{-0.03}$ & $3.08^{+0.18}_{-0.25}$ & \multirow{2}{*}{$5.78^{+0.28}_{-0.26}$} & $43.10^{+0.08}_{-0.09}$ & $-0.79^{+0.16}_{-0.16}$ \\
 & 2 & 1 &  &  &  & $5.48^{+0.03}_{-0.03}$ & $3.25^{+0.26}_{-0.23}$ &  & $42.92^{+0.10}_{-0.08}$ & $-0.93^{+0.16}_{-0.16}$ \\
\hline
\multirow{3}{*}{GSN\,069} & 1 & 1 & \multirow{3}{*}{ $20.66^{+0.03}_{-0.03}$} & \multirow{3}{*}{ $0.05^{+0.01}_{-0.01}$}   & \multirow{3}{*}{$7.16^{+0.06}_{-0.05}$} & $5.47^{+0.03}_{-0.02}$ & -- & \multirow{3}{*}{$6.45^{+0.13}_{-0.11}$} & $43.82^{+0.05}_{-0.03}$ & $-0.72^{+0.05}_{-0.08}$ \\
 & 2 & 1 &  &  &  & $5.45^{+0.02}_{-0.02}$ & $2.70^{+0.08}_{-0.07}$ &  & $43.71^{+0.06}_{-0.03}$ & $-0.82^{+0.05}_{-0.08}$ \\
 & 3 & 1 &  &  &  & $5.42^{+0.02}_{-0.02}$ & $3.07^{+0.58}_{-0.24}$ &  & $43.62^{+0.05}_{-0.03}$ & $-0.92^{+0.05}_{-0.08}$ \\
\hline
\multirow{2}{*}{AT2021ehb} & 1 & 2 & \multirow{2}{*}{$21.02^{+0.12}_{-0.22}$} & \multirow{2}{*}{$0.12^{+0.04}_{-0.05}$} & \multirow{2}{*}{$7.40^{+0.20}_{-0.16}$} & $5.52^{+0.04}_{-0.06}$ & $2.54^{+0.20}_{-0.20}$ & \multirow{2}{*}{$6.66^{+0.35}_{-0.22}$} & $44.49^{+0.17}_{-0.19}$ & $-0.30^{+0.18}_{-0.18}$ \\
 & 2 & 2 &  &  &  & $5.40^{+0.04}_{-0.06}$ & $2.49^{+0.24}_{-0.18}$ &  & $44.01^{+0.18}_{-0.22}$ & $-0.77^{+0.17}_{-0.22}$ \\
\hline
\multirow{3}{*}{ASASSN-14li} & 1 & 1 & \multirow{3}{*}{$20.11^{+0.37}_{-0.60}$} &\multirow{3}{*}{$0.01^{+0.01}_{-0.00}$}  & \multirow{3}{*}{$7.48^{+0.18}_{-0.12}$}  & $5.52^{+0.03}_{-0.03}$ & \nodata & \multirow{3}{*}{$6.44^{+0.23}_{-0.15}$} & $44.49^{+0.18}_{-0.14}$ & $-0.05^{+0.07}_{-0.11}$ \\
 & 2 & 1 &  &  &  & $5.44^{+0.02}_{-0.03}$ & $1.90^{+0.10}_{-0.15}$ &  & $44.25^{+0.23}_{-0.15}$ & $-0.29^{+0.07}_{-0.06}$ \\
 & 3 & 1 &  &  &  & $5.32^{+0.02}_{-0.03}$ & $1.95^{+0.11}_{-0.17}$ &  & $43.77^{+0.24}_{-0.15}$ & $-0.76^{+0.08}_{-0.09}$ \\
\hline
\multirow{3}{*}{AT2019azh} & 1 & 1 & \multirow{3}{*}{$<19.8$} & \multirow{3}{*}{$<0.01$} & \multirow{3}{*}{$7.48^{+0.08}_{-0.08}$ } & $5.48^{+0.02}_{-0.03}$ &\nodata & \multirow{3}{*}{ $6.52^{+0.15}_{-0.11}$} & $44.27^{+0.08}_{-0.10}$ & $-0.26^{+0.08}_{-0.11}$ \\
 & 2 & 1 &  &  &  & $5.36^{+0.02}_{-0.03}$ & $1.97^{+0.09}_{-0.09}$ &  & $43.85^{+0.07}_{-0.05}$ & $-0.67^{+0.05}_{-0.06}$ \\
 & 3 & 2 &  &  &  & $5.23^{+0.04}_{-0.04}$ & $1.99^{+0.09}_{-0.09}$ &  & $43.35^{+0.14}_{-0.17}$ & $-1.18^{+0.13}_{-0.15}$ \\
\hline
AT2022dsb & 1 & 2 & $21.45^{+0.06}_{-0.07}$ & $<0.01$$\dagger$ & $7.52^{+0.12}_{-0.13}$ & $5.56^{+0.03}_{-0.03}$ & $2.16^{+0.15}_{-0.17}$ & $6.75^{+0.27}_{-0.22}$ & $44.85^{+0.17}_{-0.14}$ & $-0.01^{+0.16}_{-0.14}$ \\
\hline
\multirow{3}{*}{AT2022lri} & 1 & 1 & \multirow{3}{*}{$<19.4$}  & \multirow{3}{*}{$<0.01$}  & \multirow{3}{*}{$6.69^{+0.07}_{-0.07}$} & $5.85^{+0.03}_{-0.02}$ & \nodata & \multirow{3}{*}{$5.79^{+0.07}_{-0.09}$} & $44.29^{+0.09}_{-0.07}$ & $0.39^{+0.13}_{-0.07}$ \\
 & 2 & 1 &  &  &  & $5.71^{+0.03}_{-0.02}$ & $2.54^{+0.10}_{-0.07}$ &  & $43.85^{+0.04}_{-0.03}$ & $-0.05^{+0.10}_{-0.06}$ \\
 & 3 & 2 &  &  &  & $5.62^{+0.03}_{-0.02}$ & $2.62^{+0.10}_{-0.08}$ &  & $43.46^{+0.04}_{-0.03}$ & $-0.43^{+0.10}_{-0.06}$ \\
\hline
\multirow{2}{*}{ASASSN-15oi} & 1 & 1 & \multirow{2}{*}{$19.78^{+0.43}_{-0.39}$} & \multirow{2}{*}{$0.01^{+0.01}_{-0.00}$} & \multirow{2}{*}{$7.40^{+0.13}_{-0.10}$} & $5.46^{+0.02}_{-0.02}$ & $1.83^{+0.11}_{-0.12}$ & \multirow{2}{*}{$6.38^{+0.19}_{-0.14}$} & $44.13^{+0.15}_{-0.08}$ & $-0.35^{+0.05}_{-0.05}$ \\
 & 2 & 2 &  &  &  & $5.15^{+0.11}_{-0.10}$ & $1.89^{+0.17}_{-0.15}$ &  & $42.95^{+0.41}_{-0.40}$ & $-1.55^{+0.39}_{-0.39}$ \\
\hline
AT2018cqh & 1 & 2 & $20.20^{+0.24}_{-0.56}$ & $0.02^{+0.01}_{-0.01}$ & $6.78^{+0.14}_{-0.12}$ & $5.56^{+0.02}_{-0.04}$ & $2.61^{+0.23}_{-0.18}$ & $5.81^{+0.20}_{-0.22}$ & $43.38^{+0.15}_{-0.16}$ & $-0.53^{+0.04}_{-0.06}$ \\
\hline
\multirow{3}{*}{AT2019dsg} & 1 & 1 & \multirow{3}{*}{$20.77^{+0.06}_{-0.06}$ } & \multirow{3}{*}{$0.07^{+0.01}_{-0.01}$} & \multirow{3}{*}{$7.64^{+0.10}_{-0.09}$} & $5.49^{+0.02}_{-0.02}$ & \nodata & \multirow{3}{*}{$6.72^{+0.19}_{-0.15}$} & $44.78^{+0.11}_{-0.13}$ & $-0.06^{+0.09}_{-0.10}$ \\
 & 2 & 1 &  &  &  & $5.43^{+0.02}_{-0.02}$ & \nodata &  & $44.47^{+0.14}_{-0.14}$ & $-0.34^{+0.09}_{-0.12}$ \\
 & 3 & 1 &  &  &  & $5.18^{+0.11}_{-0.10}$ & $2.16^{+0.21}_{-0.18}$ &  & $43.60^{+0.36}_{-0.33}$ & $-1.21^{+0.36}_{-0.41}$ \\
\hline
\multirow{2}{*}{3XMM J2150-05} & 1 & 1 & \multirow{2}{*}{$< 20.25$} & \multirow{2}{*}{$< 0.02 $} & \multirow{2}{*}{$5.24^{+0.13}_{-0.05}$} & $6.18^{+0.03}_{-0.05}$ & \nodata & \multirow{2}{*}{$4.39^{+0.17}_{-0.08}$} & $42.84^{+0.05}_{-0.03}$ & $0.35^{+0.04}_{-0.05}$ \\
 & 2 & 1 &  &  &  & $5.94^{+0.03}_{-0.05}$ & $3.68^{+0.09}_{-0.10}$ &  & $41.89^{+0.06}_{-0.05}$ & $-0.61^{+0.04}_{-0.07}$ \\
\hline
AT2023cvb & 1 & 1 & $20.27^{+0.26}_{-0.72}$ & $0.02^{+0.02}_{-0.02}$ & $7.31^{+0.23}_{-0.24}$ & $5.55^{+0.08}_{-0.06}$ & $2.45^{+0.18}_{-0.23}$ & $6.43^{+0.28}_{-0.24}$ & $44.42^{+0.21}_{-0.20}$ & $-0.08^{+0.06}_{-0.10}$ \\
\hline
\multirow{2}{*}{AT2024aepd}  & 1 & 1 &  \multirow{2}{*}{ $< 20.51$}   &  \multirow{2}{*}{  $<0.01$} & \multirow{2}{*}{  $7.50^{+0.34}_{-0.23}$ }  & $5.43^{+0.05}_{-0.09}$ & \nodata & \multirow{2}{*}{$6.65^{+0.37}_{-0.33}$}   & $44.30^{+0.50}_{-0.38}$ & $-0.46^{+0.27}_{-0.17}$ \\
 & 2 & 2 & &  &  & $5.18^{+0.11}_{-0.14}$ & $2.89^{+0.43}_{-0.49}$ &  & $43.33^{+0.31}_{-0.26}$ & $-1.32^{+0.27}_{-0.55}$ \\ \hline
\multirow{2}{*}{AT2019vcb} & 1 & 1 & \multirow{2}{*}{$<20.33 $} & \multirow{2}{*}{$<0.02$} & \multirow{2}{*}{$6.77^{+0.38}_{-0.25}$} & $5.57^{+0.08}_{-0.07}$ & \nodata & \multirow{2}{*}{$5.85^{+0.41}_{-0.25}$} & $43.45^{+0.33}_{-0.25}$ & $-0.50^{+0.13}_{-0.18}$ \\
 & 2 & 1 &  &  &  & $5.43^{+0.08}_{-0.07}$ & $> 2.23$ &  & $42.89^{+0.35}_{-0.23}$ & $-1.03^{+0.14}_{-0.20}$ \\
\hline
\multirow{2}{*}{AT2020ksf} & 1 & 1 & \multirow{2}{*}{$20.07^{+0.45}_{-0.52}$} &  \multirow{2}{*}{$ 0.02^{+0.01}_{-0.01}$}  & \multirow{2}{*}{$7.08^{+0.23}_{-0.12}$} & $5.70^{+0.03}_{-0.03}$ & \nodata & \multirow{2}{*}{$5.97^{+0.21}_{-0.13}$} & $44.33^{+0.16}_{-0.12}$ & $0.29^{+0.15}_{-0.10}$ \\
 & 2 & 1 &  &  &  & $5.59^{+0.03}_{-0.03}$ & $2.55^{+0.17}_{-0.19}$ &  & $44.06^{+0.11}_{-0.09}$ & $0.04^{+0.05}_{-0.08}$ \\ \hline
 \multirow{2}{*}{AT2020wjw}  & 1 & 1 &  \multirow{2}{*}{ $20.60^{+0.11}_{-0.11}$}  & \multirow{2}{*}{ $0.05^{+0.01}_{-0.01}$}  &  \multirow{2}{*}{$7.91^{+0.17}_{-0.09}$}   & $5.55^{+0.03}_{-0.03}$ & \nodata & \multirow{2}{*}{$7.03^{+0.17}_{-0.16}$ }  & $45.59^{+0.21}_{-0.11}$ & $0.48^{+0.09}_{-0.09}$ \\
& 2 & 1 & &  & & $5.32^{+0.04}_{-0.08}$ & $> 2.59$ &  & $44.71^{+0.15}_{-0.13}$ & $-0.37^{+0.12}_{-0.18}$ \\
\hline
\enddata
\tablecomments{(a) $\dagger$ symbol imply that E(B-V) and $N_{\rm H}$ were let to vary freely, for the remaining sources they were tied using a Galactic gas-to-dust ratio, see text for details.}
\end{deluxetable*}

The intrinsic and dynamical parameters of \texttt{kerrSED} are, as discussed above, fixed across epochs or allowed to vary freely between them, respectively. The only exception is $R_{\rm out}$ for sources whose first epoch ($E_1$) is constrained by X-ray data alone. In those cases, we impose $R_{\rm out}(E_1)\leq R_{\rm out}(E_n)$ for all $n>1$. This restriction is physically required by angular-momentum conservation (fixed-mass disks with conserved angular momentum can only expand) and also aids fitting convergence and visualization. No scientific conclusions are drawn from the posteriors of $R_{\rm out}(E_1)$ for these sources because the available data do not constrain this parameter (only UV/optical data constrain $R_{\rm out}$; see the discussion in \citealt{Guolo_Mummery2025}).

Lastly, to use as few free parameters as possible, we first attempt to fit all sources by assuming a Milky Way-like gas-to-dust ratio for the intrinsic host-galaxy column density and dust-attenuation color excess. Specifically, we tie $N_{\rm H}$ and $E(B-V)$ using $N_{\rm H}({\rm cm}^{-2})=2.21\times10^{21}\times R_V,E(B-V)$ \citep{Guver2009}, where $R_V=4.05$ for the law of \citet{Calzetti2000}. We then evaluate the fit results using standard $Q$--$Q$ residual plots \citep[see, e.g.,][]{Buchner2023}. For most sources, this assumption provides a satisfactory description of the data. However, the fits fail catastrophically for 2/16 sources. For those sources, we untie $N_{\rm H}$ and $E(B-V)$, allowing both to vary independently while remaining fixed across epochs because they are intrinsic host-galaxy properties. This approach yields successful fits for all sources, with the two affected sources requiring finite gas/X-ray absorption but negligible extinction. A detailed list of the free parameters, assumed priors (all either uniform or log-uniform), and adopted ranges is provided in Table~\ref{tab:kerrsed_params}.

\section{Results}\label{sec:results}

The results of our fitting are shown in Figs.~\ref{fig:SED_fit1} and \ref{fig:SED_fit2} and Table~\ref{tab:best_fit}. These results demonstrate that: 
(i) during the plateau phase, the full SED (X-ray spectra and UV/optical photometry) can be consistently described by a compact but otherwise standard accretion disk;
(ii) the parameters of both the black hole and the disk can be constrained with high precision; and
(iii) at early times---when the optical light curve is not produced by direct disk emission---the X-ray spectra remain consistent with the same disk. This final point is demonstrated by the feasibility of the early-time X-ray spectral fit even with $R_{\rm in}$ fixed across all epochs and only $T_p$ allowed to vary. 

More specifically, Figs.~\ref{fig:SED_fit1} and \ref{fig:SED_fit2} present the resulting SED fits. The first shows the observed fluxes (i.e., without absorption or extinction corrections) together with the unfolded X-ray spectra, while the second shows the intrinsic emission after correction for both Galactic and intrinsic gas absorption and dust attenuation. In both figures, the contours indicate the 68\% credible intervals. For epochs in which only X-ray data are fitted (e.g., E1), the model extension to lower energies is shown for illustrative purposes only, as $R_{\rm out}(E1)$ cannot be constrained by the data.

Table~\ref{tab:best_fit} lists the inferred values of the free parameters $N_{\rm H}$, $E(B-V)$, \Rin, \Tp, and \Rout, as well as those of the secondary/derived parameters ($M_{\bullet}$, $L_{\rm Bol}^{\rm disk}$, and $L_{\rm Bol}^{\rm disk}/L_{\rm Edd}$), all of which are calculated element by element from the posterior according to their definitions in Section~\ref{sec:model}. For the three main disk parameters (\Rin, \Tp, and \Rout), the posteriors for almost all  sources and epochs converge to values well within the bounds of our priors (only \Rout does not converge in 2 sources); we therefore report the medians and 68\% credible intervals. This includes the \Rout values for all epochs with simultaneous X-ray and plateau-phase UV/optical data. The only exceptions are AT2019vcb (E2) and AT2020wjw (E2), for which the UV-only detections constrain only lower limits on \Rout. Similarly, the posteriors for $N_{\rm H}$ and $E(B-V)$ show a mixture of convergence and upper limits. The table also indicates which model (1 or 2) was adopted for each epoch and whether $N_{\rm H}$ and $E(B-V)$ were tied or allowed to vary independently.

\begin{figure}
    \centering
    \includegraphics[width=1\columnwidth]{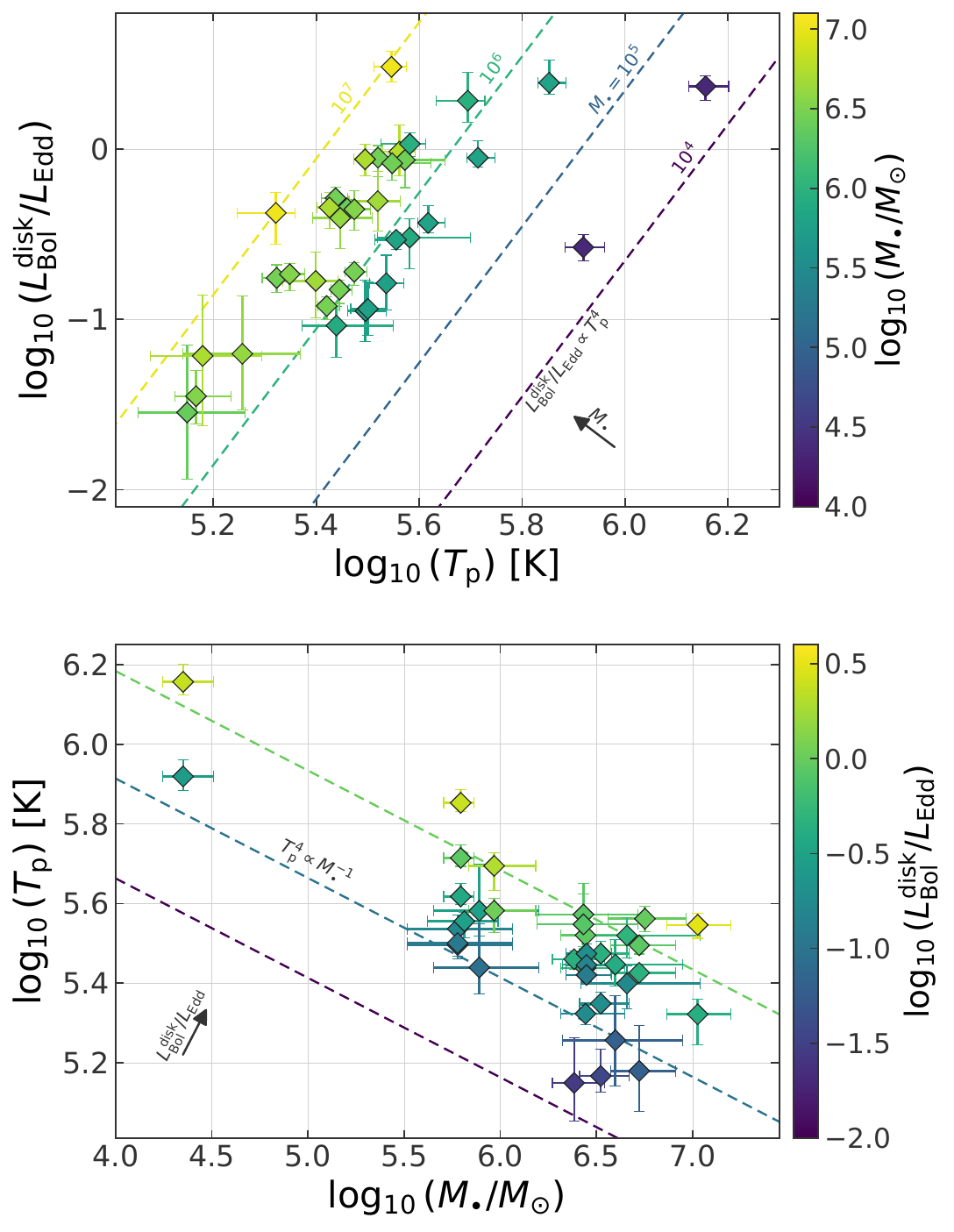}
   
    \caption{Recovered accretion correlations. Top: Eddington ratio ($L_{\rm Bol}^{\rm disk}/L_{\rm Edd}$) versus peak disk temperature ($T_p$). Lines of constant black hole mass ($M_{\bullet}$), spin ($a_{\bullet}$), and disk area trace diagonals of the form $L_{\rm Bol}/L_{\rm Edd}\propto T_p^4$. Distinct epochs of a given source lie approximately along such lines. Lines perpendicular to these diagonals correspond to different $M_{\bullet}$ values, as indicated by the color scheme showing the median of the $M_{\bullet}$ posterior. The lines in the top panel show the correlations for varying $M_{\bullet}/M_{\odot}=10^4,10^5,10^6,10^7$, with $a_{\bullet}=0$ and $R_{\rm out}=100\,r_g$. Bottom: Dependence of $T_p$ on $M_{\bullet}$; at a constant Eddington ratio, sources follow $T_p^4\propto M_{\bullet}^{-1}$. The lines in the bottom panel show the correlations for varying $L_{\rm Bol}^{\rm disk}/L_{\rm Edd}=10^{-2},10^{-1},1$, with $a_{\bullet}=0$ and $R_{\rm out}=100\,r_g$.}

    \label{fig:accretion}
\end{figure}

Importantly, even with a relatively small sample, it is clear that a wide range of parameter values can be recovered. The physical scale of the disk inner radii spans from $R_{\rm in}\sim10^5$ km to $R_{\rm in}\sim10^8$ km, which translates into inferred black hole masses between $\log(M_{\bullet}/M_{\odot})\sim4.4$ and $\log(M_{\bullet}/M_{\odot})\sim7$. When uncertainties are included, this range expands to three orders of magnitude and, notably, confirms the interpretation of 3XMM J2150-05 as an IMBH, consistent with previous studies based on X-ray-only data and alternative models \citep{Lin2018,Lin2020,Wen2021}; this is the only IMBH ($\lesssim10^5 M_{\odot}$) in this sample. Similarly, we recover a wide range of peak disk temperatures spanning $\sim1$ dex, from $\log(T_p/{\rm K})\sim5.2$ to $\log(T_p/{\rm K})\sim6.2$, and relative disk sizes ranging from $\log(R_{\rm out}/r_g)\sim{\rm few}\times10$ to $\sim{\rm few}\times1000$.

\begin{figure*}
    \centering
    \includegraphics[width=0.93\linewidth]{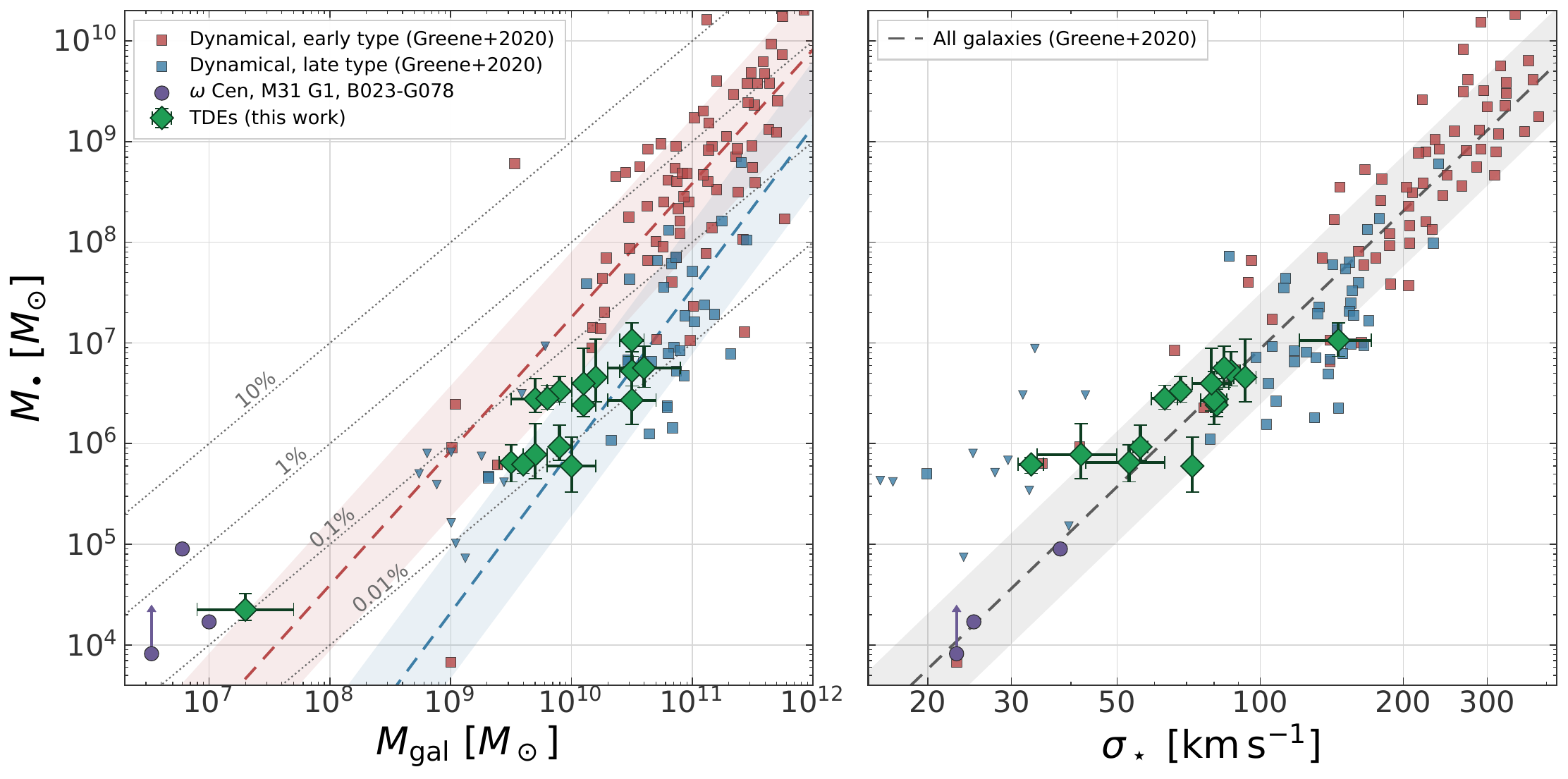} 
    \caption{Black hole–host galaxy scaling relations. Left panel:  galaxy stellar mass ($M_{\rm gal}$) versus black hole mass ($M_{\bullet}$). Right panel: nuclear stellar velocity dispersion ($\sigma_{\star}$) versus $M_{\bullet}$. Blue and red points correspond to late- and early-type galaxies, respectively, with black hole masses measured through stellar or gas dynamical modeling of nearby galaxies; data are taken from \citet{Kormendy2013} and \citet{Greene2020}. Contours represent the best-fit correlations from \citet{Greene2020}. Additional purple points indicate massive globular clusters hosting dynamically inferred IMBHs \citep{Gebhardt2005,Pechetti2022,Haberle2024}. Our TDE sample, with $M_{\bullet}$ inferred from full SED fitting, is shown as green points and lies directly on the established correlations. Diamonds and squares are measurements, and upward/downward triangles are upper/lower limits.} 
    \label{fig:host_scaling}
\end{figure*}

The black hole spin ($a_{\bullet}$) and inclination ($i$), however, act more as nuisance variables, in the sense that varying them across their allowed ranges produces only mild changes in the model fluxes, unlike the variations of many orders of magnitude induced by changing \Rin, \Tp, or \Rout\ \citep[see][for a detailed discussion of parameter dependencies in the emitted spectra]{Guolo_Mummery2025}. Consequently, $a_{\bullet}$ and $i$ rarely converge to well-defined, Gaussian-like posteriors, and reporting their median values and credible intervals could be uninformative or even misleading. Instead, we present their joint posterior distributions in Fig.~\ref{fig:a_i}, with contours marking the 68\% and 99\% credible intervals. Figure~\ref{fig:a_i} shows that (i) the ability to constrain $a_{\bullet}$ and $i$ varies substantially from source to source, depending largely on data quality; and (ii) employing a relativistic model remains valuable, as substantial regions of the $a_{\bullet}\times i$ parameter space can be excluded with high confidence for nearly all sources---a constraint that is not achievable with a Newtonian-like model. Excluding parts of the $a_{\bullet}$--$i$ space directly improves the precision of other parameters, particularly \Tp\ and, most importantly, $M_{\bullet}$ (Eq.~\ref{eq:mbh}).

These spin values, however, should be interpreted with caution, as the inferred values are sensitive to several assumptions. In particular, the results depend on the adoption of a \textit{zero-torque} (or null-stress) inner boundary condition for the disk temperature profile (Eq.~\ref{eq:T})\footnote{This assumption, introduced by \citet{Shakura1973} for simplicity, is commonly used in analytical disk models and in all but one of the fitting implementations currently available (\texttt{fullkerr}; \citealt{Mummery24Plunge}). However, general-relativistic magnetohydrodynamic (GRMHD) simulations \citep[e.g.,][]{Noble2011,Rule2025} and detailed modeling of X-ray binary spectra \citep{Mummery24Plunge,Mummery2025} show this assumption to be physically inaccurate. Current X-ray data quality for TDEs, however, does not allow disk solutions with finite-stress or plunging regions to be tested.} and on the choice of color-correction prescription.\footnote{While the use of a color-correction factor, rather than no correction, is well supported by radiative-transfer simulations of disk atmospheres \citep[e.g.,][]{Davis2006}, the commonly used analytical prescriptions \citep[e.g.,][]{Hubeny2001,Done2012} remain simplifications of the actual physics.} In practice, these factors dominate the systematic uncertainties and the error budget of spin measurements \citep{SalvesenMiller2021,mummery25_spinwrong}.

It is worth noting that we recover a wide range of black hole spins with no clear systematic preference. In a few sources, maximum or near-maximum spins are clearly excluded. This result contrasts with most previous attempts to recover $a_{\bullet}$ from TDEs---typically based on slim-disk models fitted only to X-ray data \citep{Wen2020,wen22_disk_spectrum,Wen2021}---which systematically find $a_{\bullet}\geq0.8$. The origin of this discrepancy is not yet clear and lies beyond the scope of this work. Nevertheless, the black hole masses we derive are generally consistent with those studies, at least to within an order of magnitude, including in extreme cases such as ASASSN-14li and 3XMM J2150-05.

We now analyze the results of our parameter inference, their correlations with each other, and their relations to independent host galaxy properties.

\subsection{Accretion Correlations}\label{sec:accretion}

The fact that these sources can be successfully modeled with standard accretion disk physics naturally implies the presence of correlations between key parameters. Perhaps the most fundamental is
\begin{equation}\label{eq:accretion}
L_{\rm Bol}^{\rm disk}/L_{\rm Edd} \propto T_p^4 \propto  M_{\bullet}^{-1},
\end{equation}
\noindent which expresses the well-known result that, at fixed Eddington ratio, lower-mass black holes host hotter disks, while higher-mass black holes host cooler disks. Equivalently, for a given black hole mass, higher disk temperatures correspond to higher luminosities. Figure~\ref{fig:accretion} demonstrates that these correlations are clearly present in our sample. In the top panels, diagonal tracks of the form $L_{\rm Bol}^{\rm disk}/L_{\rm Edd} \propto T_p^4$ trace the temporal evolution of individual systems (i.e., fixed $M_{\bullet}$ and $a_{\bullet}$), while offsets perpendicular to these tracks reflect differences in black hole mass between systems. 

Naturally, these correlations are exact only under specific conditions. In particular, Eq.~\ref{eq:accretion} holds in closed form \citep[e.g.,][]{Frank2002}, with exact numerical values, only for steady-state disks of constant area and the same spin. For a sample spanning a range of black hole spins, the relation will not be exact: two black holes with the same $M_{\bullet}$ and Eddington ratio but different spins (e.g., a rapidly rotating Kerr black hole versus a Schwarzschild black hole) will exhibit slightly different disk temperatures. However, the spin dependence is relatively mild, particularly in our case, where the spin constraints are very limited, and thus it primarily introduces scatter into the three-parameter space of Fig.~\ref{fig:accretion}.

Similarly, the relation $L_{\rm Bol}^{\rm disk}/L_{\rm Edd} \propto T_p^4$ holds exactly only for disks with a constant emitting area. In TDE disks, which have a finite mass supply, the emitting area must evolve and only asymptotically approach a steady-state configuration. This evolution in disk size introduces additional scatter in the correlation. Nevertheless, because the bolometric luminosity is dominated by the innermost (hottest) regions of the disk, the effect of variations in the outer disk area is modest, and the expected accretion correlations remain clearly visible, as shown in Fig.~\ref{fig:accretion}.

%Extrapolating this analysis to lower-mass black hole systems with thermally dominated X-ray spectra, one would expect—and indeed observations confirm \citep{Remillard2006}—that X-ray binaries (XRBs) in the soft state exhibit peak disk temperatures of $\sim \text{few} \times 10^6$–$10^7$ K. Such values would place them at the far right-hand side of the top panel, beyond the plotted range.

In this context, it is worth noting the recent results of \citet{Arcodia2025}, who analyzed the quiescent disk emission of quasi-periodic eruption (QPE) sources \citep[e.g.,][]{Miniutti2019,Nicholl2024}, many of which originate from TDEs. They reported no evidence for a $T_p^4\propto M_{\bullet}^{-1}$ correlation. This outcome is not unexpected, given that (i) the relation involves at least three parameters, including $L_{\rm Bol}^{\rm disk}/L_{\rm Edd}$, which was neither estimated nor accounted for, and (ii) black hole masses were inferred not from disk emission but from host-scaling relations, with assumed uncertainties of $\sim0.7$ dex. Together, these factors, combined with the limited dynamical range of $M_{\bullet}$ over which nuclear TDEs/QPEs are found, make recovery of the correlation essentially impossible. Notably, two of our sources, GSN 069 and AT2019qiz, are QPE sources and naturally follow the same trends as the rest. Extending our analysis to additional QPE sources---particularly those not clearly linked to known TDEs \citep[such as the \srge\ discoveries;][]{Arcodia2021,Arcodia2024a}---will be essential to clarify this picture.

\subsection{Black Hole vs. Host Galaxy Correlations}
\label{sec:host_scaling}

A critical reader may note that some of the results presented in the previous section primarily reflect the internal physics of accretion (i.e., Eq.~\ref{eq:flux_kerr}) and the model's success in describing the data (although we stress again that $T_p^4\propto M_\bullet^{-1}$ reflects a measurement of the system, not an intrinsic property of the model). The most pessimistic reading of these results would be that they demonstrate internal consistency rather than directly establishing intrinsic properties of the sources themselves. To address this limitation, it is important to seek independent confirmation that the derived quantities---such as the black hole masses---are not arbitrary, but are instead consistent, at the population level, with established empirical relations between black holes and their host galaxies.

In Fig.~\ref{fig:host_scaling}, we present the correlations between black hole mass and host-galaxy properties: the $M_{\bullet}$–$M_{\rm gal}$ relation (left panel) and the $M_{\bullet}$–$\sigma_{\star}$ relation (right panel). Blue points (late-type galaxies) and red points (early-type galaxies) are taken from \citet{Greene2020}, largely based on the compilation of \citet{Kormendy2013}. In these cases, black hole masses are determined through dynamical modeling of stellar and/or gas kinematics in extremely nearby galaxies. For completeness, we also include a few purple points corresponding to dynamical modeling of massive globular clusters hosting candidate IMBHs \citep[e.g.,][]{Haberle2024}. In the left panel, the $M_{\bullet}$–$M_{\rm gal}$ relation is shown separately for late- and early-type galaxies, while in the right panel we show the $M_{\bullet}$–$\sigma_{\star}$ relation for the combined sample, as both galaxy types follow essentially the same trend.

\begin{figure}
    \centering
    \includegraphics[width=1\columnwidth]{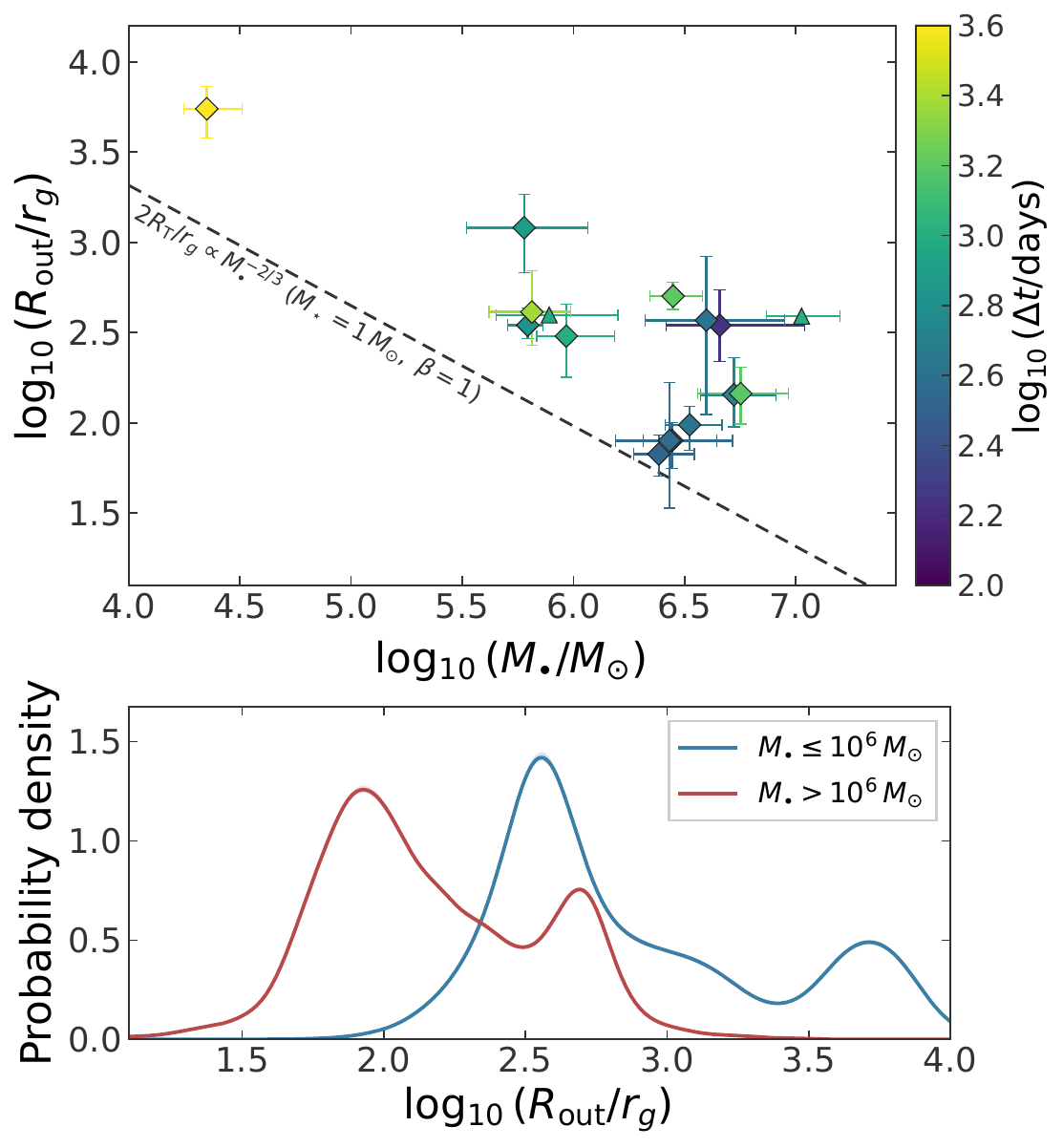}
   
    \caption{Top panel: Correlation between normalized outer disk radius ($R_{\rm out}/r_g$) and black hole mass ($M_{\bullet}$). We show one measurement (the first available) for each source. The black line indicates the circularization radius for the disruption of a $1\,M_{\odot}$ star with an impact parameter $\beta=1$. Diamonds represent measurements, and triangles represent lower limits. Bottom panel: Probability distributions of $R_{\rm out}/r_g$ for sources with $M_{\bullet}$ greater than or less than $10^{6}\,M_{\odot}$, which divides our sample roughly in half. More massive black holes clearly host relatively more compact disks, as expected.
 }
    \label{fig:Rout}
\end{figure}

Our accretion-based $M_{\bullet}$ values for the TDE sample lie directly on the established host--black hole correlations (statistical tests will be performed in a later section). In the $M_{\bullet}$--$M_{\rm gal}$ plane, the TDE hosts appear to align most closely with the late-type galaxy relation. The only notable outlier is the IMBH in 3XMM J2150$-$05, which sits above the relations. Such an offset is not unexpected because of astrophysical factors, including the fact that the host of 3XMM J2150$-$05 is a UCD. Such galaxies are thought to be tidally stripped remnants of larger galaxies that have fallen into the halos of their massive parent systems, naturally producing elevated black hole-to-galaxy mass ratios \citep[e.g.,][]{Seth2014}.

The fact that our inferred $M_{\bullet}$ values lie on the expected host--black hole scaling relations indicates that the results presented in Section~\ref{sec:accretion}, and those discussed in Section~\ref{sec:tde_disk}, are not merely internal consistencies of our assumed model, but instead reflect genuine physical properties of these systems. This strengthens the case that our inferred $M_{\bullet}$ values are not only precise but also reliable, with uncertainties comparable to, and values consistent with, those obtained from dynamical modeling of extremely nearby galaxies.

\subsection{TDE-disk Correlations}\label{sec:tde_disk}

\begin{figure}

    \centering
    \includegraphics[width=1\linewidth]{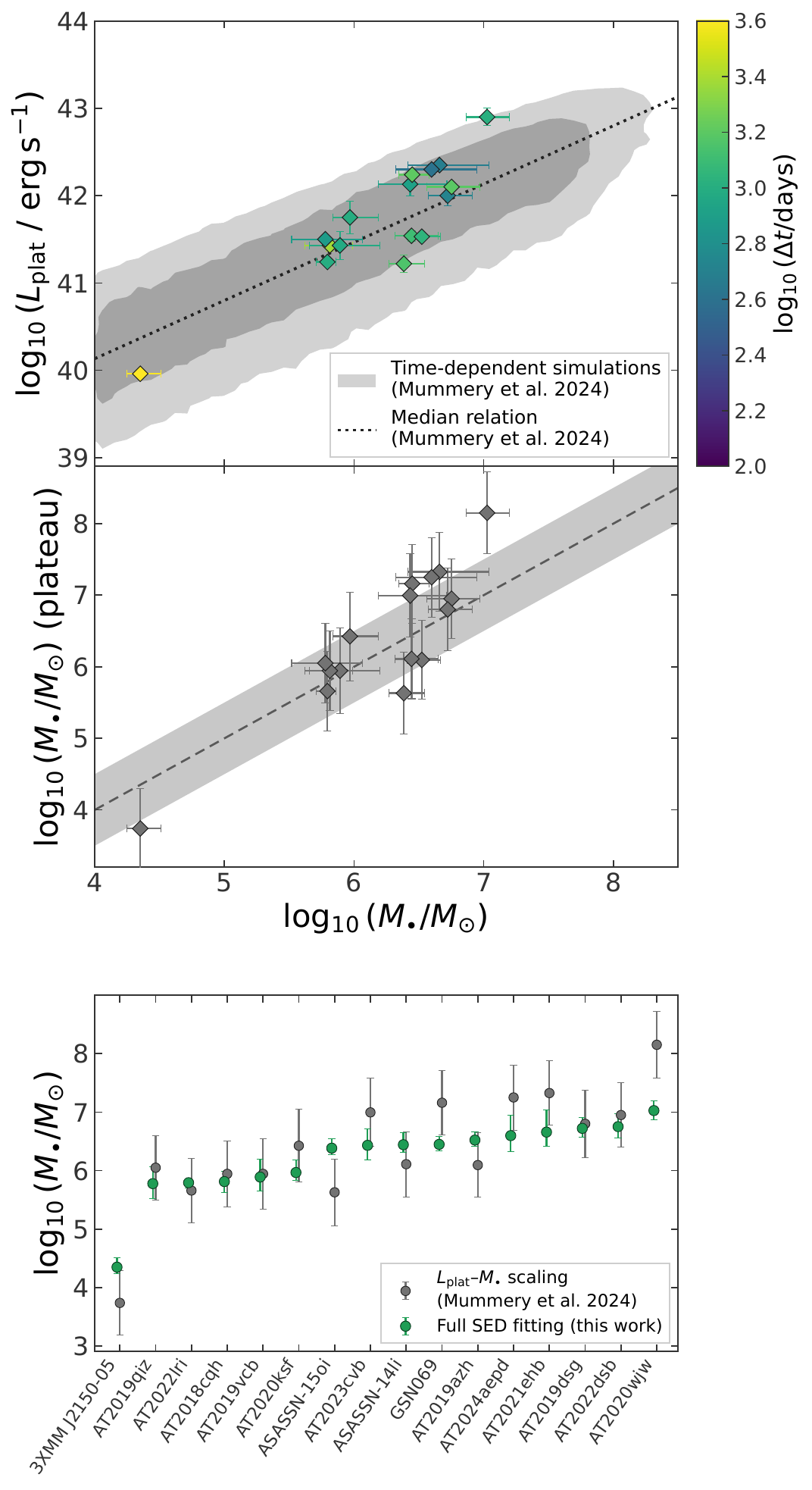}
   
    \caption{ Top: Correlation between the UV luminosity ($L_{\rm plat}$) during the plateau phase and black hole mass ($M_{\bullet}$). Colored points show our sample, computed at $\nu = 10^{15} {\rm Hz}$, at the epoch closest to $\Delta t = 1000$ days available for each source. Point colors indicate $\log_{10}(\Delta t/{\rm days})$. The grey shaded region shows the credible intervals from the time-dependent simulations of \citet{Mummery2024}, which solve the relativistic disk equations assuming half the stellar debris circularizes, the initial disk radius equals the circularization radius, and luminosities are evaluated 1000 days after disk formation. Middle: Direct comparison between the $M_{\bullet}$  obtained here from full SED fitting  and those inferred from the plateau-luminosity scaling relations \citep{Mummery2024}. The dashed line shows the $1{:}1$ relation. Bottom: Source-by-source comparison of the two mass estimates, with the corresponding uncertainties. }

    \label{fig:plateu}
\end{figure}

While the correlations discussed in \S\ref{sec:accretion} should be valid, at least approximately, for any accreting black hole system, the disks formed in the aftermath of TDEs are more constrained than the standard steady-state ``free-$\dot{M}$'' solutions. This is because many of the disk properties are constrained (or at least should be if the modeling reflects reality) by the disruption process itself.
The characteristic scale of the initial disk is set by the disruption process itself---i.e., it forms and is fed close to the black hole where the star is disrupted---rather than at large radii, as in the case of interstellar-medium (ISM)-fed active galactic nuclei (AGN) or Roche-lobe overflow in XRBs. In addition, TDE disks are supplied with only a finite mass reservoir (at most approximately one-half of the disrupted stellar mass, and in practice only the fraction that successfully circularizes), in contrast to AGN and XRB disks, for which the mass contained in the disk at any given time is only a tiny fraction of the supply available from the ISM or the companion star.  

The first obvious constraint imposed by the finite mass supply is that a TDE-fed disk will have a finite---albeit long in human terms (see Fig.~\ref{fig:lcs})---lifetime. Equally important, the peak disk temperature will necessarily decrease over time. This occurs because, in an accretion disk, the local temperature depends on the product of the local surface density, $\Sigma(r)$, and the local stress \citep{Balbus1999,Mummery2020}. Since the long-term evolution of $\Sigma(r)$ is fundamentally limited by the finite mass supply, and $\Sigma(r)$ should therefore inevitably decrease, the peak disk temperature must decrease as well.\footnote{The stress cannot decrease more rapidly than $1/\Sigma$ if the disk evolution is to remain stable \citep{lightman74_instability}---as it clearly does in TDE sources, which display smooth evolution over years to decades.} As expected, this behavior has been confirmed observationally in TDEs \citep[e.g.,][]{Guolo2024,Yao2024_22lri} and is evident in the $T_p$ values reported in Table~\ref{tab:best_fit}.

A more novel result arises from the fact that the origin of these disks should also impose a characteristic scale on their size. Not only are they expected to be compact, but, in a simplified picture of the disk formation process, they should form at approximately the so-called circularization radius,  

\begin{equation}
R_{\rm out} \propto R_{\rm circ} \;\approx\; \frac{2R_{T}}{\beta}
\;\approx\; \frac{92}{\beta } \left ( \frac{M_*}{M_{\odot}} \right)^{7/15} \left ( \frac{M_{\bullet}}{10^6 M_{\odot}} \right)^{-2/3} r_g
\end{equation}

\noindent where $R_{T}\simeq R_{*}(M_{\bullet}/M_{*})^{1/3}$ is the tidal radius, $M_{*}$ and $R_{*}$ are the stellar mass and radius, $\beta$ is the impact parameter (the ratio of the pericenter of the incoming star's orbit to the tidal radius), and we have assumed a main-sequence mass--radius relation \citep[$R_{*} \propto M_{*}^{4/5}$,][]{Clayton1968} in obtaining the final expression on the right.  

This implies that, even when accounting for a distribution of disrupted stellar masses and impact parameters, the general expectation is that lower-mass black holes should host relatively larger disks (in units of $r_{g}$) than higher-mass ones, following a relation of the form $R_{\rm out}/r_g \propto M_{\bullet}^{-2/3}$.  

In Fig.~\ref{fig:Rout}, we show that our inferred $R_{\rm out}$ values---here showing only the first epoch with a measured value---are consistent with this expectation. The top panel presents $R_{\rm out}/r_g$ as a function of $M_{\bullet}$, while the lower panel shows the probability distributions of the measured $R_{\rm out}/r_g$ for systems with $M_{\bullet}$ above and below $10^{6}\,M_{\odot}$ (effectively dividing our sample in half). It is clear that our inferred disk sizes retain information about the black hole mass and broadly follow the simple expectation that lower-mass black holes have relatively larger disks.  

In the top panel, we also compare our inferred $R_{\rm out}$ to $2R_{T}/r_{g}$ for $M_{*}=1\,M_{\odot}$ and $\beta=1$. Our values are systematically larger than the circularization radius, as expected: $R_{\rm circ}$ sets only the initial disk size. The disk must expand as it evolves in order to conserve angular momentum, initially rapidly and later more slowly (e.g., ${\rm d}R_{\rm out}/{\rm d}t \propto t^{2n -3}$, where $n \approx1.2$, \citet{Cannizzo1990}). Thus, by the epochs at which we measure $R_{\rm out}$ (see Tables~\ref{tab:SED} and \ref{tab:best_fit}), the disk should already be substantially larger than its formation scale. Nevertheless, the overall expected behavior is clearly visible in the data. Importantly, this scaling is not imposed by our modeling, but emerges naturally from the SED fits, demonstrating again that the inferred disk sizes trace genuine physical properties of these systems.  

Another important scaling relation, arising from the compact initial configuration of TDE disks, their finite mass supply, and the constraints of mass and angular-momentum conservation, is that the late-time luminosity in the disk-dominated phase of the UV/optical light curve (i.e., the plateau phase) should scale with black hole mass. This relation has been analyzed in detail by \citet{Mummery2024}, who confirmed its validity through analytical arguments, numerical simulations, and observational tests. Although the relation exhibits relatively large intrinsic scatter ($\sim0.5$ dex), reflecting additional free parameters beyond $M_{\bullet}$, \citet{Mummery2024} showed that it can be approximated as $L_{\rm plat}\propto M_{\bullet}^{2/3}$, where $L_{\rm plat}$ is the plateau $\nu L_{\nu}$ luminosity measured over a narrow wavelength range. A key result of their work was the simulation of a large ensemble ($N=10^{6}$) of relativistic, time-dependent disk models \citep{Balbus2017,Mummery2020,Mummery2023b}, assuming initial radii equal to the circularization radius and sampling the remaining parameters from probability distributions. From these simulations, the authors extracted $L_{\rm plat}$ at a characteristic time ($t\simeq1000$ days after disk formation) and constructed the $L_{\rm plat}$--$M_{\bullet}$ relation, enabling black hole mass inference in TDEs with a theoretical scatter of $\sim0.5$ dex.

The scaling relation of \citet{Mummery2024} relied on assumed distributions of stellar properties, the scale of turbulence in the disk, the orientation of the observer, and the properties of the black holes. It is therefore firmly rooted in {\it assumed} TDE physics, which could, in principle, be inaccurate if TDEs do not behave as assumed. Here, we need to make no assumptions about any distributions beyond broad priors, as we constrain the physical parameters of the system from the data. Our results therefore provide a direct test of the assumptions of \citet{Mummery2024} and the validity of their reported black hole mass scaling relation. 

In Fig.~\ref{fig:plateu}, we overplot the measured $L_{\rm plat}$ and inferred $M_{\bullet}$ for our sample on the simulated population from \citet{Mummery2024}, demonstrating excellent agreement. Consistently, the middle panel of Fig.~\ref{fig:plateu} shows that our inferred black hole masses also agree with those obtained through the $L_{\rm plat}$--$M_{\bullet}$ scaling relation (with the dashed line indicating one-to-one agreement): the vast majority are consistent within $1\sigma$, and all lie within $\leq2\sigma$. The bottom panel of Fig.~\ref{fig:plateu} presents the same comparison source-by-source. This agreement provides further confidence in the robustness of both the model developed here and the plateau scaling relation.

\begin{figure*}
    \centering

     \includegraphics[width=0.75\linewidth]{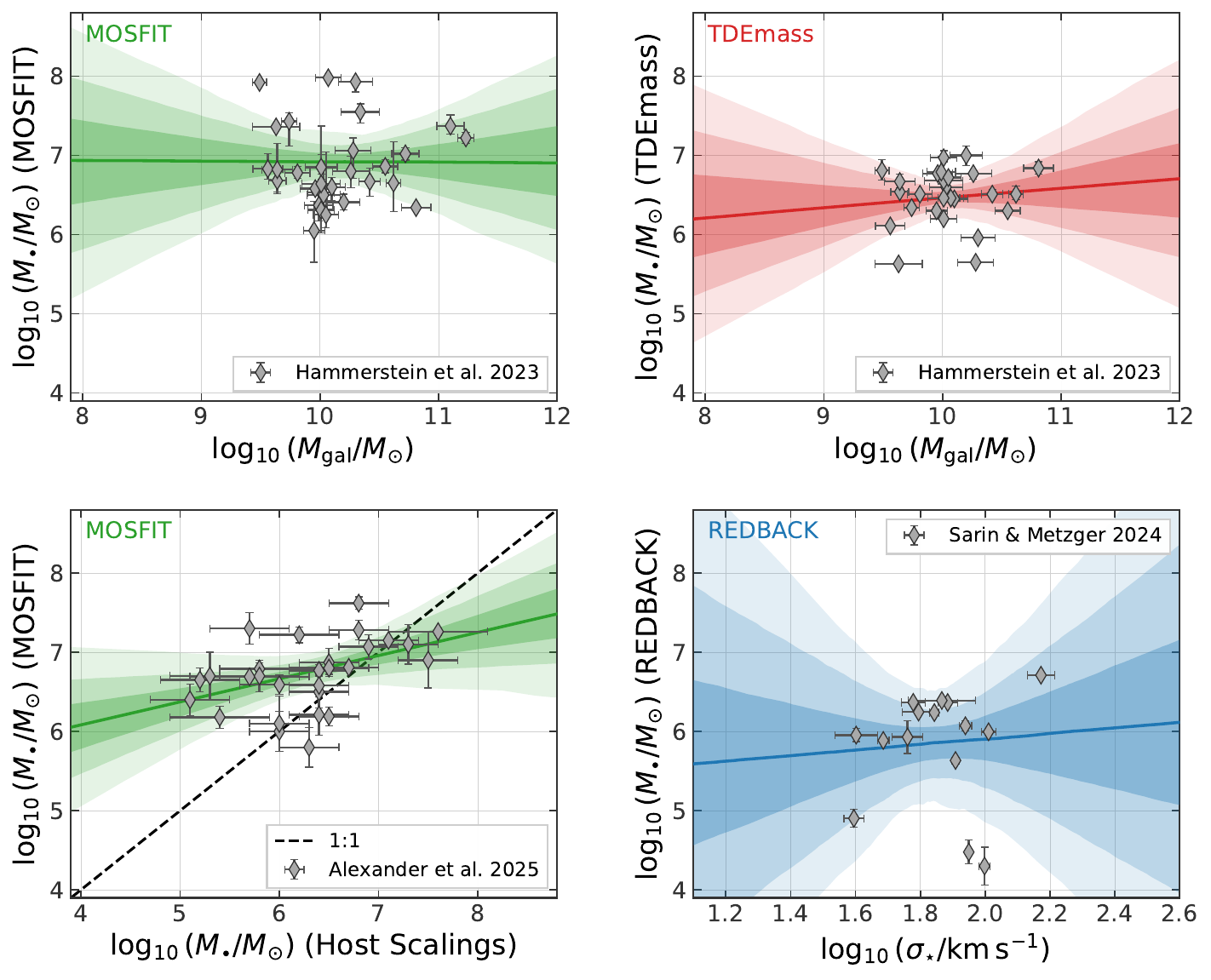}

    \caption{Black hole mass inferred using either \texttt{TDEmass} (top) or \texttt{MOSFIT} (bottom) versus galaxy mass for the first 30 ZTF-discovered TDEs \citep{Hammerstein2023a}. The colored contours show the 68\%, 95\%, and 99.7\% credible intervals of the posterior distributions from the correlation Markov chain Monte Carlo (MCMC) fits. Both cases are consistent with no correlation.}
    \label{fig:mostfit_tdemass_redback}
\end{figure*}

Naturally, the black hole masses we infer from direct SED fitting are significantly more precise. This is because the $L_{\rm plat}$--$M_{\bullet}$ relation relies on assumed probability distributions for all other parameters of the theory (describing both the black hole and the disrupted star), which introduce substantial scatter. By contrast, our method directly marginalizes over the disk and black hole parameters that can be constrained by the data themselves. In the approach presented here, parameter inference is rooted in a time-independent accretion framework, in which the key parameters are recovered from their direct imprints on the observed emission without explicitly assuming any dynamical evolution or initial conditions.

\section{Discussion}\label{sec:discussion}

\subsection{Black Hole Mass Inference in TDEs}

Having shown that full SED fitting provides a precise and reliable method for inferring $M_{\bullet}$ in TDEs---recovering the established black hole--galaxy scaling relations (\S\ref{sec:host_scaling}) and yielding internally consistent physical parameters (\S~\ref{sec:accretion} and \S\ref{sec:tde_disk})---we now compare this approach with alternative methods used in the literature.

\subsubsection{Host-Galaxy Scaling Relations}

The most common and straightforward way to estimate a black hole mass is through a host-galaxy scaling relation, in which a measured galaxy property, such as total stellar mass, bulge mass, or nuclear stellar velocity dispersion, is inserted into the corresponding empirical relation. This approach provides a simple and comparatively model-independent estimate and is often the most practical option when only limited information is available. However, it has important limitations.

First, a host relation is in fact not a  measurement of $M_{\bullet}$ for an individual source; instead it assumes that a population-level correlation applies to that particular black hole--galaxy system. Second, the intrinsic scatter in these relations can be substantial even before statistical measurement uncertainties are included. For example, the $M_{\bullet}$--$\sigma_{\star}$ relation has an intrinsic $1\sigma$ scatter as small as approximately $0.3$ dex in the high-mass regime, $\sigma_{\star}\gg100\,{\rm km\,s^{-1}}$ \citep{Kormendy2013}. Most TDE hosts, however, lie in the range $50<\sigma_{\star}/{\rm km\,s^{-1}}<100$ \citep{Hammerstein_21,Yao2023}, where the scatter increases to $\gtrsim0.5$ dex \citep{Greene2020}. The scatter is even larger for relations such as $M_{\bullet}$--$M_{\rm gal}$ in the mass range relevant to TDE hosts \citep[see also Fig.~\ref{fig:host_scaling}]{Greene2020}.

These statistical limitations become particularly important when host-based masses are used to derive secondary quantities, such as Eddington ratios, for individual sources. In practice, the literature commonly follows one of two approaches. The first is to adopt the central value of the host-based estimate and treat it as the true mass. Although convenient, this can be statistically misleading. Consider, for example, a Gaussian-like constraint $\log_{10}(M_{\bullet}/M_{\odot})=6.0\pm0.5$. The probability that the true value lies within $\pm0.1$ dex of the central estimate is only approximately $16\%$, and the probability of lying within $\pm0.05$ dex is approximately $8\%$; naturally, this probability approaches zero as the interval narrows. Thus, there is no statistical justification for treating $10^6\,M_{\odot}$ as the exact mass without propagating its uncertainty into \textit{all} derived quantities.

The second approach is to propagate the full uncertainty into quantities such as the Eddington ratio. This is statistically appropriate, but the resulting constraints may have limited physical discriminating power. With a $0.5$-dex uncertainty in $M_{\bullet}$, for example, a source radiating at $50\%$ of its Eddington luminosity may be statistically indistinguishable from one radiating at $5\%$. Such ambiguity directly limits the physical interpretation of the accretion state and multiwavelength emission of an individual system.

More fundamentally, relying exclusively on host-galaxy scaling relations risks overlooking one of the central opportunities offered by TDEs: using radiation produced near the black hole to constrain independently the demographics of otherwise quiescent black holes. Host-based estimates effectively assume that the large-scale properties of the galaxy contain more useful information about the black hole than the emission generated in its immediate vicinity during and after a disruption. While host relations remain valuable, the TDE emission provides an independent and physically direct source of information that should be exploited whenever the data permit.

This need is especially relevant for TDEs in dwarf galaxies, such as 3XMM J2150$-$05 and the recently discovered EP240222a \citep{Jin2025}. At $M_{\rm gal}\sim10^7\,M_{\odot}$, host--black hole scaling relations are essentially unconstrained, and extrapolating relations calibrated at much higher masses is unjustified. Emission-based estimates of $M_{\bullet}$ are therefore indispensable in this regime. The same is true for off-nuclear TDEs without a clearly detected stellar counterpart, including AT2024tvd \citep{Yao2025,Guolo2026}, AT2025abcr \citep{stein2026}, and eRASS J1421$-$29 \citep{Grotova2025}, for which host correlations are not applicable.

The number of dwarf-host and/or off-nuclear events should increase substantially with wide-field time-domain facilities such as the Vera C. Rubin Observatory's Legacy Survey of Space and Time \citep[LSST;][]{Ivezic2019} and Einstein Probe \citep{Yuan2018}. Moving beyond host-based estimates will therefore be essential if TDEs are to populate, extend, and ultimately refine black hole--host scaling relations independently.

\begin{deluxetable*}{lcCCCCCCc}
\label{tab:mass_stats}
\tabletypesize{\scriptsize}
\tablecaption{Summary Review of $M_{\bullet}$ Inference Methods and Their Ability to Reproduce Black-Hole vs. Host-Galaxy Correlations \label{tab:methods}}
\tablehead{
\multicolumn{2}{c}{\textbf{Method}} &
\colhead{Correlation} & 
\colhead{$\Delta X$ (dex)} & 
\colhead{N} & 
\colhead{$\beta$} & 
\colhead{$\epsilon$} & 
\colhead{$P(\beta > 0)$} & 
\colhead{Data References}
}
\startdata
\multirow{5}{*}{\parbox[c]{2.5cm}{\centering Accretion Disk\\ Emission
}} 
 & \multirow{2}{*}{\parbox[c]{2.5cm}{\centering Plateau Scaling \\ Relation}} & M_{\bullet}$-$M_{\rm gal}  & 1.9 & 49 &1.26^{+0.16}_{-0.16} & 0.23^{+0.12}_{-0.14}& \geq0.9999995 \ (\geq5\sigma) & \multirow{2}{*}{\citet{Mummery2024}}  \\
 & &   M_{\bullet}-\sigma_{\star}  & 0.8 & 34 &2.57^{+0.45}_{-0.45} & 0.27^{+0.15}_{-0.16}  & \geq0.9999995 \ (\geq5\sigma) & \\
  & &  M_{\bullet}$-$M_{\rm bulge}  & 2.2 & 40 & 1.22^{+0.17}_{-0.18} & 0.23^{+0.13}_{-0.14} & \geq 0.9999995 \ (\geq5\sigma)  & \citet{Ramsden2025}\\
 \cline{2-9}  % <-- draws a horizontal line across the middle columns
 & \multirow{3}{*}{\parbox[c]{2.5cm}{\centering Full SED \\ Fitting}} &M_{\bullet}$-$M_{\rm gal}  & 3.3 & 16 & 0.75^{+0.14}_{-0.12} & 0.18^{+0.09}_{-0.08} & \geq0.9999995 \ (\geq5\sigma) & \multirow{3}{*}{This Work}  \\
 & & M_{\bullet}$-$M_{\rm gal} (a) & 1.1 & 15 & 0.92^{+0.23}_{-0.24} & 0.17^{+0.10}_{-0.09} & 0.9994 \ (3.4\sigma)  &\\
& & M_{\bullet}-\sigma_{\star} & 0.6 & 15 & 1.98^{+0.36}_{-0.33} & 0.10^{+0.09}_{-0.06} & 0.9999 \ (4.0\sigma)& \\
\hline
\multirow{5}{*}{\parbox[c]{2.8cm}{\centering Optical Flare \\ Scaling or Modeling}}   
 & \texttt{TDEMass} & M_{\bullet}$-$M_{\rm gal} & 1.3 & 26 & 0.12^{+0.23}_{-0.25} & 0.37^{+0.07}_{-0.05} & 0.68 \ (1.0\sigma) & \citet{Hammerstein2023a}  \\
\cline{2-9}  % <-- draws a horizontal line across the middle columns
 &  \multirow{3}{*}{\texttt{MOSFIT}}  & M_{\bullet}$-$M_{\rm gal} & 1.7 & 30 & $0.01^{+0.24}_{-0.22}$ & 0.51^{+0.08}_{-0.07} & 0.48 \ (0.7\sigma)  & \citet{Hammerstein2023a} \\
 &  & M_{\bullet}$-$M_{\rm bulge} & 2.1 & 29 & 0.23^{+0.15}_{-0.14} & 0.16^{+0.09}_{-0.09} & 0.95 \ (2.0\sigma)  &  \citet{Ramsden2022}\\
    \cline{3-9} 
   &  & M_{\bullet}$-$M_{\bullet} {\rm (host)} & 2.5 & 28 & 0.30^{+0.12}_{-0.11} & 0.35^{+0.08}_{-0.06} & 0.991 \ (2.6\sigma) &  \citet{Alexander2025}\\
   \cline{2-9} 
   & \texttt{REDBACK} & M_{\bullet}-\sigma_{\star} & 0.6 & 15 & 0.35^{+1.30}_{-1.39} & 0.76^{+0.19}_{-0.14} & 0.60 \ (0.9\sigma) & \citet{Sarin2024}\\
% \cline{2-8} 
%  & & L_{\rm peak}- M_{\rm gal} & 49 &0.83^{+0.11}_{-0.11} & 0.46^{+0.04}_{-0.03}   &  \geq0.9999995 \ (\geq5\sigma) &\multirow{2}{*}{\citet{Mummery2024}}  \\ 
%    & & L_{\rm peak}- \sigma_{\star} & 33&1.67^{+0.28}_{-0.28} & 0.47^{+0.05}_{-0.04} &  \geq0.9999995 \ (\geq5\sigma) & 
\enddata
\tablecomments{$N$ is the number of sources used in the fitting. $\Delta X$ is the dynamical range of the $X$ axis for each dataset fitted. (a) 3XMM J2105-05 was excluded on this fitting.}
\end{deluxetable*}
\subsubsection{Early-time Optical Flare Scaling and Modeling}\label{sec:mosfit_mass}

The discovery of optically selected TDEs with prominent optical flares, enabled by modern wide-field time-domain surveys, has motivated extensive theoretical work aimed at determining the physical origin of this emission. These studies have generally converged on two main competing scenarios. In a simplified picture, the optical emission may arise either from (i) stream shocks near apocenter \citep[e.g.,][]{Ryu2020a,Ryu2023} or (ii) the reprocessing of high-energy, fallback-powered radiation \citep[e.g.,][]{Dai2018,Mockler2019}.

These scenarios have also motivated publicly available tools for inferring the physical parameters of TDEs. \texttt{TDEmass} \citep{Ryu2020} uses the peak blackbody luminosity and UV/optical color temperature to infer the black hole and disrupted stellar masses through analytic expressions calibrated against global simulations of the disruption process. By contrast, \texttt{MOSFIT} \citep{Mockler2019} assumes that the luminosity directly traces the fallback rate,
$L(t)=\eta c^2\dot{M}_{\rm fb}(t)$,
where the efficiency parameter $\eta$ converts the fallback rate into an observable luminosity, which is subsequently passed through reprocessing functions to produce the optical emission. Both methods were originally applied to relatively small samples of observed TDEs, and their inferred black hole masses were suggested to be consistent with black hole--host galaxy scaling relations. However, those studies were limited to a handful of sources and, in most cases, did not present a formal statistical quantification of this consistency.

The rapid increase in the TDE discovery rate \citep{Hammerstein2023a,Yao2023}, particularly through ZTF, now makes population-level tests of these methods possible. Here, we assess whether current implementations of early-time optical-flare models recover statistically significant correlations between their inferred black hole masses and independently measured host-galaxy properties. Our aim is not to determine whether the methods reproduce the exact slope and normalization of a particular host-scaling relation, but first to establish whether they recover any significant correlation with host properties at all.

\paragraph{Statistical framework.}

In this and the following sections, we follow \citet{Greene2020} and \citet{Mummery2024} and fit power-law relations of the form
\begin{equation}\label{power_law_fit}
\log_{10}(Y)=\alpha+\beta\log_{10}(X),
\end{equation}
where
\begin{equation}
Y\equiv\frac{M_{\bullet}}{M_{\odot}}
\end{equation}
and $X$ is a normalized host-galaxy scaling variable. We account for intrinsic scatter in the host relation by introducing an additional scatter parameter $\epsilon$, such that
\begin{equation}
\left(\delta\log_{10}Y\right)^2
\;\to\;
\left(\delta\log_{10}Y\right)^2+\epsilon^2,
\end{equation}
where $\delta\log_{10}Y\equiv\delta\log_{10}(M_{\bullet}/M_{\odot})$ is the measurement uncertainty on the logarithm of the inferred black hole mass. We use \texttt{emcee} \citep{Foreman-Mackey_13} to sample the posterior associated with the log-likelihood
\begin{multline}\label{eq:likehood}
{\cal L}=-\frac{1}{2}\sum_i\Bigg[
\frac{\left(\log_{10}Y_i-\alpha-\beta\log_{10}X_i\right)^2}
{\left(\delta\log_{10}Y_i\right)^2+
\left(\beta\,\delta\log_{10}X_i\right)^2+\epsilon^2}\\
+\ln\!\left(2\pi\left[
\left(\delta\log_{10}Y_i\right)^2+
\left(\beta\,\delta\log_{10}X_i\right)^2+\epsilon^2
\right]\right)\Bigg],
\end{multline}
where the sum runs over all pairs $(X_i,Y_i)$. Equation~\ref{eq:likehood} differs from the likelihoods adopted by \citet{Greene2020} and \citet{Mummery2024} because it includes the uncertainties in $X_i$ (the galaxy properties), whereas those studies account only for the uncertainties in $Y_i$ ($M_{\bullet}$).

We quantify the evidence for a positive relation between the inferred black hole mass and a given host property using the marginalized posterior probability
\begin{equation}
P(\beta>0)\equiv P(\beta>0\mid\mathcal{D},M)
=\int_{0^+}^{\infty}p(\beta\mid\mathcal{D},M)\,d\beta,
\end{equation}
where $M$ is the model in Equation~\ref{power_law_fit}, $\mathcal{D}=\{(X_i,Y_i)\}$ is the dataset, and $p(\beta\mid\mathcal{D},M)$ is the marginalized posterior distribution of $\beta$. We convert this probability to a Gaussian-equivalent significance using $\Phi^{-1}[P(\beta>0)]$, where $\Phi^{-1}$ is the inverse cumulative distribution function of the standard normal distribution. In this and the following section, the colored regions in the figures show credible intervals derived from the marginalized MCMC samples; the posterior medians, credible intervals, and statistical properties are listed in Table~\ref{tab:mass_stats}. Full posterior distributions are presented in Appendix~\ref{app:stats}.

\paragraph{Population-level tests of early-time flare models.}

We first consider the analysis of \citet{Hammerstein2023a}, who applied both \texttt{TDEmass} and \texttt{MOSFIT} to the first 30 TDEs discovered within the ZTF survey footprint during its first phase of operations. The top two panels of Fig.~\ref{fig:mostfit_tdemass_redback} compare their inferred $M_{\bullet}$ values with the reported total galaxy masses, for which we adopt $X=M_{\rm gal}/(3\times10^{10}\,M_{\odot})$. Consistent with the conclusions of \citet{Hammerstein2023a}, neither method yields a clear correlation between $M_{\bullet}$ and $M_{\rm gal}$: the posterior slopes are consistent with zero. Quantitatively, $P(\beta>0)=0.68$ for \texttt{TDEmass} and $P(\beta>0)=0.48$ for \texttt{MOSFIT}, corresponding to reported statistical significances of $1.0\sigma$ and $0.7\sigma$, respectively.

\begin{figure*}
    \centering
    \includegraphics[width=0.8\textwidth]{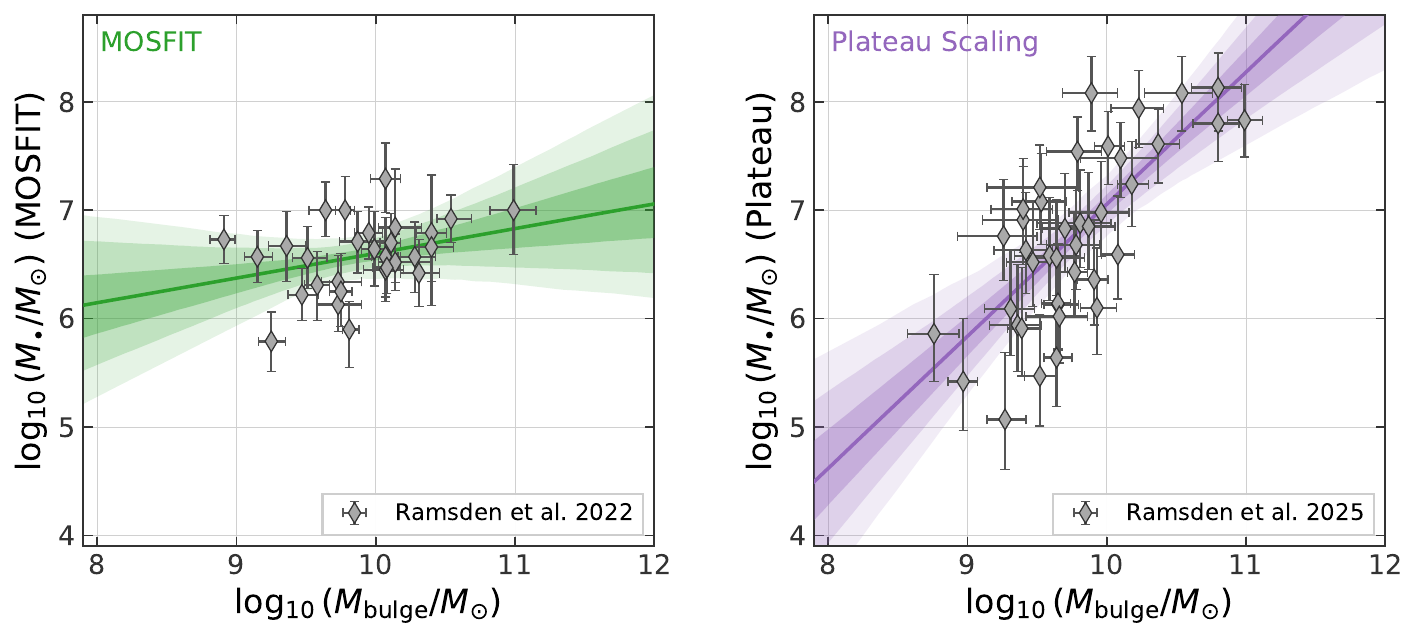}
    \caption{Black hole mass ($M_{\bullet}$)---estimated either with \texttt{MOSFIT} by \citet{Nicholl2022} (left) or from the plateau scaling relation of \citet{Mummery2024} (right)---versus host-galaxy bulge mass ($M_{\rm bulge}$), as measured by \citet{Ramsden2022,Ramsden2025}. The contours have the same meaning as in Fig.~\ref{fig:mostfit_tdemass_redback}. The significances of a non-zero correlation are $2.0\sigma$ for \texttt{MOSFIT} and $\geq5\sigma$ for the plateau scaling relation.}
    \label{fig:cor_bulge}
\end{figure*}

We next examine the larger population-level analysis of \citet{Alexander2025}, who compared black hole masses inferred with \texttt{MOSFIT} against host-based estimates $M_{\bullet,{\rm host}}$, derived primarily from the stellar velocity dispersion $\sigma_{*}$ using galaxy scaling relations \citep{Kormendy2013}. Their reported values are shown in the bottom-left panel of Fig.~\ref{fig:mostfit_tdemass_redback}. Applying the same MCMC analysis with $X=M_{\bullet,{\rm host}}/(10^6\,M_{\odot})$, we obtain $\beta=0.30\pm0.12$, corresponding to a $2.6\sigma$ significance for a non-zero correlation (Table~\ref{tab:mass_stats}).

Although this correlation is stronger than that obtained for the \citet{Hammerstein2023a} sample, its slope is inconsistent with the expectation for a comparison of two estimates of the same quantity. A reliable $M_{\bullet}$--$M_{\bullet}$ comparison should yield a slope statistically consistent with $\beta\simeq1$, together with scatter $\epsilon\gtrsim0.3$ dex \citep[comparable to the scatter measured by][]{Kormendy2013}. Instead, the inferred slope is approximately $4.5\sigma$ away from unity. The shallow slope indicates a systematic mass-dependent bias: at the low-mass end, \texttt{MOSFIT} overestimates $M_{\bullet}$ relative to host-galaxy expectations, whereas at the high-mass end it underestimates $M_{\bullet}$. This behavior is directly visible in Fig.~\ref{fig:mostfit_tdemass_redback}: every source with $\log_{10}(M_{\bullet,{\rm host}}/M_{\odot})\lesssim6$ is assigned $\log_{10}(M_{\bullet}/M_{\odot})\gtrsim6$ by \texttt{MOSFIT}.

As a further test of \texttt{MOSFIT}, we compare the $M_{\bullet}$ values inferred by \citet{Nicholl2022} with the host-galaxy bulge masses measured by \citet{Ramsden2022}. The results are shown in the left panel of Fig.~\ref{fig:cor_bulge} and summarized in Table~\ref{tab:mass_stats}. This comparison yields a $2.0\sigma$ significance for a non-zero correlation. For context, the right panel of Fig.~\ref{fig:cor_bulge} shows black hole masses inferred from the plateau scaling relation of \citet{Mummery2024} for the same host quantity; that accretion-based estimator produces a correlation significant at $\geq5\sigma$.

Finally, we test the black hole masses obtained using the implementation in \texttt{REDBACK} \citep{Sarin2024} of the ``cooling envelope'' model \citep{Metzger2022}. We compare these masses with the host-galaxy stellar velocity dispersions $\sigma_{\star}$ in the bottom-right panel of Fig.~\ref{fig:mostfit_tdemass_redback}. The inferred slope is again consistent with zero, with $P(\beta>0)=0.6$, corresponding to a $0.9\sigma$ significance for a non-zero correlation.
\vspace{-0.15cm}
\paragraph{Connection to observed peak-luminosity correlations.}

The preceding results raise the question of whether any directly observed property of the early optical flare correlates with $M_{\rm gal}$, $\sigma_{*}$, or $M_{\rm bulge}$. \citet{Mummery2024} found that both the observed peak $g$-band luminosity, $L_{\rm g,peak}$, and the peak blackbody luminosity, $L_{\rm BB,peak}$, correlate approximately linearly with $M_{\rm gal}$ and $\sigma_{*}$. Applying the same statistical tests used here, the recent review of early-time optical-flare models by \citet{Mummery_Calorimetry} found correlations between $L_{\rm peak}$ and $M_{\rm gal}$, $\sigma_{*}$, and $M_{\rm bulge}$ at $\geq5\sigma$ significance, with intrinsic scatter $\epsilon\simeq0.5$--$0.6$ dex.

As discussed in detail by \citet{Mummery_Calorimetry}, this positive empirical correlation between $L_{\rm peak}$ and host properties (which serve as proxies for $M_{\bullet}$), may explain why the model-based methods fail to recover reliable black hole--host scaling relations. Both \texttt{TDEmass} and \texttt{MOSFIT} assume that the peak luminosity decreases as the black hole mass increases. Specifically, \texttt{TDEmass} adopts
\begin{equation}
L_{\rm peak}\propto M_{\bullet}^{-1/6}\,\Xi(M_{\bullet})^{5/2},
\end{equation}
where $\Xi(M_{\bullet})$ is itself a decreasing function of $M_{\bullet}$, producing the approximate scaling $L_{\rm peak}\propto M_{\bullet}^{-3/8}$. \texttt{MOSFIT}, in turn, assumes
\begin{equation}
L(t)\simeq\eta c^2\dot{M}_{\rm fb}(t),
\end{equation}
and therefore $L_{\rm peak}\propto\eta\dot{M}_{\rm fb,peak}\propto\eta M_{\bullet}^{-1/2}$. Both assumptions are in direct tension with the observed, highly significant positive correlation between $L_{\rm peak}$ and proxies for $M_{\bullet}$.

The cooling-envelope model implemented in \texttt{REDBACK} differs in this respect because it assumes $L_{\rm peak}=L_{\rm Edd}\propto M_{\bullet}$ and therefore predicts a positive relation between peak luminosity and black hole mass. Nevertheless, its application to observed sources still fails to recover a statistically significant host-scaling relation. Inspection of Fig.~\ref{fig:mostfit_tdemass_redback} suggests that this failure is driven largely by AT2021ehb and iPTF16fnl, for which \texttt{REDBACK} infers intermediate-mass black holes with $M_{\bullet}\sim\mathrm{few}\times10^4\,M_{\odot}$, even though their host properties---particularly $\sigma_{\star}$, but also $M_{\rm gal}$ and $M_{\rm bulge}$---indicate substantially larger black hole masses. Understanding why the method fails so strongly for these two sources, while returning more reasonable masses for the others despite assuming a positive $L_{\rm peak}$--$M_{\bullet}$ relation, lies beyond the scope of this work.

\paragraph{Summary.}

Our analysis is not intended to distinguish among the detailed physical assumptions of the early-time flare models, which are discussed extensively by \citet{Mummery_Calorimetry}, nor to determine which model is most consistent with global hydrodynamical simulations. Instead, our goal is simpler and more practical: to determine whether applying these models to observed sources recovers the well-established correlations between black hole mass and host-galaxy properties and can therefore provide reliable estimates of $M_{\bullet}$. We emphasize that this is a deliberately minimal test: we do not require the recovery of the exact slope and intercept of a particular host relation, but only evidence for any positive and statistically significant correlation between the inferred $M_{\bullet}$ and an independent host property. Nevertheless, the results presented above show that black hole masses inferred from currently available physical models of the early-time optical emission do not exhibit any statistically significant correlation with host-galaxy properties, implying that these models do not recover reliable black hole masses.

\begin{figure*}
    \centering
    \includegraphics[width=0.75\textwidth]{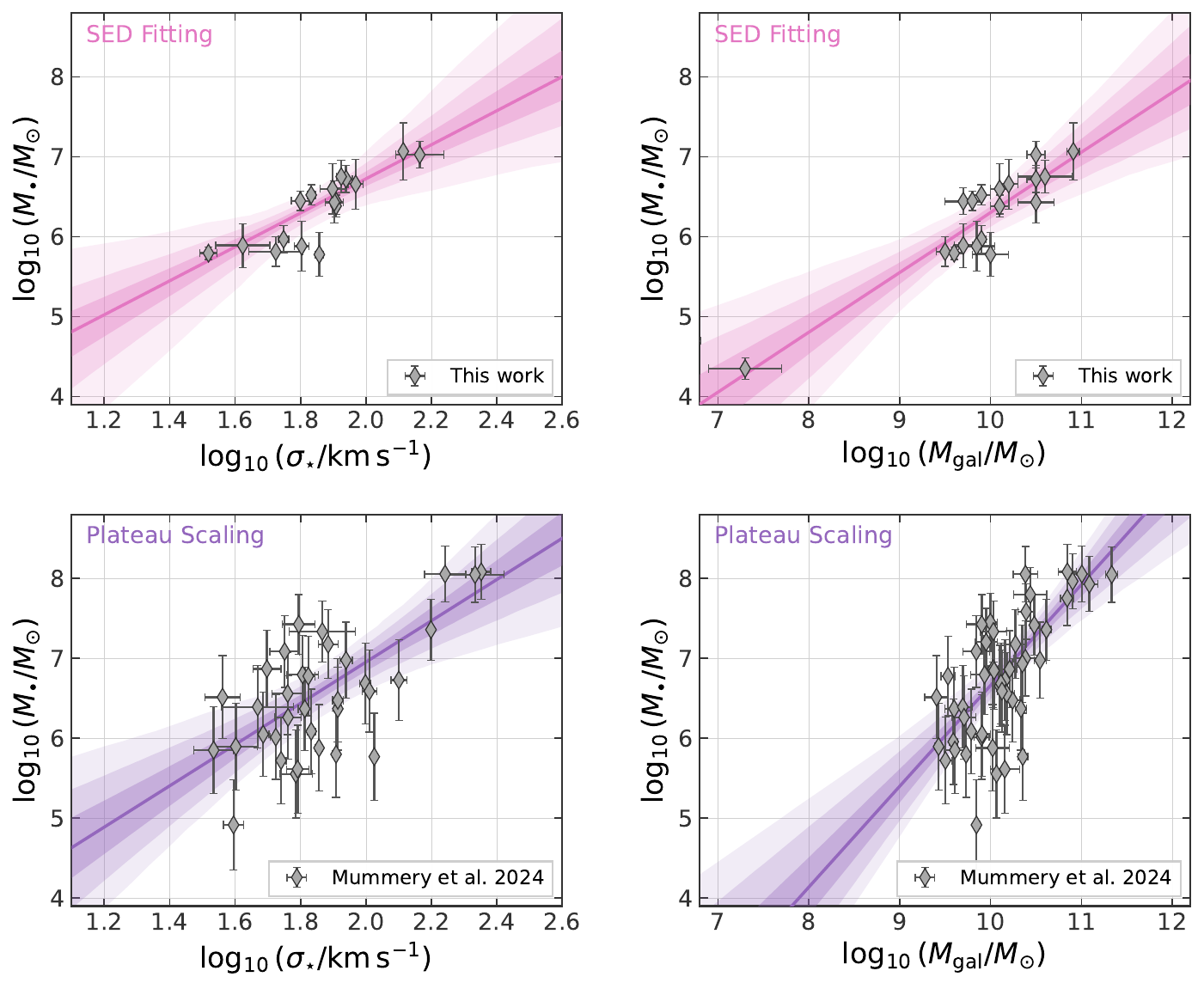}
    \caption{Black hole masses inferred using accretion-based methods---either full SED fitting (top; this work) or the plateau scaling relation (bottom; \citealt{Mummery2024})---versus galaxy mass (right panels) and nuclear stellar velocity dispersion (left panels). The contours have the same meaning as in Fig.~\ref{fig:mostfit_tdemass_redback}. All four panels show statistically significant correlations, ranging from $3.8\sigma$ to $\geq5\sigma$; see Table~\ref{tab:mass_stats}.}
    \label{fig:combined_sed_mummery}
\end{figure*}
\subsubsection{Accretion-Based Measurements}

\paragraph{Full SED Fitting versus Plateau Scaling.}

We now consider methods that infer $M_{\bullet}$ from accretion-disk emission, focusing on their ability to recover black hole--host galaxy scaling relations.

In \S\ref{sec:tde_disk}, we compared our full SED-fitting approach with the plateau-luminosity relation of \citet{Mummery2024}, outlining their common physical basis as well as their different observational requirements. Full SED fitting uses multiwavelength information to constrain both the inner disk, through quantities such as its peak temperature and black hole spin, and the outer disk. It can therefore yield relatively precise estimates of $M_{\bullet}$, but requires high-quality, contemporaneous multiwavelength coverage and is naturally restricted to nearby, well-observed TDEs. The plateau relation is less precise for an individual source but requires only a single luminosity measurement, making it applicable to much larger samples. The choice between these approaches therefore depends primarily on the available data and the aim of the analysis; because both are rooted in the same underlying accretion physics, they are expected to produce mutually consistent results when applied appropriately.

Full SED fitting also has an important limitation at the high-mass end. For $M_{\bullet}\gg10^7\,M_{\odot}$, the accretion disk is cooler and is therefore unlikely to produce detectable thermal X-ray emission (\S\ref{sec:accretion}). Such systems may exhibit only hard X-ray emission from a corona \citep[e.g.,][]{mummery_hard,Yao2026}, or no detectable X-ray emission at all. Although hard X-ray spectra could in principle be incorporated into the fits, the absence of a measurable peak disk temperature, $T_p$, reduces the precision of the inferred parameters. Consequently, the tight constraints achieved for thermally dominated systems should not be expected to extend directly to higher-mass, non-thermal sources.

\paragraph{Results of the Statistical Tests.}

Following the statistical framework introduced in \S\ref{sec:mosfit_mass}, we test whether the black hole masses inferred with these two accretion-based methods reproduce the established $M_{\bullet}$--$M_{\rm gal}$ and $M_{\bullet}$--$\sigma_{\star}$ relations. We adopt $X=M_{\rm gal}/(3\times10^{10}\,M_{\odot})$ for the galaxy-mass relation and $X=\sigma_{\star}/(160\,{\rm km\,s^{-1}})$ for the velocity-dispersion relation. The results are presented in Table~\ref{tab:mass_stats} and Fig.~\ref{fig:combined_sed_mummery}, with the full posterior distributions shown in Appendix~\ref{app:stats}.

For the full SED-fitting sample, the $M_{\bullet}$--$M_{\rm gal}$ relation is recovered with a Gaussian-equivalent significance of $\geq 5\sigma$. Because 3XMM J2150-05 substantially extends the dynamical range of the sample, we repeat the analysis excluding this source and still recover the correlation at a significance of approximately $3.4\sigma$. Despite the relatively small number of sources, the $M_{\bullet}$--$\sigma_{\star}$ relation is independently recovered at a significance of approximately $4.0\sigma$. The plateau-scaling masses yield even stronger results because of the larger sample size and dynamical range: both the $M_{\bullet}$--$M_{\rm gal}$ and $M_{\bullet}$--$\sigma_{\star}$ relations are recovered with significances of $\geq 5\sigma$, consistent with \citet{Mummery2024}. Thus, both accretion-based approaches recover the expected host-galaxy trends. Their principal differences lie in the required data quality, the uncertainties in individual mass measurements, and the intrinsic scatter, $\epsilon$, of the resulting correlations (Table~\ref{tab:mass_stats}).

We extend this test to the $M_{\bullet}$--$M_{\rm bulge}$ relation using the bulge-mass measurements and plateau-based black hole masses presented by \citet{Ramsden2025}. The right panel of Fig.~\ref{fig:cor_bulge} and Table~\ref{tab:mass_stats} show a non-zero correlation at $\geq5\sigma$. This is substantially stronger than the approximately $2.0\sigma$ correlation obtained by \citet{Ramsden2022} when using black hole masses inferred with \texttt{MOSFIT}.

Taken together, the correlations with $M_{\rm gal}$, $\sigma_{\star}$, and $M_{\rm bulge}$ demonstrate that accretion-based mass estimates consistently recover independently measured host-galaxy trends. These results also sheds light on the the pitfalls of the methods based on the early-time flare: the failure of \texttt{TDEmass}, \texttt{MOSFIT}, and \texttt{REDBACK} to recover black hole--host scaling relations cannot be attributed simply to TDE hosts intrinsically departing from the established galaxy population \citep[as argued by some authors, e.g.][]{Angus2026}. Figures~\ref{fig:host_scaling}, \ref{fig:cor_bulge}, and \ref{fig:combined_sed_mummery} show that TDE hosts do show the expected relations when $M_{\bullet}$ is inferred using accretion-based methods. The contrast therefore points instead to intrinsic limitations in the early-time optical-modeling methods themselves.

\begin{figure}
    \centering
    \includegraphics[width=\columnwidth]{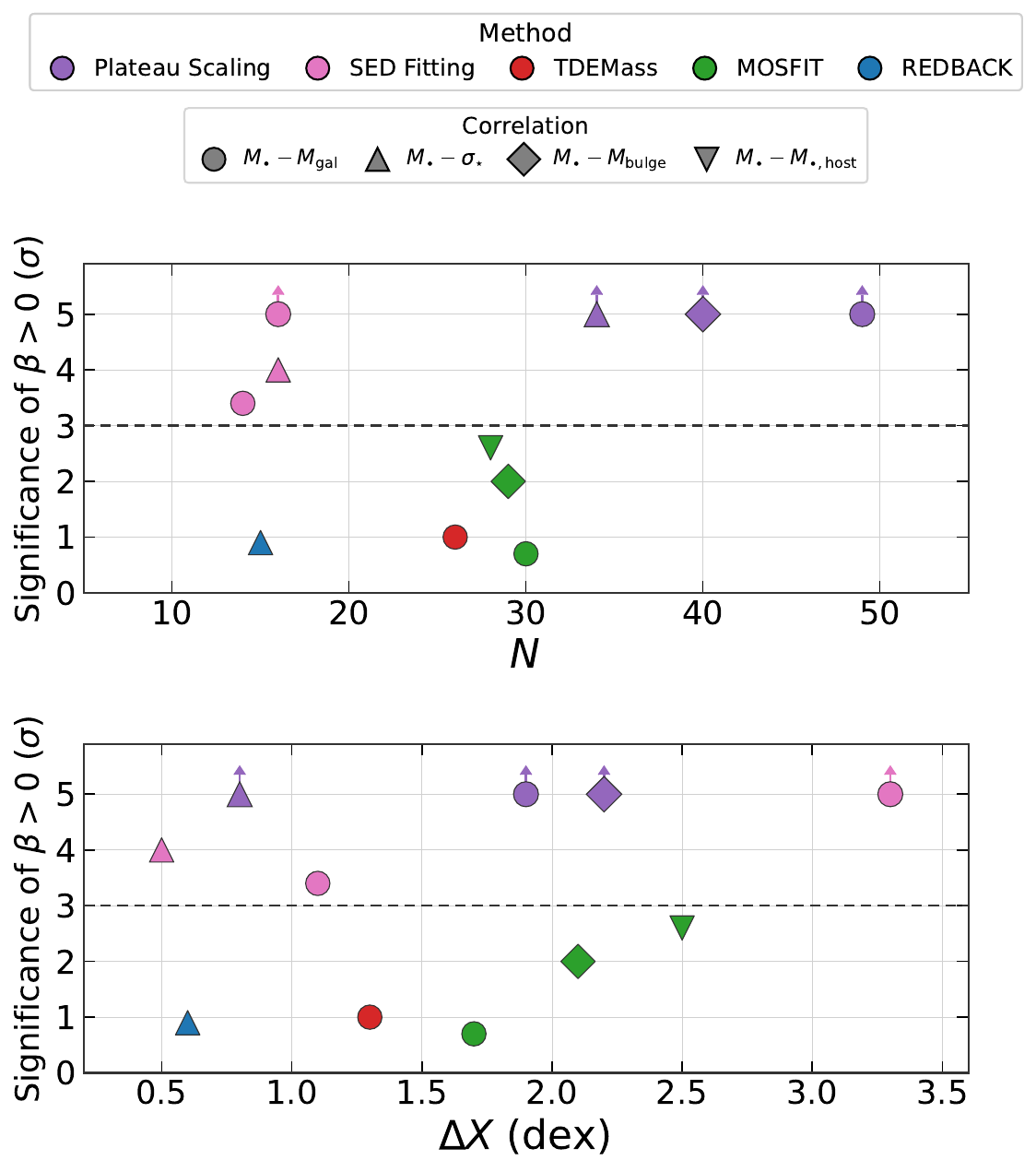}
    \caption{Gaussian-equivalent significance of the correlation between inferred black hole mass ($M_{\bullet}$) and host-galaxy properties as a function of sample size ($N$) and dynamical range in the host property ($\Delta X$) for the inference methods reviewed in this work. Within the samples considered here, the ability to recover a statistically significant ($>3\sigma$) black hole--host relation shows no apparent dependence on either $N$ or $\Delta X$. Accretion/disk-emission methods consistently recover significant correlations, whereas methods based on physical models of the early-time optical flare (\texttt{TDEmass}, \texttt{MOSFIT}, and \texttt{REDBACK}) do not.}
    \label{fig:sigma_method}
\end{figure}

Figure~\ref{fig:sigma_method} provides a direct comparison across all of the methods considered here. Within the assembled samples, the significance with which a black hole--host correlation is recovered shows no dependence on either the number of sources, $N$, or the dynamical range of the host property, $\Delta X$. Instead, the results separate according to the inference method: all accretion/disk-emission approaches recover correlations above $3\sigma$, whereas \texttt{TDEmass}, \texttt{MOSFIT}, and \texttt{REDBACK} remain below this threshold. The recovery of significant correlations by accretion-based methods even in relatively small samples further indicates that sample size alone does not explain the null results from the early-time optical models.

\paragraph{Implications for TDE Demographics.}

A natural conclusion from the results presented here is that the accretion-based framework provides a physically transparent description of TDE disk emission that can reproduce multi-wavelength observations while also recovering established black hole--host correlations. This highlights the potential of TDEs as an independent probe of precise black hole demographics. This interpretation is in agreement to existing attempts to reconstruct the black hole mass function with TDEs, which have relied on masses inferred either from host-galaxy scaling relations \citep{Yao2023} or from the plateau relation \citep{Mummery_vanVelzen2024}, but not from models of the early-time optical flare.

As the sample of TDEs with reliable accretion-based $M_{\bullet}$ estimates grows, these measurements may enable more detailed investigations of the black hole--host galaxy relations. In particular, TDEs can extend these relations into the low-mass regime, where direct dynamical measurements of $M_{\bullet}$ remain scarce. Larger samples could therefore determine whether the $M_{\bullet}$--$M_{\rm bulge}$ and $M_{\bullet}$--$\sigma_{\star}$ relations remain single power laws or change slope at low masses. Breaks in these relations have been predicted theoretically \citep[e.g.,][]{Cirasuolo2005,Fontanot2015,Hopkins2022}, and recent analyses of TDE samples have provided tentative, though not yet conclusive, observational evidence for such behavior \citep{Ramsden2025}.

\subsection{Implications for TDE Physics and Modeling}

\paragraph{Origin of the diverse X-ray emission.}
The fitting results and parameter correlations presented in \S\ref{sec:results} have several implications for the physical interpretation of TDE emission. First, the ability of a simple thin-disk model to reproduce both the X-ray spectra and the UV/optical photometry during the plateau strongly supports a common origin in direct disk emission. The same model, with fixed \Rin\ and $a_{\bullet}$, also describes the X-ray spectra at earlier epochs, when the optical emission is still dominated by a separate component, simply by allowing the peak disk temperature $T_p$ to evolve. This consistency indicates that the accretion disk powers the X-ray emission throughout the observed evolution.

In systems for which the early X-ray evolution can be reproduced by varying only $T_p$ while holding the intrinsic disk parameters fixed (Fig.~\ref{fig:lcs} and Appendix~\ref{app:data}), the X-rays cannot be strongly absorbed and reprocessed into optical emission by a spherical outflow surrounding the disk. If spherical reprocessing dominated, the direct disk emission would not remain visible with the same intrinsic disk parameters while the early optical flare was present. This conclusion is consistent with the independent luminosity-function analysis of \citet{Mummery_vanVelzen2024}.

A related result, first noted by \citet{Guolo2024} using host-based estimates of $M_{\bullet}$ and confirmed here using accretion-based masses, is that prompt and delayed X-ray TDEs show no clear difference in black hole mass. ASASSN-14li and AT2019azh, for example, have nearly identical black hole masses within their small uncertainties, yet the X-rays in ASASSN-14li peaked promptly, whereas those in AT2019azh peaked several months later, after the early optical component had nearly disappeared. Differences in $M_{\bullet}$ therefore cannot be the primary cause of delayed X-ray emission.

An alternative to spherical reprocessing is a geometric scenario in which the obscuring material occupies a fixed or time-dependent solid angle around the equatorial plane. In a high-inclination system, such material could reprocess high-energy photons to lower energies, suppressing the observed early X-rays while powering the optical flare \citep[e.g.,][]{Dai2018}. \citet{Guolo2024} proposed this possibility to explain why standard disk fits to some rising X-ray spectra yield unphysical parameters, such as unrealistically small inner radii. The inclination constraints obtained here, however, do not reveal the expected orientation dependence: both prompt and delayed X-ray emission occur in systems inferred to have low and high inclinations. ASASSN-14li (prompt) and AT2020ksf (delayed) are both nearly face-on, whereas AT2019dsg (prompt) and AT2011ehb (delayed) are likely viewed at higher inclinations. Moreover, any reprocessing model in which the luminosity tracks the fallback rate, including orientation-dependent versions of the model proposed by \citet{Dai2018}, retains the difficulty discussed above of reproducing the observed positive correlation between $L_{\rm peak}$ and $M_{\bullet}$. Together, these results disfavor both black hole mass and viewing angle as the sole drivers of the diversity in X-ray rise times.

\begin{figure*}
    \centering
    \includegraphics[width=0.9\linewidth]{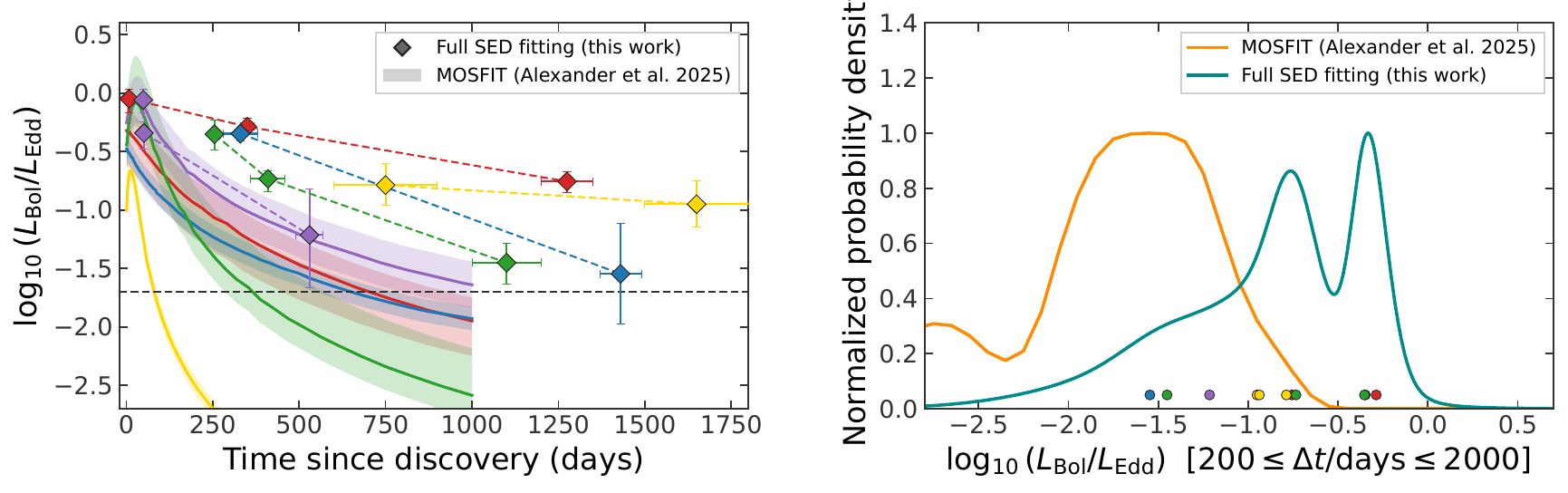}
    \caption{Comparison between the late-time Eddington ratios, $L_{\rm Bol}/L_{\rm Edd}$, obtained from full SED fitting and the ``accretion rates'' inferred with \texttt{MOSFIT} for the five sources common to this work and \citet{Alexander2025}. Colors identify the sources: AT2019qiz (yellow), ASASSN-14li (red), ASASSN-15oi (blue), AT2019azh (green), and AT2019dsg (purple). The left panel shows $L_{\rm Bol}/L_{\rm Edd}$ as a function of time, and the right panel shows the probability-density distributions over $200\leq\delta t/{\rm days}\leq2000$. \texttt{MOSFIT} systematically underestimates the Eddington ratio, with the discrepancy increasing over time. At $\delta t=1000$ days, \texttt{MOSFIT} assigns all five sources Eddington ratios $\leq3\%$ (dotted line on left panel), whereas the full SED fits imply that only AT2019dsg (purple) lies below this threshold.}
    \label{fig:eddcomp}
\end{figure*}

Nevertheless, the principal conclusion of \citet{Guolo2024} remains valid: X-ray-selected and optically selected TDEs are drawn from the same underlying black hole population. This result has since received independent support from TDE luminosity functions \citep{Mummery_vanVelzen2024}. Our sample provides additional evidence because it contains both X-ray- and optically selected events, all of which can be described within the same physical framework without an evident dependence on $M_{\bullet}$.

A natural explanation for the range of X-ray rise times is instead a diversity of viscous timescales among TDE disks \citep[an argument introduced by][and made more specific here]{Mummery2021b}. The viscous timescale of an accretion flow scales as
\begin{equation}
t_{\rm visc}\propto\frac{\sqrt{G M_{\bullet}r_0^3}}{W^r_{\phi}},
\end{equation}
where $W^r_{\phi}$ is the turbulent stress tensor \citep{Balbus1999}. Under the $\alpha$ prescription of \citet{Shakura1973}, $W^r_{\phi}=G M_{\bullet}(h/r)^2\alpha$, where $h/r$ is the disk aspect ratio. Because the tidal radius scales as $R_T\simeq R_{\star}(M_{\bullet}/M_{\star})^{1/3}$, the viscous timescale of a flow formed near $R_T$ becomes
\begin{multline}
t_{\rm visc,TDE}\simeq\alpha^{-1}(h/r)^{-2}
\sqrt{\frac{R_T^3}{G M_{\bullet}}}\\
\simeq\alpha^{-1}(h/r)^{-2}
\sqrt{\frac{R_{\star}^3}{G M_{\star}}},
\end{multline}
and is therefore independent of $M_{\bullet}$ to leading order. A flow formed near $R_T$ and propagating inward should reach its X-ray peak after $\delta t\sim t_{\rm visc}$.\footnote{Here, $\delta t$ is the interval between debris circularization and the X-ray peak. It should not be confused with $\Delta t$, used previously for the interval between discovery and a given observation.} Observed TDE viscous timescales span at least approximately two orders of magnitude \citep{Guolo2025b}, plausibly because of variations in stellar density, $t_{\rm visc}\propto\rho_{\star}^{-1/2}$, and in the nuisance parameters $\alpha(h/r)^2$. The absence of a relation between $M_{\bullet}$ and prompt or delayed X-ray emission is therefore consistent with simple viscous-disk theory and merits further investigation.

\paragraph{Bolometric luminosities and accretion rates}

Our results also have important implications for estimates of bolometric luminosity and, consequently, of accretion rate or Eddington ratio. The bolometric luminosity is the total emission integrated over all wavelengths, but observations usually cover only a limited part of the SED. A common approach is to fit the UV/optical emission with a single-temperature blackbody and identify its integrated luminosity, $L_{\rm BB}$, with the bolometric luminosity, as is done in \texttt{MOSFIT}. This estimate necessarily excludes the observed X-ray emission (which routinely reaches $L_X\sim10^{43}\,{\rm erg\,s^{-1}}$) as well as assumes a shape (a single temperature blackbody) for the extreme-ultraviolet emission, which is not measured.

Another common prescription is to integrate an X-ray spectral model over a finite band,
\begin{equation}
L_X=4\pi D^2\int_{0.3\,{\rm keV}}^{10\,{\rm keV}}F_E\,{\rm d}E,
\end{equation}
where $F_E$ is the fitted spectral model, and then define $L_{\rm Bol}=L_X+L_{\rm BB}$. This prescription is also incomplete because there is no physical reason for truncating the disk emission at $0.3$ keV. A typical observed TDE SED rises toward the Lyman limit in the UV and is already declining when it emerges in the observed soft X-ray band. Its peak must therefore lie in the unobserved extreme-ultraviolet region between these bands (Figs.~\ref{fig:SED_fit1} and \ref{fig:SED_fit2}). A physically motivated model that describes the UV, extreme-ultraviolet, and X-ray emission self-consistently is thus required to estimate $L_{\rm Bol}$ reliably. This is the approach adopted here. Methods that omit the unobserved spectral peak will underestimate the bolometric luminosity and bias the inferred system properties.

Accurate Eddington ratios are particularly important when testing whether black hole accretion is scale invariant. One outstanding question is whether supermassive black hole disks undergo state transitions near $\dot m\sim0.02$, as observed in X-ray binaries \citep[e.g.,][]{Fender2004}. TDEs provide an unusually powerful means of addressing this question because their accretion flows evolve on observable timescales. Such a test, however, requires both a physical model of the accretion flow and a reliable estimate of its accretion rate.

Figure~\ref{fig:eddcomp} compares the Eddington-normalized bolometric luminosities obtained from our full SED fits with the ``accretion rates'' inferred by \citet{Alexander2025} using \texttt{MOSFIT}.\footnote{\texttt{MOSFIT} does not contain an accretion disk or model the physics of disk accretion. It follows the rate at which debris returns to pericenter, $\dot M_{\rm fb}$, which is not equivalent to the rate at which matter reaches the black hole, $\dot M_{\rm acc}$, because the returning material has angular momentum. \texttt{MOSFIT} includes a viscous-delay parameter intended to account for this distinction, but \citet{Alexander2025} find it to be consistent with zero for most sources.} The former are derived from emission produced directly by the accretion flow---the late-time UV/optical plateau and X-rays---whereas the latter are inferred from fits to the early optical flare.

Beyond $\Delta t\sim250$ days, the \texttt{MOSFIT} values differ from our full-SED estimates by at least an order of magnitude for every source except the rapidly evolving AT2019dsg. AT2019qiz provides a particularly striking example: its \texttt{MOSFIT}-inferred peak Eddington ratio is lower than the value measured from the full SED at 1750 days, an epoch at which its SED is well constrained (Figs.~\ref{fig:SED_fit1} and \ref{fig:SED_fit2}; see also \citealt{Nicholl2024}). At 1000 days, \citet{Alexander2025} infer Eddington ratios below $3\%$ for every source in their sample, whereas only one of the 16 sources studied here, AT2019dsg, falls below that value.

This discrepancy is directly relevant to the search by \citet{Alexander2025} for a relation between radio properties and the proposed $\dot m\sim0.02$ state transition. Because their inferred rates are derived from emission that is not produced by the disk, using a model without accretion physics and black hole masses that do not reproduce host-galaxy scaling relations, the absence of a correlation cannot provide a meaningful test of the proposed transition.

The full SED fits further show that the bolometric luminosity of TDE disks declines much more slowly than the observed X-ray luminosity (Fig.~\ref{fig:SED_fit2}), owing to the exponential dependence of X-ray luminosity on disk temperature \citep{Mummery2020,Mummery2023}. Consequently, the correction from observed X-ray luminosity to bolometric luminosity grows exponentially with time. Combined with disks that remain luminous for at least a decade, this behavior naturally resolves the TDE ``missing-energy problem'' \citep[e.g.,][]{Mummery2021b,Mummery_vanVelzen2024,Guolo_Mummery2025,Guolo2025b}.

%Finally, allowing $E(B-V)$ to vary in the full SED fits yields non-zero host-galaxy extinction for several sources (Table~\ref{tab:best_fit}), whereas many TDE studies assume zero host extinction. Although the inferred values are small, extinction increases rapidly toward shorter UV wavelengths and may therefore affect measurements of early-time optical-flare luminosities and color temperatures. This possibility warrants further investigation.

\section{Conclusions}\label{sec:conclusions}

We present a multi-wavelength analysis of 16 TDEs with available X-ray spectra during the UV/optical plateau phase of their evolution, using full spectral energy distribution fitting as a method for black hole and disk parameter inference, alongside a comparative review of the methods used to estimate black hole masses in TDEs. Our main conclusions are as follows:

\begin{itemize}
    \item During the late-time ``plateau phase'' of TDE  evolution the entire SED from optical to X-ray wavelengths can be described by an evolving accretion flow, with no need for any additional spectral components. 
    \item The very same disk models, simply with a larger peak disk temperature, reproduce the X-ray emission observed from TDEs at all times. This means that the disks observed at late times in the UV/optical are the same accretion systems which produce X-rays at all observational epochs. 
    \item Full SED fitting in the disk-nominated phase provides a robust and accurate (typical uncertainty of $\pm 0.2$ dex) way of measuring black hole masses in TDEs. These masses independently recovers correlations with host properties at high significance ($>3\sigma$), despite the small sample size. Showing our method is not only precise, but also reliable. In the case of the $M_{\bullet}-\sigma_{\star}$ correlation, our method recovers a correlation with a scatter of only $\epsilon \sim \pm 0.1$ dex.
    \item Detailed modeling of late-time TDE disks recovers the expected accretion-disk scaling relations, $L_{\rm bol}^{\rm disk}/L_{\rm Edd} \propto T_p^4 \propto M_{\bullet}^{-1}$. Characteristic TDE correlations are also recovered, including the UV plateau scaling $L_{\rm plat} \propto M_{\bullet}^{2/3}$ and, for the first time, the compact-disk size scaling $R_{\rm out}/r_g \propto M_{\bullet}^{-2/3}$. This second result confirms the TDE interpretation of all sources in our sample.
    \item Both the general accretion correlations and those specific to TDE disks are, for the first time, self-consistently extended—through a homogeneous modeling and sample analysis—into the IMBH regime, using the off-nuclear TDE 3XMM~J2150$-$05.
    \item We have reviewed different techniques for inferring black hole masses in TDEs, using a uniform statistical analysis, we show that approaches based on models of the early-time UV/optical emission (i.e., those which are the base of \texttt{TDEmass}, \texttt{MOSFIT}, and \texttt{REDBACK}) are not able to recover black hole–host galaxy scalings, implying that parameter inference with these techniques is unreliable.
    \item We have demonstrated that the accretion rates ($L_{\rm Bol}/L_{\rm Edd}$) in TDEs cannot be estimated from the fallback-based modeling of the early-time optical flare. Given that this approach is based on: inferred $M_{\bullet}$ that cannot reproduce host-scaling relations, and $L_{\rm Bol}$ that cannot explain their full data set (UV/optical plateau and X-ray at any time).
\end{itemize}
    
The results presented in this paper highlight the value of detailed multi-wavelength modeling of TDE disks, particularly at late times. This powerful observational approach will be especially important as we enter the era of LSST, with its anticipated capability to discover large numbers of TDEs, and as future missions at complementary wavelengths—such as UVEX \citep{Kulkarni2021}, CASTOR \citep{Cote2012} and AXIS \citep{Reynolds2023}—enabling combined multi-wavelength studies.

\textit{Acknowledgments} -- 
MG acknowledges support from NASA through XMM-Newton grant 80NSSC24K1885.  A.M. acknowledges support from the Ambrose Monell Foundation, the W.M. Keck Foundation and the John N. Bahcall Fellowship Fund at the Institute for Advanced Study. 
MN is supported by the European Research Council (ERC) under the European Union’s Horizon 2020 research and innovation programme (grant agreement No.~948381).

\clearpage

\appendix
\section{Data Reduction and Basic Analysis}\label{app:data}

\subsection{New $\sigma_{\star}$ Measurements}

\begin{deluxetable*}{lcccccc}[htbp!]
\tabletypesize{\small}
\tablecaption{Log of medium-resolution optical spectroscopy with Keck-II ESI. \label{tab:spec_medres}}
\tablehead{
    \colhead{IAU Name}
    & \colhead{Start Date MJD}  
	& \colhead{Exp. (s)} 
        & \colhead{$r_{\rm extract}$ (pixel)\tablenotemark{a}} 
        & \colhead{S/N}
        & \colhead{$\sigma_\ast$}
	}
\startdata
AT2019vcb & 2021-12-28.53  & 1200 & -- & 4.6 & $41.86^{+8.70}_{-8.09}$\\
AT2022dsb & 2023-04-15.59  & 900 & 6.5 & 20.0 & $84.07^{+3.82}_{-3.45}$\\ 
AT2023cvb & 2022-11-27.6   & 1500 & 6.5 & 18.1 & $79.98^{+5.92}_{-5.01}$ \\ 
\enddata 
\tablenotetext{a}{The radius used for extracting the spectrum. $r_{\rm extract}$ can be converted to angular scale using a conversion factor of 0.154$^{\prime\prime}$ per pixel. For AT2019vcb, due to the low S/N of the observation, we extracted the spectrum from the full trace.}
\end{deluxetable*}

\begin{figure}[htbp!]
    \centering
    \includegraphics[width=\columnwidth]{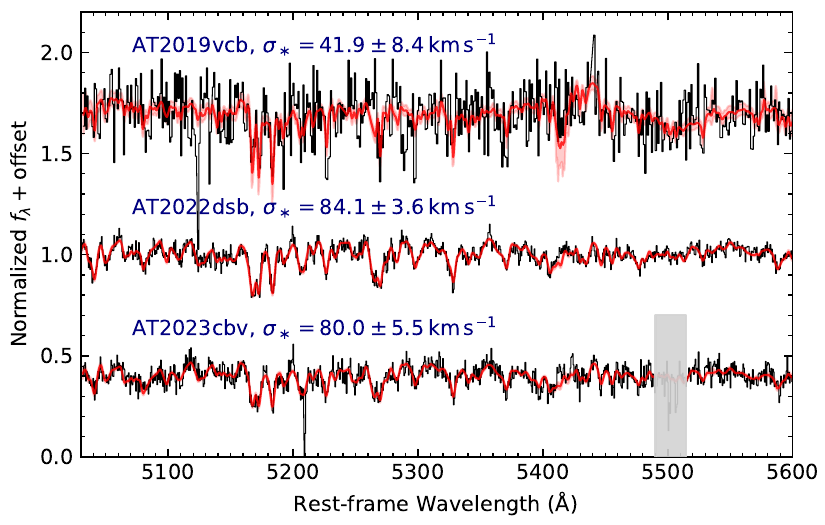}
      \includegraphics[width=0.9\columnwidth]{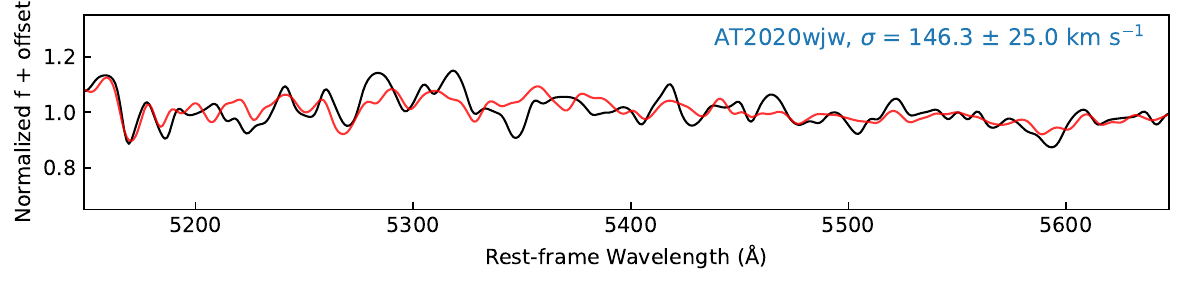}
    \caption{Top: ESI spectra of the host galaxies of TDEs AT2019vcb, AT2022dsb, and AT2024cvb (black) and the best-fit models (red). Bottom: Same as above for WiFeS spectrum of AT2020wjw/eRa J2344-35.}
    \label{fig:new_sigma}
\end{figure}

The host galaxies of AT2019cvb, AT2022dsb, and AT2023cvb were observed by the Echellette Spectrograph and Imager (ESI; \citealt{Sheinis2002}) on the Keck II telescope (see Table~\ref{tab:spec_medres} for a log). The slit width for all observations is 0.75\arcsec, corresponding to $\sigma_{\rm inst} = 23.7\,{\rm km\,s^{-1}}$. We follow the same procedures as outlined in \citet{Yao2023} to reduce the data and measure $\sigma_\ast$ by fitting the rest-frame 5030--5600\,\AA\ spectrum with the penalized pixel-fitting (\texttt{pPXF}) software \citep{Cappellari2004, Cappellari2017}. The data and best-fit models are shown in the top panel of Fig.~~\ref{fig:new_sigma}.

The host galaxy of AT2020wjw/eRA J2344-35 was observed with the Wide-Field Spectrograph (WiFeS; \citealt{Dopita2010}) on the ANU 2.3\,m telescope at Siding Spring Observatory and made publicly available by \citet{Grotova2025b}. The R3000 grating was used, providing a spectral resolution of $R = 3000$ at 7600\,\AA, corresponding to $\sigma_{\rm inst} = 42.4\,{\rm km\,s^{-1}}$ (assuming $R = \lambda/{\rm FWHM}$). We measure the stellar velocity dispersion $\sigma_{\star}$ by fitting the rest-frame 5100--5500\,\AA\ spectrum with \texttt{pPXF}. The data and best-fit model are shown in the bottom panel of Fig.~\ref{fig:new_sigma}.

\subsection{Sources of Data}

X-ray data from \xmm/EPIC-pn, \nicer/XTI, \swift/XRT, and \chandra/ACIS were used in this work. The \xmm\ data were reduced following the procedures described in \citet{Guolo2024}. The \swift/XRT data were processed using the \textit{Swift}/UK automated online tool\footnote{\url{https://www.swift.ac.uk/user_objects}} \citep{Evans2009}. The \nicer/XTI data were reduced as in \citet{Guolo2024b}, except for the background modeling: instead of using \texttt{3C50} \citep{Remillard2021}, we employed the \texttt{SCORPION} model by setting \texttt{bkgmodeltype=scorpeon bkgformat=file} in \texttt{nicerl3-spec}. The \chandra data utilized in this work comprise only the ACIS-S observation (Obs-ID 17862) of 3XMM J2150-05. We performed all the data reduction using {\tt CIAO 4.17} and {\tt CALDB 4.12.0}, starting with reprocessing the data using the {\tt chandra\_repro} tool. We then extracted the spectral files, including the ARFs and RMFs, using the {\tt specextract} tool in CIAO, with a circular source region of radius 2.5'' and an annular background region with inner and outer radii equal to 4'' and 9''. 

For AT2019qiz, AT2021ehb, ASASSN-14li, AT2019azh, AT2022dsb, AT2022lri, ASASSN-15oi, AT2018cqh, and AT2023vcb, host-subtracted UV/optical photometry was obtained from the \texttt{ManyTDE} library, whose reduction procedures are described in \citet{Mummery2024}. From this library, we use both \swift/UVOT and ZTF light curves.

For AT2019vcb and AT2020ksf, individual \texttt{ManyTDE} observations do not yield significant ($\geq 3\sigma$) detections of the plateau phase. For these sources, we instead stacked the \swift/UVOT images over the time intervals listed in Table~\ref{app:data}, and adopted the stacked fluxes for epochs detected at $\geq 3\sigma$ above the host level. For display purposes only, we also include the ATLAS \citep{Tonry2022_discovery_report} light curve of AT2020ksf, the only instrument that captured its optical peak; the reduction of these data is described in \citet{Wevers2024}.

For GSN~069, we use the two epochs of binned, host-subtracted UV \textit{HST}/STIS spectroscopy, processed as detailed in \citet{Guolo2025} and \citet{Guolo2025b}.

For 3XMM~J2150$-$05, we use both pre-transient (host) and post-transient photometry as reported in \citet{Lin2018, Lin2020}. Host subtraction for this source follows the methods described in Mummery \& Guolo et al., in prep.

\subsection{Median SED computation}

We compute the mean UV/optical SED in bins during the plateau phase as shown in Table \ref{sec:data}. Within each bin, the mean of $N$ measurements of the flux $F_i$ is computed using inverse-variance weighting: 
\begin{equation}
    \bar{F} = \frac{\sum_i^N \, F_i\sigma_{F_i}^{-2}} {\sum_i^N \sigma_{F_i}^{-2}}
\end{equation}
with $\sigma_{F}$ the measurement uncertainty of the flux. The variance of this mean flux is given by
\begin{equation}
   \sigma_{\bar{F}}^2 = \frac{1}{{\sum_i \sigma_{F_i}^{-2}}} \label{eq:statunc}
\end{equation}
If the flux measurement contains a source of systematic uncertainty, this will be apparent from the sample variance, $\sigma_{\rm sample}^2 = (N-1)^{-1} \sum_i^N (F_i-\bar{F})^2 $. Systematic uncertainty dominates when the square root of the sample variance is larger than the typical measurement uncertainty. For bins with $N>5$ data points, we therefore also compute the uncertainty on the mean flux under the assumption that true uncertainty of each observation follows from the sample variance: 
\begin{equation}
    \sigma^2_{\bar{F}, \rm sample} = \frac{\sigma^2_{\rm sample}}{N}\quad.\label{eq:sampleunc}
\end{equation}
If the statistical uncertainty (Eq.~\ref{eq:statunc}) is smaller than the sample-variance based estimate (Eq.~\ref{eq:sampleunc}), we use the latter in our likelihood function.

\begin{figure*}[h]
\centering
    \includegraphics[width=\textwidth]{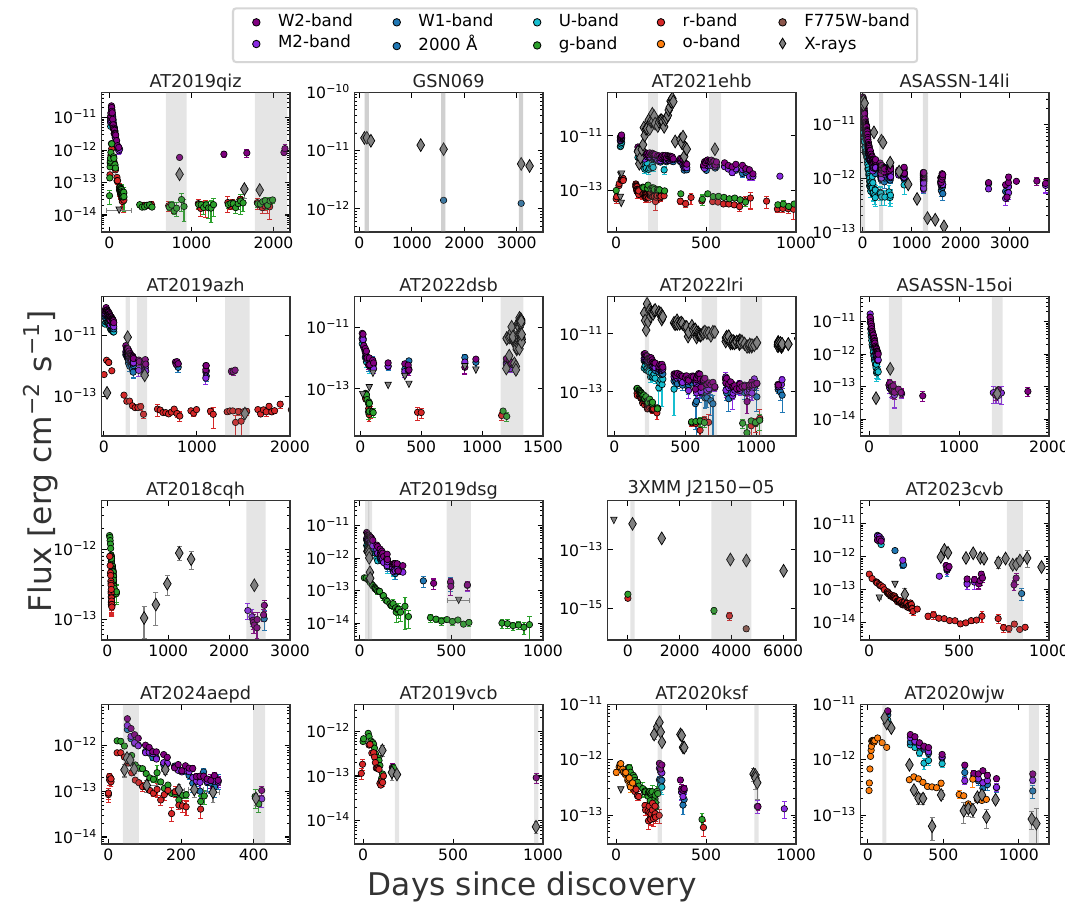}
       \caption{Host subtracted multi-wavelength light curves for all sources, shown as observed with no correction for extinction Galactic or intrinsic. X-ray light curves show absorbed fluxes.  }
\end{figure*}

\section{Statistical Supplements}\label{app:stats}

\begin{figure*}[h]
     
    \centering
    \includegraphics[width=0.45\textwidth]{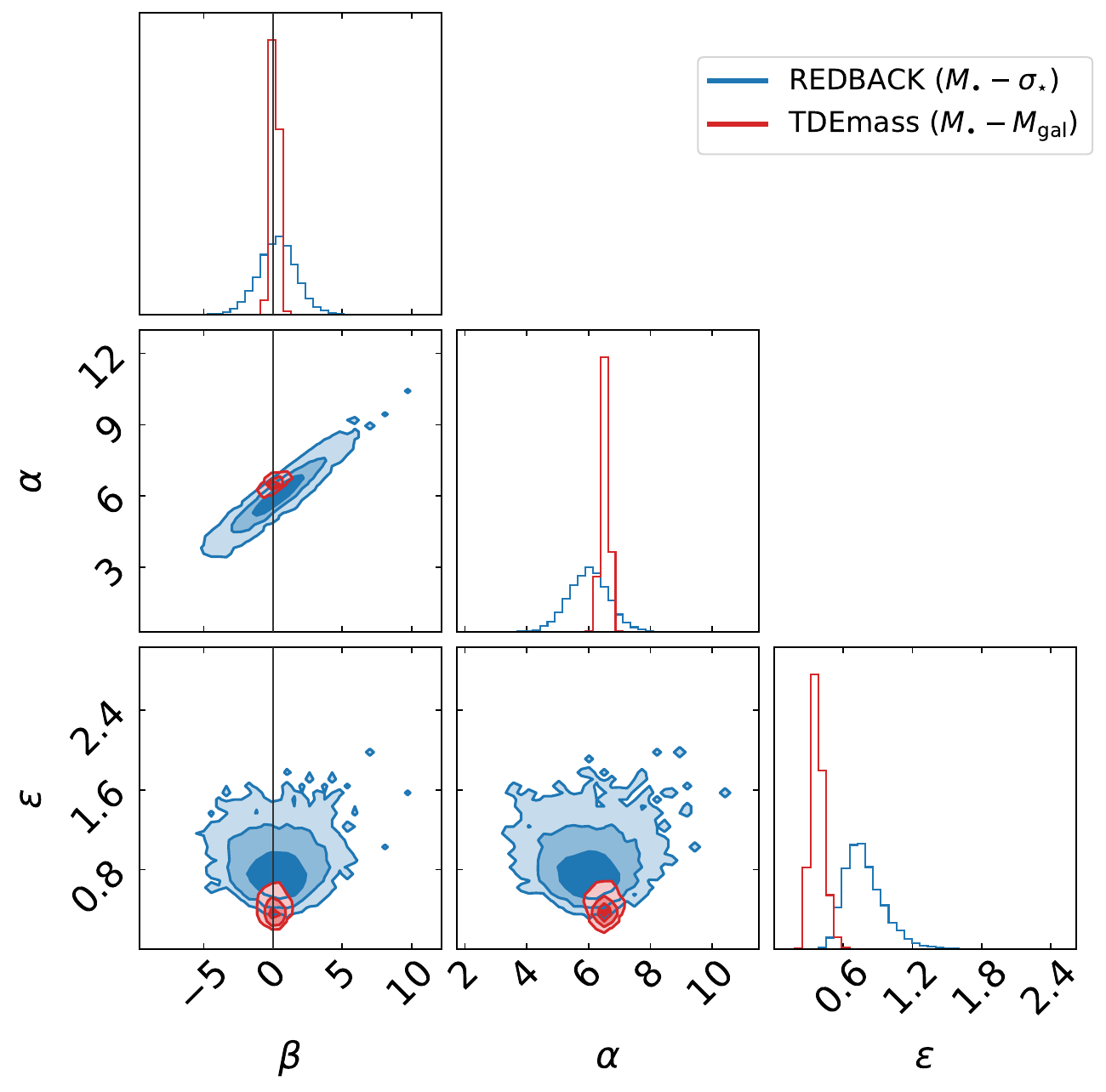}
     \includegraphics[width=0.45\textwidth]{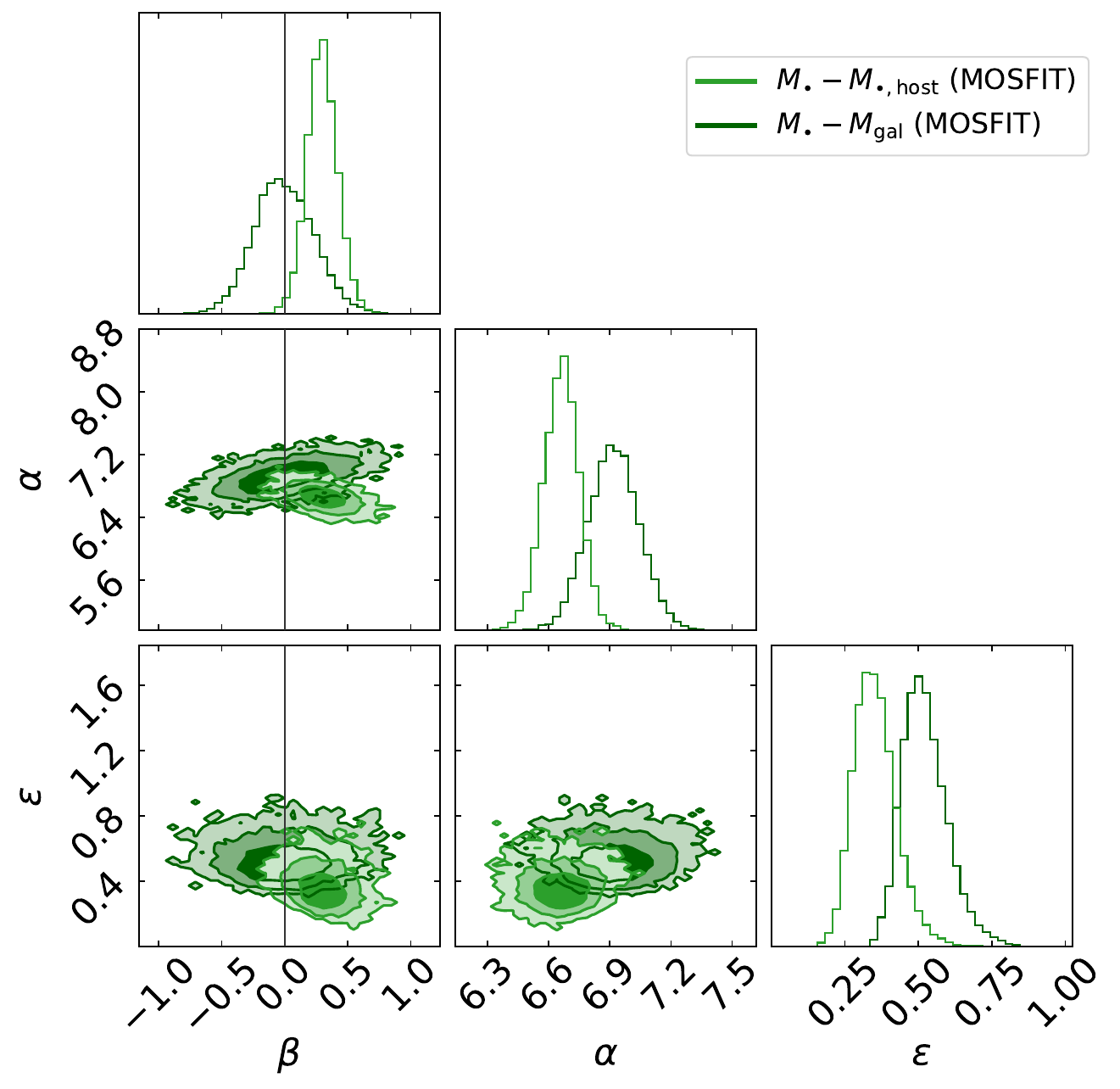}
           \caption{Posterior distributions corresponding to the MCMC fits presented in Figs~\ref{fig:mostfit_tdemass_redback}. In both panels black line shows the demarcation for non-correlation, i.e., $\beta=0$.}
        \label{fig:post_optical}
\end{figure*}

\begin{figure*}
  
    \centering
     \includegraphics[width=0.45\textwidth]{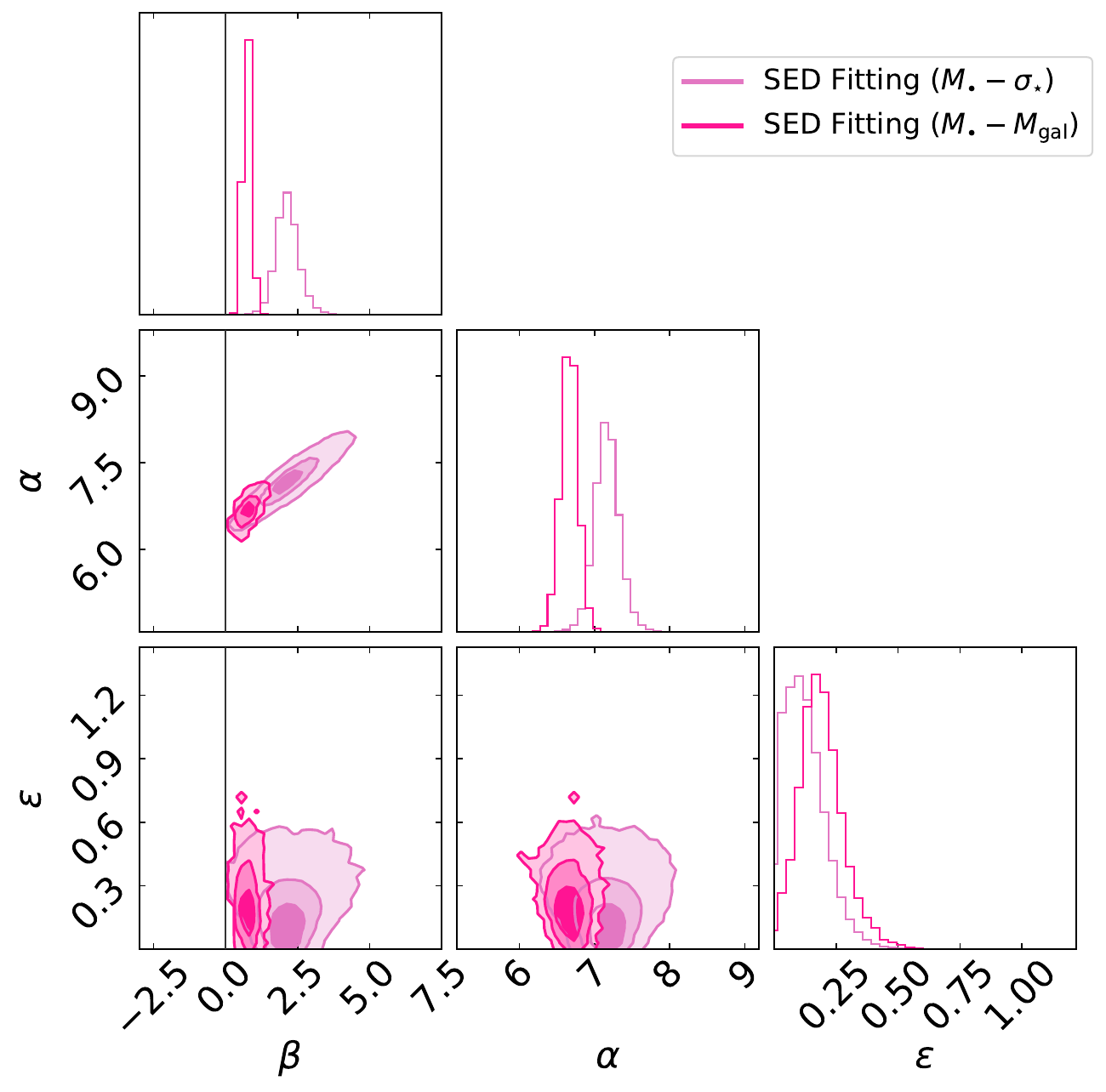}
      \includegraphics[width=0.47\textwidth]{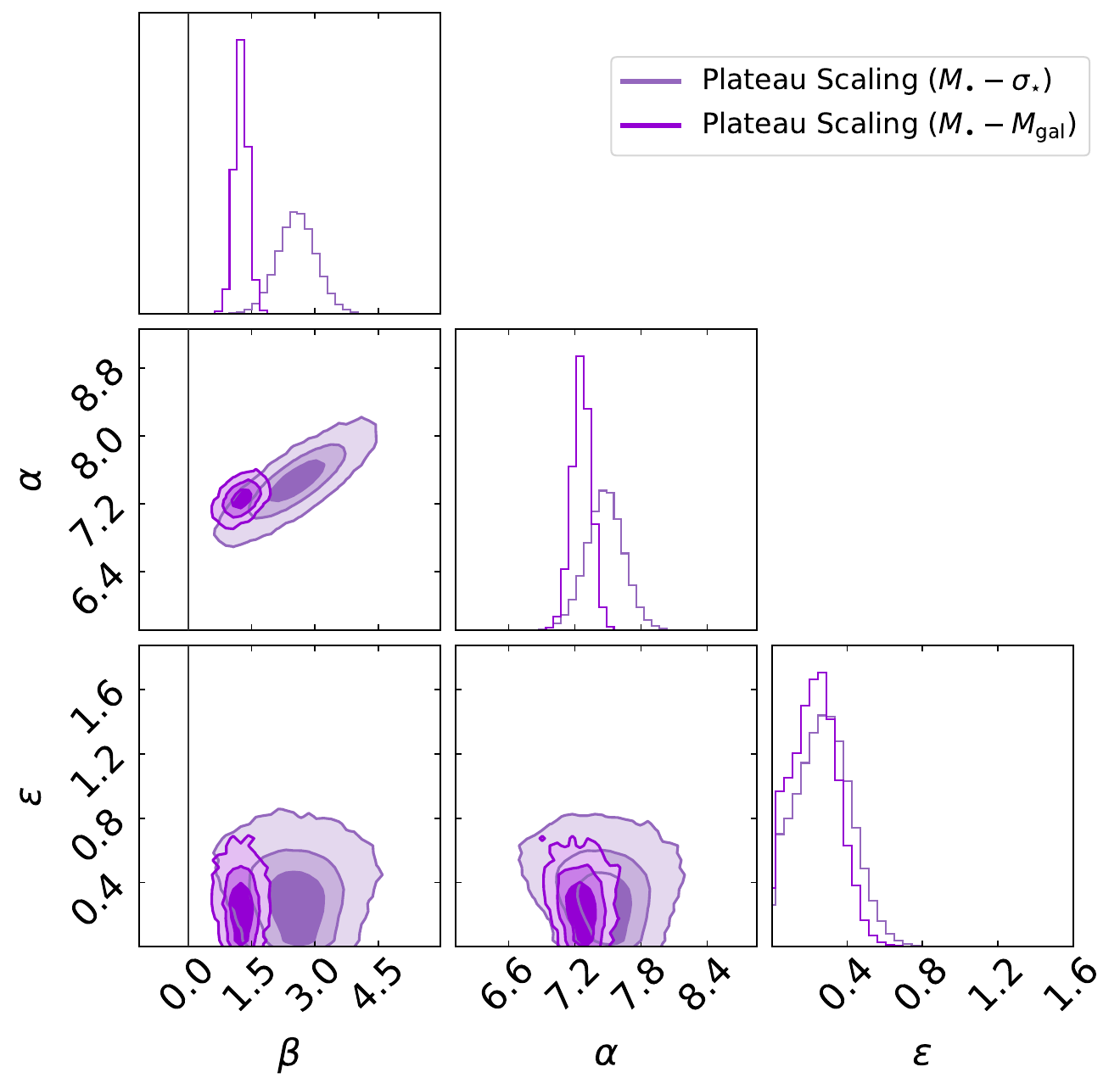}
        \caption{Posterior distributions corresponding to the MCMC fits presented in Fig~\ref{fig:combined_sed_mummery}. In both panels black line shows the demarcation for non-correlation, i.e., $\beta=0$.}
        \label{fig:post_MG}
\end{figure*}

\begin{figure*}

    \centering
    \includegraphics[width=0.47\textwidth]{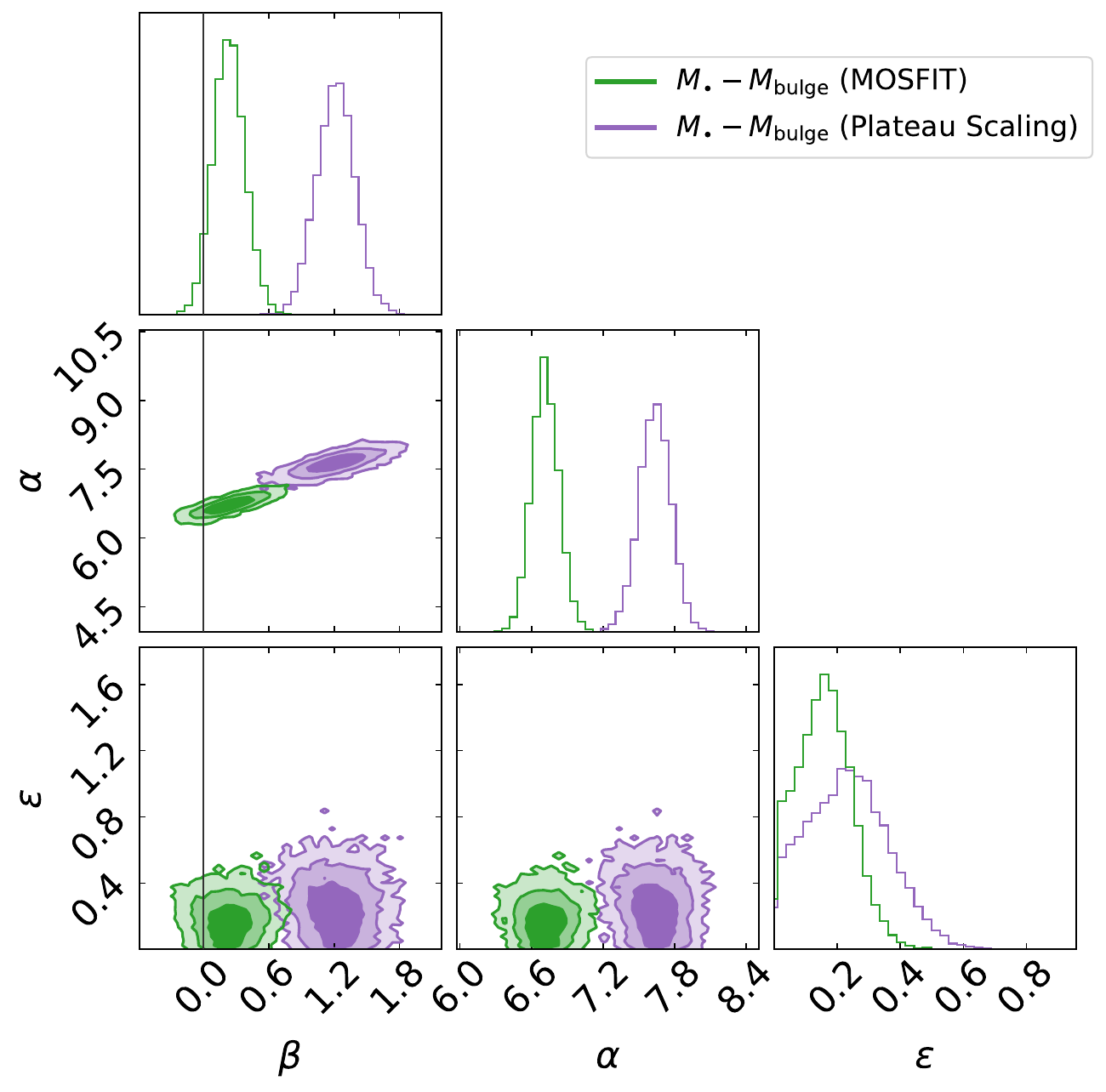}\\
    \caption{Posteriors of correlations (or lack thereof) between black hole mass ($M_{\bullet}$) —estimated either with \texttt{MOSFIT} (green) or from the plateau scaling relation of \citet{Mummery2024} (purple) — and host galaxy bulge mass ($M_{\rm bulge}$), as measured and presented in \citet{Ramsden2022, Ramsden2025}. Black line shows the demarcation for non-correlation, i.e., $\beta=0$.}
    \label{fig:post_paige}
\end{figure*}

Here we present additional statistical results. Figure~\ref{fig:post_optical} shows the posterior distributions corresponding to the MCMC fits presented in Fig.~\ref{fig:mostfit_tdemass_redback}, while Fig.~\ref{fig:post_MG} shows the posterior distributions for the fits shown in Fig.~\ref{fig:combined_sed_mummery}. Figure~\ref{fig:post_paige} presents the posterior distributions of the correlations shown in Fig.~\ref{fig:cor_bulge}.

\bibliography{tde}{}
\bibliographystyle{aasjournal}

\end{document}